\documentclass[11pt,onecolumn,draftclsnofoot,oneside,letterpaper]{IEEEtran}

\makeatletter

\let\proof\@undefined
\let\endproof\@undefined
\makeatother

\makeatletter

\let\labelindent\@undefined
\makeatother

\usepackage[dvips]{graphics}
\usepackage[dvips]{color}
\usepackage{amssymb}
\usepackage{amsmath}
\usepackage{amsthm}
\usepackage{enumitem}
\usepackage{url}
\usepackage[dvips]{geometry}
\usepackage{setspace}

\theoremstyle{plain}
\newtheorem{theorem}{Theorem}[section]
\newtheorem{lemma}{Lemma}[section]
\newtheorem{prop}{Proposition}[section]
\newtheorem{coro}{Corollary}[section]

\theoremstyle{definition}
\newtheorem{remark}{Remark}[section]

\def\tr{\mathop{\rm tr}\nolimits}%
\def\var{\mathop{\rm var}\nolimits}%
\def\diag{\mathop{\rm diag}\nolimits}%
\def\Re{\mathop{\rm Re}\nolimits}%
\def\rank{\mathop{\rm rank}\nolimits}%
\def\essinf{\mathop{\rm ess\,inf}}%
\def\sgn{\mathop{\rm sgn}\nolimits}%
\def\erfc{\mathop{\rm erfc}\nolimits}%
\def\clin{\mathop{\rm clin}\nolimits}%
\def\cov{\mathop{\rm cov}\nolimits}%
\def\inff{\mathop{\rm \vphantom{p}inf}}%

\newcommand{\iid}{\text{ i.i.d. }}

\newcommand{\geql}{\succeq}

\newcommand{\Complex}{\mathbb{C}}
\newcommand{\Real}{\mathbb{R}}
\newcommand{\eit}{e^{i\theta}}
\newcommand{\dth}{\,\frac{d\theta}{2\pi}}
\newcommand{\intt}{\int_{-\pi}^{\pi}}
\newcommand{\half}{\frac{1}{2}}

\newenvironment{eqenumerate}
  {
  \setenumerate{topsep=.75em, partopsep=0em, leftmargin=3.6\parindent,
  rightmargin=0\parindent, labelsep=1em, itemsep=3pt}
   \begin{enumerate}
   \setenumerate{label=(\arabic*), leftmargin=*}
   \setcounter{enumi}{\value{equation}}}
  {\end{enumerate}\setcounter{equation}{\value{enumi}} }

\begin{document}

\title{Feedback Capacity of Stationary Gaussian Channels}

\author{Young-Han Kim%
\thanks{This research was supported in part by NSF Grant CCR-0311633.}
\\Stanford University
}
\maketitle

\begin{abstract}
The feedback capacity of additive stationary Gaussian noise channels
is characterized as the solution to a variational problem.  Toward
this end, it is proved that the optimal feedback coding scheme is
stationary.  When specialized to the first-order autoregressive
moving average noise spectrum, this variational characterization
yields a closed-form expression for the feedback capacity.  In
particular, this result shows that the celebrated Schalkwijk--Kailath
coding scheme achieves the feedback capacity for the first-order
autoregressive moving average Gaussian channel, positively answering a
long-standing open problem studied by Butman, Schalkwijk--Tiernan,
Wolfowitz, Ozarow, Ordentlich, Yang--Kav\v{c}i\'{c}--Tatikonda, and
others.  More generally, it is shown that a $k$-dimensional
generalization of the Schalkwijk--Kailath coding scheme achieves the
feedback capacity for any autoregressive moving average noise spectrum
of order $k$.  Simply put, the optimal transmitter iteratively refines
the receiver's knowledge of the intended message.
\end{abstract}


\section{Introduction}
\label{sec:intro}
We consider a communication scenario in which one wishes to
communicate a message index $W \in \{1,\ldots,2^{nR}\}$ over the additive
Gaussian noise channel $Y_i = X_i + Z_i,\enspace i=1,2,\ldots,$ where
the additive Gaussian noise process $\{Z_i\}_{i=1}^\infty$ is
stationary with $Z^n = (Z_1,\ldots,Z_n) \sim N_n(0, K_{Z,n})$ for each
$n = 1,2,\ldots.$ For block length $n$, we specify a $(2^{nR}, n)$
feedback code with codewords $X^n(W,Y^{n-1}) = (X_1(W), X_2(W,Y_1),
\ldots,$ $X_n(W,Y^{n-1})),\enspace W=1,\ldots,2^{nR},$ satisfying the
average power constraint
\[
\frac{1}{n} \sum_{i=1}^n E X_i^2(W,Y^{i-1}) \le P
\]
and decoding function $\hat{W}_n: \Real^n \to \{1,\ldots,2^{nR}\}.$
The probability of error $P_e^{(n)}$ is defined as
\begin{align*}
P_e^{(n)} &:= \frac{1}{2^{nR}} \sum_{w=1}^{2^{nR}}
\Pr\{\hat{W}_n(Y^n) \ne w | X^n = X^n(w, Y^{n-1}) \} \\
&= \Pr\{\hat{W}_n(Y^n) \ne W\}
\end{align*}
where the message $W$ is uniformly distributed over
$\{1,2,\ldots,2^{nR}\}$ and is independent of $Z^n$.  We say that the
rate $R$ is achievable if there exists a sequence of $(2^{nR}, n)$
codes with $P_e^{(n)} \to 0$ as $n\to\infty$.  The feedback capacity
$C_\textit{FB}$ is defined as the supremum of all achievable rates.  We
also consider the case in which there is no feedback, corresponding to
the codewords $X^n(W) = (X_1(W), \ldots, X_n(W))$ independent of the
previous channel outputs.  We define the nonfeedback capacity $C$, or
the capacity in short, in a manner similar to the feedback case.

It is well known that the nonfeedback capacity is characterized by
water-filling on the noise spectrum, which is arguably one of the most
beautiful results in information theory.  More specifically, the
capacity $C$ of the additive Gaussian noise channel $Y_i = X_i + Z_i,
\enspace i=1,2,\ldots,$ under the power constraint $P$, is given by
\begin{equation}
\label{eq:c}
C = \intt \half \log
\frac{\max\{S_Z(\eit), \lambda\}}{S_Z(\eit)}\dth
\end{equation}
where $S_Z(\eit)$ is the power spectral density of the stationary
noise process $\{Z_i\}_{i=1}^\infty$, i.e., the Radon-Nikodym
derivative of the spectral distribution of $\{Z_i\}_{i=1}^\infty$ (with
respect to Lebesgue measure), and the water level $\lambda$ is chosen
to satisfy
\begin{equation}
\label{eq:p}
P = \intt \max\{0, \lambda -
S_Z(\eit)\}\dth.
\end{equation}
Although \eqref{eq:c} and \eqref{eq:p} give only a parametric
characterization of the capacity $C(\lambda)$ under the power
constraint $P(\lambda)$ for each parameter $\lambda \ge 0$, this
solution is considered simple and elegant enough to be called
\emph{closed-form}.  Just like many other fundamental developments in
information theory, the idea of water-filling comes from
Shannon~\cite{Shannon1949}, although it is sometimes attributed to
Holsinger~\cite{Holsinger1964} or Ebert~\cite{Ebert1966}.

For the case of feedback, no such elegant solution exists.  Most
notably, Cover and Pombra~\cite{Cover--Pombra1989} characterized the $n$-block
feedback capacity $C_{\textit{FB},n}$ for arbitrary time-varying
Gaussian channels via the asymptotic equipartition property (AEP) for
arbitrary nonstationary nonergodic Gaussian processes as
\begin{equation}
\label{eq:cnfb}
C_{\textit{FB},n} = \max_{K_{\!V\!,n}, B_n} \half \log
\frac{\det(K_{V,n}\!+\!
(B_n\!+I)K_{Z,n}(B_n\!+I)')^{1/n}}{\det(K_{Z,n})^{1/n}}
\end{equation}
where the maximum is taken over all positive semidefinite matrices
$K_{V,n}$ and all strictly lower triangular $B_n$ of sizes $n\times n$
satisfying $\tr(K_{V,n} + B_n K_{Z,n} (B_n)') \le nP$.  
When
specialized to a stationary noise process, the Cover--Pombra
characterization gives the feedback capacity as a limiting expression
\begin{align}
\nonumber
C_\textit{FB} &= \lim_{n\to\infty} C_{\textit{FB}, n}\\
&= \lim_{n\to\infty} \max_{K_{\!V\!,n}, B_n} \half \log
\frac{\det(K_{V,n} +
(B_n+I_n)K_{Z,n}(B_n+I_n)')^{1/n}}{\det(K_{Z,n})^{1/n}}.
\label{eq:cfb}
\end{align}

Despite its generality, the Cover--Pombra formulation of the feedback
capacity falls short of what we can call a closed-form solution.  It
is very difficult, if not impossible, to obtain an analytic expression
for the optimal $(K_{V,n}^{\star}, B_n^{\star})$ in \eqref{eq:cnfb}
for each $n$.  Furthermore, the sequence of optimal
$\{K_{V,n}^{\star}, B_n^{\star}\}_{n=1}^\infty$ is not necessarily
consistent, that is, $(K_{V,n}^{\star}, B_n^{\star})$ is not
necessarily a subblock of $(K_{V,n+1}^{\star}, B_{n+1}^{\star})$.
Hence the characterization \eqref{eq:cnfb} in itself does not give
much hint on the structure of optimal $\{K_{V,n}^{\star},
B_n^{\star}\}_{n=1}^\infty$ achieving $C_{\textit{FB},n}$, or more
importantly, its limiting behavior.

In this paper, we make one step forward by first 
characterizing the Gaussian feedback capacity $C_\textit{FB}$
in Theorem~\ref{thm:fb-capacity} as
\begin{equation}
\label{eq:cfb_var}
C_\textit{FB} = \sup_{S_V(\eit), B(\eit)}
\intt \half
\log
\frac{S_V(\eit) + |1 + B(\eit)|^2 S_Z(\eit)}{S_Z(\eit)}\dth
\end{equation}
where $S_Z(\eit)$ is the power spectral density of the noise process
$\{Z_i\}_{i=1}^\infty$ and the supremum is taken over all power
spectral densities $S_V(\eit) \ge 0$ and all strictly causal finite
impulse response filters $B(\eit) = \sum_{k=1}^m b_k e^{ik\theta}$
satisfying the power constraint 
$\intt (S_V(\eit) + |B(\eit)|^2
S_Z(\eit)) \dth \le P.
$
Roughly speaking, this characterization shows the asymptotic
optimality of the stationary solution $(K_{V,n}^\star, B_n^\star)$ in
\eqref{eq:cnfb} and hence it can be viewed as the justification for
interchange of the order of limit and maximum in \eqref{eq:cfb}.

Since our characterization is in a variational form, we will
subsequently find in Propositions~\ref{prop:cfb-necessary} and
\ref{prop:fb-var-suff} necessary and sufficient conditions for the
optimal $(S_V^{\star}(\eit), B^{\star}(\eit))$ from Lagrange duality
theory and additional information theoretic arguments.  This result,
when specialized to the first-order autoregressive (AR) noise spectrum
$ S_Z(\eit) = |1 + \beta\eit|^{-2}, -1<\beta<1,$ yields a closed-form
expression for feedback capacity as
\[
C_\textit{FB} = - \log x_0
\]
where $x_0$ is the unique positive root of the fourth-order polynomial
\[
P\,x^2 = \frac{(1-x^2)}{(1+|\beta|x)^2},
\]
establishing the long-standing conjecture by Butman~\cite{Butman1969,
Butman1976}, Tiernan--Schalkwijk \cite{Tiernan--Schalkwijk1974,
Tiernan1976}, and Wolfowitz~\cite{Wolfowitz1975}.  In fact, we will
obtain an explicit feedback capacity formula for the first-order
autoregressive moving average (ARMA) noise spectrum in
Theorem~\ref{thm:arma1}, which generalizes the result in
\cite{Kim2004} and confirms a recent conjecture by Yang,
Kav\v{c}i\'{c}, and Tatikonda~\cite{Yang--Kavcic--Tatikonda2004}.  As
we will see later, our result shows that the celebrated
Schalkwijk--Kailath coding scheme~\cite{Schalkwijk--Kailath1966,
Schalkwijk1966} achieves the feedback capacity.  

More generally, we will show in Theorem~\ref{thm:armak-capacity} that
a $k$-dimensional generalization of the Schalkwijk Kailath coding
scheme achieves the feedback capacity for any autoregressive moving
average noise spectrum of order $k$.

The literature on Gaussian feedback capacity is vast.  Instead of
trying to be complete, we sample the results that are closely related
to our discussion.  A more complete survey can be found in
\cite{Kim2004}.  The standard literature on the Gaussian feedback
channel and associated simple feedback coding schemes traces back to
Elias's 1956 paper~\cite{Elias1956} and its sequels~\cite{Elias1961,
Elias1967}.  Schalkwijk and Kailath~\cite{Schalkwijk--Kailath1966,
Schalkwijk1966} made a major breakthrough by showing that a simple
linear feedback coding scheme achieves the feedback capacity of the
additive white Gaussian noise channel with doubly exponentially
decreasing probability of decoding error.  More specifically, the
transmitter sends a real-valued information bearing signal at the
beginning of communication and subsequently refines the receiver's
knowledge by sending the error of the receiver's estimate of the
message.  This simple coding scheme, or no coding in a sense, achieves
the capacity of the Gaussian channel and the resulting error
probability of the maximum likelihood decoding decays
doubly-exponentially in the duration of the communication.  This
fascinating result has been extended in many directions, for example,
by Pinsker~\cite{Pinsker1968}, Omura~\cite{Omura1968},
Wyner~\cite{Wyner1968}, Schalkwijk~\cite{Schalkwijk1968},
Kramer~\cite{Kramer1969}, Zigangirov~\cite{Zigangirov1970}, Schalkwijk
and Barron~\cite{Schalkwijk--Barron1971}, and Ozarow and
Leung-Yan-Cheong~\cite{Ozarow--Leung-Yan-Cheong1984, Ozarow1984}.

Following these results on the white Gaussian noise channel, the focus
naturally shifted to the feedback capacity of the nonwhite Gaussian
noise channel.  Butman~\cite{Butman1969, Butman1976} extended the
Schalkwijk--Kailath coding scheme to autoregressive noise channels.
Subsequently, Tiernan and Schalkwijk~\cite{Tiernan--Schalkwijk1974,
Tiernan1976}, Wolfowitz~\cite{Wolfowitz1975}, and
Ozarow~\cite{Ozarow1990a, Ozarow1990b} studied the feedback capacity
of finite-order autoregressive moving average additive Gaussian noise
channels and obtained many interesting upper and lower bounds.
Recently, Yang, Kav\v{c}i\'{c}, and
Tatikonda~\cite{Yang--Kavcic--Tatikonda2004} (see also Yang's
thesis~\cite{Yang2004}) revived the control-theoretic approach
(\text{cf.} Omura~\cite{Omura1968}, Tiernan and
Schalkwijk~\cite{Tiernan--Schalkwijk1974}) to the finite-order
autoregressive moving average Gaussian feedback capacity problem.
After reformulating the feedback capacity problem as a stochastic
control problem, Yang et al.\@ used dynamic programming for the
numerical computation of $C_{\textit{FB},n}$ and offered a conjecture
that $C_\textit{FB}$ can be characterized as a solution of another
maximization problem, the size of which depends only on the order of
the noise process.

With a more general line of attack, Cover and
Pombra~\cite{Cover--Pombra1989} obtained the $n$-block capacity for
the arbitrary nonwhite Gaussian channel with or without feedback,
using an AEP theorem for nonstationary nonergodic Gaussian processes.
(Recall \eqref{eq:cnfb} for the feedback case; the nonfeedback case
corresponds to taking $B \equiv 0$.)  They also showed that feedback
does not increase the capacity much; namely, feedback at most doubles
the capacity (a result obtained by Pinsker~\cite{Pinsker1969} and
Ebert~\cite{Ebert1970}), and feedback increases the capacity at most
by half a bit.  The extensions and refinements of the Cover--Pombra
result abound.  Ihara obtained a coding theorem for continuous-time
Gaussian channels with feedback~\cite{Ihara1980, Ihara1994} and showed
that the factor-of-two bound on the feedback capacity is tight by
considering cleverly constructed nonstationary channels for both
discrete~\cite{Ihara1988} and continuous~\cite{Ihara1990} cases.
Dembo~\cite{Dembo1989} studied the upper bounds on $C_{\textit{FB},n}$
and showed that feedback does not increase the capacity at very low
signal-to-noise ratio or very high signal-to-noise ratio. (See
Ozarow~\cite{Ozarow1990a} for a minor technical condition on the
result for very low signal-to-noise ratio.)
Ordentlich~\cite{Ordentlich1994} examined the properties of the
optimal solution $(K_{V,n}, B_n)$ for $C_{\textit{FB},n}$ in
\eqref{eq:cnfb} for a fixed $n$ and showed that the optimal $K_{V,n}$
water-fills the new noise spectrum $(I_n+B_n) K_{Z,n} (I_n + B_n)'$
and that the optimal filter $B_n$ makes the input signal orthogonal to
the past output.  Based on these two crucial observations, he also
found that the optimal $K_{V,n}$ has rank $k$ for moving average noise
processes of order $k$.  Yanagi and Chen~\cite{Yanagi1994,
Chen--Yanagi1999, Chen--Yanagi2000} studied Cover's
conjecture~\cite{Cover1987} that the feedback capacity is at most as
large as the non-feedback capacity under twice the power, and also
made several refinements on the upper bounds by Cover and Pombra.
Recently a counterexample to Cover's conjecture was found by the
author~\cite{Kim2006}.  Thomas~\cite{Thomas1987}, Pombra and
Cover~\cite{Pombra--Cover1994}, and Ordentlich~\cite{Ordentlich1996}
extended the factor-of-two bound result to the colored Gaussian
multiple access channels with feedback.

Despite many developments on the nonwhite Gaussian channels, the exact
characterization of the feedback capacity has been open, even for
simple special cases.  In~\cite{Kim2004}, the author obtained the closed-form
capacity formula for the special case in which the noise process has
the first-order moving average spectrum, establishing the feedback
capacity for the first time.  Thanks to the special structure of the
noise spectrum, the maximization problem in \eqref{eq:cnfb} can be
solved analytically under the modified power constraint on each input
signal $X_i,$ $i=1,2,\ldots.$ Then, a fixed-point theorem exploiting
the convexity of the problem is deployed to show the asymptotic
optimality of the uniform power allocation over time.  This result
confirms the common belief that the stationary Schalkwijk--Kailath
linear coding scheme achieves the feedback capacity.  A similar
argument also shows that the uniform power allocation is
asymptotically optimal for the Schalkwijk--Kailath coding scheme if
the noise process has the first-order \emph{autoregressive} spectrum.

Our approach in this paper is different from the one taken in
\cite{Kim2004} and is geared towards the general case.  As is hinted
in the similarity between the Cover--Pombra characterization of the
Gaussian feedback capacity in~\eqref{eq:cfb} and the variational
characterization~\eqref{eq:cfb_var}, our development starts from the
$n$-block capacity formula~\eqref{eq:cnfb}.  The variational
formula~\eqref{eq:cfb_var}, however, certainly has the flavor of
spectral analysis, in the context of which we will derive properties
of the optimal solution $(S_V^{\star}(\eit), B^{\star}(\eit))$.  This
optimal solution will be then linked to the asymptotic behavior of the
linear coding scheme by Schalkwijk and Kailath, and its generalization by
Butman.  Thus in a sense our development goes in a full circle through
the literature cited above.

We will make parallel developments of both nonfeedback and feedback
cases, especially because the well-trodden nonfeedback capacity
problem provides a test bed for new techniques.  Hence, we revisit the
Gaussian nonfeedback capacity problem in Section~\ref{sec:nonfeedback}
and derive the water-filling capacity formula~\eqref{eq:c} in a rather
nontraditional manner.  In Section~\ref{sec:feedback}, we go through
similar steps for the feedback case to establish~\eqref{eq:cfb_var}.
Naturally, we will encounter a few technical difficulties that do not
arise in the nonfeedback case.  Section~\ref{sec:optimal} deals with
sufficient and necessary conditions on the optimal solution
$(S_V^\star, B^\star)$ to the variational problem~\eqref{eq:cfb_var}.
As a corollary of this result, we obtain the closed-form feedback
capacity formula for the first-order ARMA Gaussian channel.  We will
then interpret this result in the context of the Schalkwijk--Kailath
coding scheme.  We will also discuss the general finite-order ARMA
channels in Section~\ref{sec:armak}. The next section recalls
necessary results from various branches of mathematics.

\vspace*{2em}
\section{Mathematical Preliminaries}
\label{sec:prelim}
\subsection{Toeplitz Matrices, Szeg\H{o}'s Limit Theorem, and Entropy Rate}
\label{subsec:toeplitz}



We first review a few important results on spectral properties of
stationary Gaussian processes, which we will use heavily for the
variational characterization of feedback capacity.

Let $R(k) = R(-k) = E Z_1 Z_{k+1}, \enspace k = 0, 1, 2,\ldots,$ be
the covariance sequence of a stationary Gaussian process
$\{Z_i\}_{i=1}^\infty$.  Then, as the elegant answer to the classical
trigonometric moment problem shows (see, for example,
Akhiezer~\cite{Akhiezer1965} and Landau~\cite{Landau1987a}), there
exists a positive measure $\mu$ on $[-\pi,\pi)$, sometimes called the
\emph{power spectral distribution} of the process
$\{Z_i\}_{i=1}^\infty$, such that
\[
R(k) = \frac{1}{2\pi}\intt e^{-ik\theta} d\mu(\theta)
\]
for all $k$.  From the Lebesgue decomposition theorem, we can write
$\mu$ as a sum $\mu = \mu_{ac} + \mu_s$, where $\mu_{ac}$ is
absolutely continuous with respect to Lebesgue measure and $\mu_s$ is
singular.  The Radon-Nikodym derivative of $\mu_{ac}$ (with respect to
Lebesgue measure), called the \emph{power spectral density} of
$\{Z_i\}_{i=1}^\infty$, exists almost everywhere and can be written as
a function of $\eit$, or more specifically, we have $d\mu_{ac} =
S(\eit)d\theta = \Re F(\eit)d\theta$ for some function $F(z)$ analytic
on the unit disc $\mathbb{D} = \{z \in \Complex: |z| < 1\}$ with $F(0)
> 0$ and $\Re F(z) > 0$ on $\mathbb{D}$.

Conversely, given a nontrivial (i.e., supported by infinitely many
points) positive measure $d\mu = S(\eit)d\theta + d\mu_s$, the
Toeplitz matrix $K_n$ of size $n\times n$ given by
\[
K_n(j,k) = 
\frac{1}{2\pi}\intt e^{-i(k-j)\theta} {d\mu(\theta)},\qquad 1\le j,k \le n
\]
is positive definite Hermitian.  Hence, $K_n$ has $n$ positive
eigenvalues $\lambda_1(K_n), \ldots, \lambda_n(K_n)$, counting
multiplicity.  In his famous limit theorem~\cite{Szego1915,
Szego1920-1921}, Szeg\H{o} proved an elegant relationship between the
asymptotic behavior of the eigenvalues of $K_n$ and the associated
spectral distribution $\mu$.  This result lies at the heart of many
different fields, including operator theory, time-series analysis,
quantum mechanics, approximation theory, and, of course, information
theory.  Here we recall a fairly general version of Szeg\H{o}'s limit
theorem, which can be found in Simon~\cite[Theorem 2.7.13]{Simon2005}.

\begin{lemma}[Szeg\H{o}'s Limit Theorem]
\label{lemma:szego}
Let $f$ be a continuous function on $[0,\infty)$ such that
\[
\lim_{x\to\infty} \frac{f(x)}{x} = c < \infty.
\]
Then,
\[
\lim_{n\to\infty} \frac{1}{n} \sum_{i=1}^n f(\lambda_i(K_n))
= \intt f(S(\eit)) \dth + \frac{c}{2\pi} \intt d\mu_s(\theta).
\]
\end{lemma}

The above limit theorem is sometimes called the first Szeg\H{o}
theorem, in order to be distinguished from the second-order
asymptotics often called the \emph{strong Szeg\H{o} theorem} and
obtained by Szeg\H{o} himself after a 38-year gap~\cite{Szego1952}.
Refer to Grenander and Szeg\H{o}~\cite[Chapter
5]{Grenander--Szego1958}, B\"ottcher and Silbermann~\cite[Chapter
5]{Bottcher--Silbermann1999}, Gray~\cite{Gray2002}, and Barry Simon's
recent two-part tome on orthogonal polynomials on the unit
circle~\cite{Simon2005} for different flavors of Szeg\H{o}'s theorem
under different levels of generality.

As a canonical application of Szeg\H{o}'s limit theorem, the
following variational statement, attributed to Szeg\H{o},
Kolmogorov~\cite{Kolmogorov1941}, and Krein~\cite{Krein1945a,
Krein1945b}, connects the entropy rate, the spectral distribution, and
the minimum mean-square prediction error of a stationary Gaussian
process.

\begin{lemma}[Szeg\H{o}--Kolmogorov--Krein Theorem]
\label{lemma:kolmo}
Let $\{Z_i\}_{i=-\infty}^\infty$ be a stationary Gaussian process with
a nontrivial spectral distribution $d\mu = S(\eit)d\theta + d\mu_s$.
Then the minimum mean-squared prediction error $E_\infty = E (Z_0 -
E(Z_0|Z_{-\infty}^{-1}))^2$ of $Z_0$ from the entire past
$Z_k,$ $k < 0,$ is given by
\begin{align*}
\nonumber
E_\infty 
&= \inf_{\{a_k\}} \frac{1}{2\pi} \intt \big|1-\sum_{k=1}^\infty a_k e^{ik\theta}
\big|^2 {\,d\mu(\theta)}\\
\nonumber
&= \exp \bigg(\intt \log S(\eit) \dth\bigg)\\
&= \frac{1}{2\pi e} e^{2h(\mathcal{Z})}
\end{align*}
where $h(\mathcal{Z}) = \lim_{n\to\infty} n^{-1} h(Z_1,\ldots,Z_n)$
denotes the differential entropy rate of the process $\{Z_i\}$.
\end{lemma}

The proof of this result follows almost immediately from Szeg\H{o}'s
limit theorem with $f(x) = \log x$.  Note that the prediction error
depends only on the absolutely continuous part of the spectral
measure; this is no surprise for us, since $\lim_{x\to\infty} (\log
x)/x = 0$.  (The fact that the prediction error is independent of the
singular part of the spectral distribution can be also proved from
somewhat deeper results on shift operators and Wold--Kolmogorov
decomposition.  See, for example, Nikolski~\cite{Nikolski2002} and 
references therein.)  We stress the relationship between the entropy
rate of a stationary Gaussian process $\{Z_i\}$ and its spectral
density $S(\eit)$ in the following familiar expression:
\begin{equation}
\label{eq:kolmo}
h(\mathcal{Z}) = \intt \half \log(2\pi e S(\eit)) \dth.
\end{equation}

Throughout this paper, in order to exclude the trivial case of
unbounded capacity, we will assume that the power spectral
distribution $\mu$ of the additive Gaussian noise process
$\{Z_i\}_{i=1}^\infty$ is nontrivial (equivalently, $K_n$ is positive
definite for all $n$), and that the power spectral density $S_Z(\eit)$
satisfies the so-called \emph{Paley--Wiener condition}:
\begin{equation}
\label{eq:paley-wiener}
\intt |\log S_Z(\eit)| \dth < \infty,
\end{equation}
which is equivalent to having prediction error $E_\infty > 0$.  Unless
noted otherwise, we will also assume that the power spectral
distribution $\mu$ of the noise process has an absolutely continuous
part only, i.e., $\mu_s = 0$, which is justified in part by
Szeg\H{o}--Kolmogorov--Krein theorem (i.e., we can filter out the
deterministic part of the noise to arbitrary accuracy by sending a
pilot sequence) and in part by \emph{physical reality} (i.e., the
mathematical model of the singular noise spectrum may have no
counterpart in physical communication systems; see, for example,
Slepian's Shannon Lecture~\cite{Slepian1976}).

\subsection{Hardy Spaces, Causality, and Spectral Factorization}
We review some elementary results on Hardy spaces (see, for example,
Duren~\cite{Duren1970}, Koosis~\cite{Koosis1980}, Rudin~\cite[Chapter
17]{Rudin1987}) that are needed for analysis of optimal feedback
filters.  Our exposition loosely follows two monographs by
Partington~\cite{Partington1997, Partington2004}.

Let $f(z) = \sum_{n=0}^\infty c_n z^n$ be an analytic function on
$\mathbb{D} = \{z\in\mathbb{C}: |z| < 1\}$.  We say that $f(z)$
belongs to the class $H_p,$ $1 \le p < \infty$, if
\[
\|f\|_{H_p} = \left(\intt |f(r \eit)|^p \dth\right)^{1/p}
\]
is bounded for all $r < 1$.  Similarly we say that $f(z)$ belongs to
the class $H_\infty$ if 
\[
\|f\|_{H_\infty} = \sup_{|z| < 1} |f(z)|
\]
is bounded.  We can easily check that $H_p$ is a Banach space for $1
\le p \le \infty$.

It is well-known that $f \in H_p$ can be extended to $\mathbb{T} =
\partial\mathbb{D} = \{z\in\mathbb{C}: |z| = 1\}$ by taking the
pointwise radial limit
\[
\tilde{f}(\eit) = \lim_{r \uparrow 1} f(r\eit)
\]
which exists for almost all $\theta$.  The extended function
$\tilde{f}$ belongs to the standard Lebesgue space $L_p$ on
$\mathbb{R}/[-\pi,\pi) \simeq \mathbb{T}$ with the same norm
$\|\tilde{f}\|_p = \|f\|_{H_p}$, so that we can consider $H_p$ as a
closed (and thus complete) subspace of $L_p$.  Therefore, we will
identify $f \in H_p$ with its radial extension $\tilde{f} \in L_p$ and
use the same symbol $f$ for both $f$ and $\tilde{f}$ throughout.  More
specifically, when we say that a function $f(\eit)$ for $\theta \in
[-\pi, \pi)$ belongs to $H_p$, we implicitly mean that $f(z)$ is also
well-defined and analytic on $\mathbb{D}$.  Also we will use $f(z)$
and $f(\eit)$ interchangeably if the context is clear.  Recall the
following set inclusion relationship between important classes of
functions on $\mathbb{T}$:
\begin{align*}
H_p &\subset  L_p, \qquad 1 \le p \le \infty,\\
H_\infty &\subset  H_2 \subset H_1,\\
\intertext{and}
L_\infty &\subset  L_2 \subset L_1.
\end{align*}

Let $f(\eit) \in L_p, \enspace 1 \le p \le \infty$. We say that $f$ is
\emph{causal} if its Fourier coefficients
\[
c_n = \intt f(\eit) e^{-i n \theta} \dth,\qquad n = 0, \pm 1, \pm 2, \ldots,
\]
satisfy $c_n = 0$ for $n < 0.$ We also say that $f$ is \emph{strictly
causal} if $c_n = 0$ for $n \le 0$, or equivalently, $f(z) = z g(z)$
for some causal $g \in L_p$.  By reversing the direction of the time
index, we also define \emph{anticausality} and \emph{strict
anticausality} in a similar way.

If $f \in H_p$, then $f$ can be easily shown to be causal.  (See
Lemma~\ref{lemma:causality} below.)  Conversely, if $f \in L_p$ is
causal, then $\sup_n |c_n| < \infty$ so that $f$ is analytic on
$\mathbb{D}$ with
\begin{equation}
\label{eq:f-conv}
f(z) = \sum_{n=0}^\infty c_n z^n,
\end{equation}
where the series on the right-hand side converges pointwise on
$\mathbb{D}$.  Therefore, we can identify the class $H_p$ with the
class of causal $L_p$ functions, which gives an alternative definition
of the $H_p$ space.

When $f \in H_\infty$, we have the pointwise convergence of the
infinite series in \eqref{eq:f-conv} on $\mathbb{T} = \{e^{i\theta}:
\theta \in [-\pi,\pi)\}$ for almost all $\theta$.  Hence, $f \in
H_\infty$ preserves the causality when acting on $L_1$ by
multiplication.  For later use, we stress this simple fact in the
following statement, the proof of which easily follows from the
dominated convergence theorem.
\begin{lemma}
\label{lemma:causality}
Let $f \in H_\infty$ and let $g \in L_1$ be causal.  Then, $fg \in
L_1$ is causal.  If, in addition, $f$ is strictly causal, then $fg \in
L_1$ is strictly causal and 
\[
\intt f(\eit)g(\eit) \dth = 0.
\] 
\end{lemma}

We recall a few important factorization theorems.  The first set of
results deals with the factorization of $H_p$ functions.  Suppose
$f(\eit) \in H_p,\enspace 1 \le p \le \infty,$ is not identically
zero.  Then, $f$ has a factorization $f(z) = g(z)u(z)$ that is unique
up to a constant of modulus 1, where $g(z)$ is an inner function
(i.e., $g(z)$ is an $H_\infty$ function with $g(\eit) = 1$ almost
everywhere) and $u(z)$ is an $H_p$ outer function given by
\[
u(z) = \exp\left(\intt \frac{\eit + z}{\eit - z} \log |f(\eit)|
\dth\right).
\]
Consequently, the zeros of $f$ (inside the unit circle) coincide with
the zeros of $g$, and $\|f\|_p = \|u\|_p$.

We define the \emph{(infinite) Blaschke product} $b(z)$ formed with the
zeros of $f(z)$ as
\[
b(z) = z^k \prod_{|z_n| \ne 0} \frac{\overline{z}_n}{|z_n|}
\frac{z_n - z}{1-\overline{z}_n z},
\]
where $\{z_n\}$ are the zeros of $f$, listed according to their
multiplicity, $k$ of them being at $0$.  It is easy to check that
$b(z)$ is well-defined in the sense that $b(z)$ converges uniformly on
compact sets to an $H_\infty$ function.  Also, $b(z) \le 1$ and
$|b(\eit)| = 1$ almost everywhere.  As a refinement of the above
{inner-outer factorization theorem}, F.\@ Riesz showed that $f$ has a
factorization $f(z) = b(z)s(z)u(z)$ that is unique up to a constant of
modulus 1, where $b$ is the Blaschke product of the zeros of $f$, $s$
is a singular inner function (without zeros), and $u$ is an outer
function.  Again $\|f\|_p = \|u\|_p$.

For our purposes, it is more convenient to introduce a normalized
variant of the Blaschke product as
\[
\hat{b}(z) = z^k \prod_{|z_n| \ne 0} \frac{1 - z_n^{-1} z}{1 - z_n z}.
\]
Then, $|\hat{b}(\eit)| = \prod_{|z_n| \ne 0} (1/|z_n|)$ almost
everywhere.  This normalized Blaschke product is often called an
\emph{all-pass filter} in the signal processing literature if
$\{z_n\}$ is finite and $k=0$.

If $f \in H_2$ and $f(0) = 1$, then $f$ has the unique factorization
$f(z) = \hat{b}(z)\hat{u}(z)$ where $\hat{b}(z)$ is the normalized
Blaschke product formed with zeros $\{z_n\}$ of $f$ and $\hat{u}(z)$
does not have any zero inside the unit circle.  In particular,
$\hat{b}(0) = \hat{u}(0) = 1$.  Now Jensen's formula states that, if
$g(z) \in H_2$ with $g(0) = 1,$ then
\[
\intt \log |g(r\eit)| \dth = \log \prod_{k=1}^p \frac{r}{|\alpha_k|}
\]
where $\alpha_1,\ldots,\alpha_p$ denote the zeros of $g(z)$ within the
circle of radius $r$.  Therefore,
\begin{equation}
\label{eq:jensen}
\intt \log |f(\eit)| \dth
= \intt \log |\hat{b}(\eit)| \dth + \intt \log |\hat{u}(\eit)| \dth
= \log \prod \frac{1}{|z_n|}.
\end{equation}
As a trivial corollary, if $f$ is rational of the form
\[
f(z) = \frac{P(z)}{Q(z)} = 
\frac{1+\sum_{n=1}^k p_n z^n}{1+\sum_{n=1}^k q_n z^n}
= \frac{\prod (1 - \beta_n^{-1} z)}{\prod (1 - \gamma_n^{-1} z)}
\]
with all zeros $\{\gamma_n\}$ of $Q(z)$ strictly outside the unit
circle, then
\[
\intt \log |f(\eit)| \dth = \log \prod_{j=1}^p \frac{1}{|\beta_j|}
\]
where $\beta_1,\ldots, \beta_p$ denote the zeros of $P(z)$ inside the
unit circle.

Our last factorization theorem is concerned with the factorization of
positive $L_1$ functions and is usually called the \emph{canonical
factorization theorem}.  Suppose $f(\eit) \in L_1$.  Then, $f(\eit) =
|g(\eit)|^2$ for some $g(\eit) \in H_2$ if and only if $f(\eit) \ge 0$
almost everywhere and the Paley--Wiener condition
\eqref{eq:paley-wiener} is satisfied.  In the light of the
aforementioned factorization theorem due to F.\@ Riesz, we can always
take the canonical factor $g$ with no zeros inside the unit circle and
$g(0) > 0$.

\subsection{Discrete Algebraic Riccati Equations}
\label{subsec:dare}
Discrete algebraic Riccati equations (DAREs) often play a crucial role
in many estimation and control problems.  Our problem is no exception,
especially the characterization of ARMA($k$) feedback capacity in
Section~\ref{sec:armak}.

Here we focus on a very special class of Riccati equations and review
a few properties of them.  Since the necessary results are somewhat
scattered in the literature, we also provide short proofs along with
probabilistic interpretations; some of these might be new.  Whenever
possible, however, we will refer to standard references.  For a more
general treatment, refer to Kailath, Sayed, and
Hassibi~\cite{Kailath--Sayed--Hassibi2000} and Lancaster and
Rodman~\cite{Lancaster--Rodman1995}.

Given matrices $F \in \mathbb{R}^{k\times k}$ and $H \in \mathbb{R}^{1
\times k}$, we study the following discrete algebraic Riccati equation:
\begin{equation}
\label{eq:dare0}
\Sigma = F \Sigma F' - \frac{(F \Sigma H') (F \Sigma H')'}
{1 + H \Sigma H'}.
\end{equation}
For each $k\times k$ Hermitian matrix $\Sigma$, define
\[
\Gamma = \Gamma(\Sigma) = \frac{F\Sigma H'}{1+H\Sigma H'}.
\]
We are concerned with solutions of \eqref{eq:dare0}, especially the
ones with stable $F - \Gamma H$.

\begin{lemma}[DARE]
Suppose $F$ has no unit-circle eigenvalue and $\{F,H\}$ is detectable,
that is, there exists $G \in \mathbb{R}^{1\times k}$ such that $F-GH$
is stable (i.e., every eigenvalue of $F-GH$ lies inside the unit
circle).  Then, the following statements hold.

\begin{enumerate}[label=(\roman*)]

\item $\Sigma \equiv 0$ is a solution to \eqref{eq:dare0}.

\item 
\label{property:uniqueness}
There is a unique solution $\Sigma = \Sigma_+$ to
\eqref{eq:dare0} such that $F - \Gamma H$ is stable.  Furthermore,
$\Sigma_+ \geql \Sigma$ for any other $\Sigma$ satisfying
\eqref{eq:dare0}.  In particular, $\Sigma_+$ is positive semidefinite.

\item If $F$ is invertible, then $F - \Gamma H$ is invertible 
for each solution $\Sigma$ and
\[
1 + H \Sigma H' = \frac{\det(F)}{\det(F-\Gamma H)}.
\]
\label{property:entropy}

\item Let $\Gamma_{\!+} = \Gamma(\Sigma_+)$.  If $F$ has eigenvalues
$\lambda_1, \ldots, \lambda_k$ with $|\lambda_1| \ge \ldots
|\lambda_j| > 1 > |\lambda_{j+1}| \ge \ldots |\lambda_k|$, then $F -
\Gamma_{\!+} H$ has eigenvalues $1/\lambda_1, \ldots, 1/\lambda_j,
\lambda_{j+1}, \ldots, \lambda_k$.
\label{property:zero-canc}

\item If every eigenvalue of $F$ lies inside the unit circle, then the
stabilizing solution $\Sigma_+$ is identically zero.  Thus, $\Sigma_+
= 0$ is the unique positive semidefinite solution to \eqref{eq:dare0}.

\item If every eigenvalue of $F$ lies outside the unit circle, then
$\Sigma_+ \succ 0$.

\item \label{property:rank}
More generally, suppose $F$ has $j$ eigenvalues outside the unit
circle and $k-j$ eigenvalues inside the unit circle.  Then,
$\rank(\Sigma_+) = j.$
\end{enumerate}
\label{lemma:dare}
\end{lemma}

\begin{proof}

\begin{enumerate}[label=(\roman*)]
\item Trivial.

\item Refer to \cite[Theorem E.5.1]{Kailath--Sayed--Hassibi2000}.

\item Note that
$\det(1 + H\Sigma H') = \det(I+\Sigma H' H).
$
Now simple algebra reveals that 
$
(F-\Gamma H)(I + \Sigma H' H) = F.
$

\item For simplicity, we assume that $F$ is invertible.
We can easily check that
\[
\left[
\begin{matrix}
F^{-1} & 0\\
-H'HF^{-1} & F'
\end{matrix}
\right]
=
\left[
\begin{matrix}
I & \Sigma \\
0 & I
\end{matrix}
\right]
\left[
\begin{matrix}
(F-\Gamma(\Sigma)H)^{-1} & 0\\
-H'HF^{-1} & (F-\Gamma(\Sigma)H)'
\end{matrix}
\right]
\left[
\begin{matrix}
I & \Sigma \\
0 & I
\end{matrix}
\right]^{-1}
\]
for any solution $\Sigma$, which implies that the eigenvalues of $\{
(F-\Gamma H)', (F - \Gamma H)^{-1}\}$ coincides with those of $\{F',
F^{-1}\}$. Now the desired result follows from the fact that $F -
\Gamma(\Sigma_+)$ is stable.

\item Refer to \cite[Theorem E.6.1]{Kailath--Sayed--Hassibi2000}.

\item Refer to \cite[Theorem E.6.2]{Kailath--Sayed--Hassibi2000}.

\item For simplicity, suppose $F$ can be diagonalized; the general
case can be proved by using the generalized eigenvectors associated
with the Jordan canonical form of $F$.  Take each
eigenvalue-eigenvector pair $(\lambda, x)$ of $F$ with $|\lambda| >
1$.  Suppose $x \Sigma_+ = 0$.  Then, we can easily check that $x (F-
\Gamma_{\!+} H) = x F = \lambda x,$ which violates the stability of $F -
\Gamma_{\!+} H$.  Thus, $x \Sigma_+ \ne 0,$ which implies $\rank(\Sigma_+)
\ge j$.

On the other hand, take each eigenvalue-eigenvector pair $(\lambda,
x)$ of $F$ with $|\lambda| < 1$.  From \eqref{eq:dare0}, we have
\[
x \Sigma_+ x' = |\lambda|^2 x \Sigma_+ x' - 
\frac{F \Sigma_+ H' H \Sigma_+ F'}{1+H\Sigma_+ H'},
\]
or equivalently,
\[
(1-|\lambda|^2) x \Sigma_+ x' +
\frac{F \Sigma_+ H' H \Sigma_+ F'}{1+H\Sigma_+ H'} = 0.
\]
Since both terms of the above sum are nonnegative, we must have $x
\Sigma_+ = 0$, which implies $\rank(\Sigma_+) \le j$.\qedhere
\end{enumerate}
\end{proof}

Algebraic Riccati equations naturally arise from asymptotic behaviors
of recursive filters (e.g., Kalman filters).  In the following lemma,
we collect a few results on the convergence of the Riccati recursion.
\begin{lemma}[Discrete Riccati recursion]
\label{lemma:riccati-recursion}
Under the same assumption on $\{F, H\}$ as in Lemma~\ref{lemma:dare},
suppose $\{\Sigma_n\}_{n=1}^\infty$ is defined as
\begin{equation}
\label{eq:riccati-recursion}
\Sigma_{n+1} = F \Sigma_n F' - \frac{(F \Sigma_n H') (F \Sigma_n H')'}
{1 + H \Sigma_n H'}
\end{equation}
for some $\Sigma_0$.  Then, the following statements hold:

\begin{enumerate}[label=(\roman*)]

\item If $\Sigma_0 = 0$, then $\Sigma_n = 0$ for all $n$.

\item If $\Sigma_0 \geql 0$, then $\Sigma_n \geql 0$ for all $n$.

\item 
If $\Sigma_0 \geql \tilde{\Sigma}_0 \geql 0$, then
$\Sigma_n \geql \tilde{\Sigma}_n \geql 0$ for all $n$.

\item If $\Sigma_0 \succ 0$, then $\Sigma_n \to \Sigma_+,$ where
$\Sigma_+ \geql 0$ is the unique stabilizing solution to the DARE
\eqref{eq:dare0}.
\label{property:convergence}

\end{enumerate}

\end{lemma}

\begin{proof}
\begin{enumerate}[label=(\roman*)]
\item Trivial.
\item
Write \eqref{eq:riccati-recursion} as
\[
\Sigma_{n+1} = (F-\Gamma(\Sigma_n))\Sigma_n (F-\Gamma(\Sigma_n))' +
\Gamma(\Sigma_n) \Gamma(\Sigma_n)'.
\]

\item
Refer to Caines~\cite[Theorem 3.5.1]{Caines1988}.

\item 
Let $\Pi \geql 0$ be the unique solution of
the Lyapunov equation
\begin{equation}
\label{eq:lyap}
\Pi = (F-\Gamma_{\!+}H)'\Pi(F-\Gamma_{\!+}H) + \frac{H' H}{1+H\Sigma_+ H'}.
\end{equation}
(Lemma~\ref{lemma:dare} guarantees the stability of $F-\Gamma_{\!+}H$
and hence there exists a unique positive semidefinite $\Pi$ satisfying
\eqref{eq:lyap}.)  Take any $\epsilon > 0 $ such that $\Sigma_0 \geql
\epsilon I$ and $I + (\epsilon I - \Sigma_+) \Pi$ is nonsingular.  Now
from Lemma 14.5.7 in \cite{Kailath--Sayed--Hassibi2000}, we have
\[
I + \bigl(\Pi^{1/2}\bigr)' (\Sigma_0 - \Sigma_+) \Pi^{1/2} \succ 
I + \bigl(\Pi^{1/2}\bigr)' (\epsilon I - \Sigma_+) \Pi^{1/2} \succ 0,
\]
which implies the exponential convergence of $\Sigma_n$ to $\Sigma_+$
by Theorem 14.5.2 in \cite{Kailath--Sayed--Hassibi2000}.\qedhere
\end{enumerate}
\end{proof}

Although our approach so far has been mostly algebraic, we can give
probabilistic interpretations to the above results in the context of
linear stochastic systems.  Since $\{F, H\}$ is detectable, we will
take some $G$ such that $F - GH$ is stable.  Consider the following
state-space representation (see, for example,
Kailath~\cite{Kailath1980}) of a stationary Gaussian process
$\{Y_n\}_{n=-\infty}^\infty$:
\begin{align}
\label{eq:state-y1}
\begin{array}{r@{\;}l}
S_{n+1} &= (F - GH) S_n - G U_n\\[.25em]
Y_n &= H S_n + U_n
\end{array}
\end{align}
where $\{U_n\}_{n=-\infty}^\infty$ are independent and identically
distributed zero-mean unit-variance Gaussian random variables, and the
\emph{state} $S_n$ is independent of $U_n$ for each $n$.  It is easy
to see that $\{Y_n\}_{n=-\infty}^\infty$ corresponds to the filter
output of the input process $\{U_n\}_{n=-\infty}^\infty$ through a
linear-time invariant filter with transfer function
\begin{equation}
\label{eq:transfer}
f(z) = \frac{\det(I - zF)}{\det(I - z(F-GH))}.
\end{equation}

Consider the state-space representation for the innovations
$\tilde{Y}_n = Y_n - E(Y_n|Y_{-\infty}^{n-1})$.  Write $\tilde{S}_n =
S_n - E(S_n|Y_{-\infty}^{n-1})$ and $\Sigma_+ =
\cov(S_n|Y_{-\infty}^{n-1}) = \cov(\tilde{S}_n)$.  Define $\Gamma_{\!+}
= \Gamma(\Sigma_+)$ as before.  Then, we can check through a little
algebra that
\begin{align*}
\begin{array}{r@{\;}l}
\tilde{S}_{n+1} &= (F - \Gamma_{\!+}H) \tilde{S}_n - \Gamma_{\!+} U_n\\[.25em]
\tilde{Y}_n &= {H} \tilde{S}_n + U_n
\end{array}
\end{align*}
which implies that
\begin{align*}
\Sigma_+ &= (F-\Gamma_{\!+} H) \Sigma_+  
(F-\Gamma_{\!+} H)' + \Gamma_{\!+}\Gamma_{\!+}'\\
&= F \Sigma_+ F' - \frac{(F\Sigma_+ H')(F\Sigma_+ H')'}
{1+H\Sigma_+ H'}.
\end{align*}
Clearly, there must be a unique solution $\Sigma_+$ to the above
equation that makes the above state-space representation well-defined;
this implies Lemma~\ref{lemma:dare}\ref{property:uniqueness}.

Note that $\{\tilde{Y}_n\}_{n=-\infty}^\infty$ is the output of
$\{U_n\}_{n=-\infty}^\infty$ via the filter
\[
g(z) = \frac{\det(I-zF)}{\det(I-z(F-\Gamma_{\!+}H))}.
\]
On the other hand, the innovations process
$\{\tilde{Y}_n\}_{n=-\infty}^\infty$ is white.  Therefore, $g(z)$
should be a normalized Blaschke product (all-pass filter), which
implies Lemma~\ref{lemma:dare}\ref{property:zero-canc}.  Furthermore,
since $\var(\tilde{Y}_n) = 1 + H\Sigma_+ H'$, applying Jensen's
formula, we have a stronger version of
Lemma~\ref{lemma:dare}\ref{property:entropy}.  The rank condition on
$\Sigma_+$ (Lemma~\ref{lemma:dare}\ref{property:rank}) can be viewed
as how many ``modes'' of the state can be causally determined by
observing the output.  Our development also gives a special case of
Szeg\H{o}--Kolmogorov--Krein theorem.  For example, if $F$ is
invertible,
\begin{align*}
h(\mathcal{Y}) &= \half \log (2\pi e (1 + H \Sigma_+ H')) \\
&= \half \log (2 \pi e) + \half \log \left(\frac{\det(F)}{\det(F - \Gamma_{\!+}H)}\right) \\ 
&= \half \log (2\pi e) + 
\half \intt \log \left| \frac{\det(I - \eit F)}{\det(I - \eit (F - GH))}
\right|^2 \dth,
\end{align*} 
where the last inequality can be justified by the inner-outer
factorization theorem and Jensen's formula.

Now we consider a slightly nonstationary Gaussian process
$\{Y'_n\}_{n=1}^\infty$, recursively defined with the same state-space
equation \eqref{eq:state-y1}, but under the initial condition $S_0 = 0$
and $U_0 = 0$.  Let $T_n$ denote the linear transformation from
$(U_1,\ldots, U_n)$ to $(Y'_1, \ldots, Y'_n)$ that corresponds to our
state-space model.  It is easy to see that $T_n$ is Toeplitz (with
respect to the natural basis on $(U_1,\ldots, U_n)$) and, in fact,
\begin{equation}
\label{eq:toeplitz}
T_n(j,k) = \intt f(\eit) e^{-i(j-k)\theta} \dth
\end{equation}
where $f(z)$ is the very transfer function in \eqref{eq:transfer}.
Since $T_n$ is lower triangular with diagonal entries equal to $1$ and
thus $\det(T_n) = 1$ for all $n$, the entropy rate of $\{Y'_n\}$ is
given as
\[
h(\mathcal{Y}') = 
\lim_{n\to\infty}\frac{h(Y'_1,\ldots, Y'_n)}{n} = 
\lim_{n \to\infty} \frac{h(U_1,\ldots, U_n)}{n}
 = h(\mathcal{U}) = \half \log (2\pi e),
\]
which is strictly less than the entropy rate $h(\mathcal{Y}) = \half
\log (2\pi e(1 + H\Sigma_+H'))$ of the stationary process $\{Y_n\}$
under the same state-representation \eqref{eq:state-y1}, provided that
$F$ has an eigenvalue outside the unit circle.

The nonzero gap between the entropy rate $h(\mathcal{Y})$ of the
stationary process $\{Y_n\}_{n=1}^\infty$ and the entropy rate
$h(\mathcal{Y}')$ of its nonstationary version $\{Y'_n\}_{n=1}^\infty$
can be understood from a beautiful result on Toeplitz operators by
Widom; see B\"ottcher and Silbermann~\cite[Proposition 1.12,
Proposition 2.12, and Example 5.1]{Bottcher--Silbermann1999}.  We use
the notation $T(f)$ to denote the Toeplitz operator associated with
\emph{symbol} $f$ as in \eqref{eq:toeplitz} and $T_n(f) \in
\mathbb{R}^{n\times n}$ to denote the finite truncation of $T(f)$.
Since the power spectral density of the stationary process $\{Y_n\}$
is $|f(\eit)|^2$, our previous discussion on Toeplitz matrices and the
trigonometric moment problem shows that the covariance matrix of
$(Y_1,\ldots,Y_n)$ is simply $T_n(|f|^2)$.  On the other hand, from
our construction of the nonstationary process $\{Y'_n\}$, the
covariance matrix of $(Y'_1, \ldots, Y'_n)$ is given as
$T_n(f)(T_n(f))'$.  Now Widom's Theorem shows
\[
T(|f|^2) = T(f)(T(f))' + (H(f))^2
\]
where $H = H(f)$ is the Hankel operator associated with symbol $f$ and
is given by 
\[H(j,k) = \intt f(\eit) e^{-i (j+k-1)\theta} \dth.\]  (This
result should not be confused with the Wiener--Hopf factorization
$T(|f|^2) = (T(f))'T(f)$; see \cite[Section
1.5]{Bottcher--Silbermann1999}.)  Thus, the Hankel adjustment term
$H^2(f)$ contributes to the strict gap between the entropy rates.  We
can represent $Y_n = Y'_n + V_n$ for some nonstationary process
$\{V_n\}$ with infinite covariance matrix $H^2(f)$ such that $\sum_n E
V_n^2 < \infty$.  Roughly speaking, the perturbation process $\{V_n\}$
with bounded total power causes a strict boost in entropy rate.
(Although our $f$ is rational, this phenomenon generalizes to any $f$
in Krein algebra, in which case $H^2(f)$ is a trace class
operator~\cite[Section 5.1]{Bottcher--Silbermann1999}.)

Finally we remark that our previous discussion on the Riccati
recursion implies a much stronger result on the boost of entropy rate
due to small perturbation.  Consider $Y''_n = Y'_n + V'_n$ where
$(V'_1,\ldots,V'_k)$ has a positive definite covariance matrix and
$V'_n \equiv 0$ for all $n > k$.
Lemma~\ref{lemma:riccati-recursion}\ref{property:convergence} shows
that the entropy rate of $\{Y''_n\}$ is $\half \log (2\pi e(1 + H
\Sigma_+ H'))$, and hence any tiny perturbation to the nonstationary
process results in the entropy rate of the stationary version.
Later,
this phonomenon gives an alternative interpretation of the role of
message-bearing signals in feedback communication.

The following example illustrates our point. 
Define $\{Y'_n\}_{n=1}^\infty$ as
\begin{align*}
Y'_1 &= U_1\\
Y'_n &= U_n + \alpha U_{n-1}\qquad n = 2,3,\ldots,
\end{align*}
where $\alpha$ is a constant with $|\alpha| > 1$.  Then, the entropy
rate of the process $\{Y'_i\}_{i=1}^\infty$ is $\half \log (2\pi e)$,
although $\{Y_2, Y_3, \ldots\}$ is stationary with entropy rate $\half
\log (2\pi e \alpha^2)$.  Now define $\{Y''_n\}_{n=1}^\infty$ as
\begin{align*}
Y''_1 &= U_1 + \epsilon V\\
Y''_n &= U_n + \alpha U_{n-1}\qquad n = 2,3,\ldots,
\end{align*}
where $\epsilon > 0$ is an arbitrary constant and $V \sim N(0,1)$ is
independent of $\{U_n\}_{n=1}^\infty$.  Then, the entropy rate of the
perturbed process is $\half \log (2 \pi e \alpha^2)$.  Evidently, the
entropy rate is discontinuous at $\epsilon = 0$ and any tiny
perturbation results in the same boost in the entropy rate.

\subsection{Matrix Inequalities}
We recall the following facts on positive semidefinite Hermitian
matrices.  Proofs can be found in standard references on matrix
analysis (see, for example, Gantmacher~\cite{Gantmacher1959} and Horn
and Johnson~\cite{Horn--Johnson1985}) or can be derived easily from
the related results therein.

\begin{lemma}
\label{lemma:psd}
Suppose a Hermitian matrix $K$ is partitioned as
\[
K = \left[
\begin{matrix}
A & B\\
B' & C
\end{matrix}
\right]
\]
where $A$ and $C$ are Hermitian.  Further suppose $C$ is positive
definite.  Then $K$ is positive semidefinite if and only if $A -
B' C^{-1} B$ is positive semidefinite.
\end{lemma}

\begin{lemma}
\label{lemma:logdet}
Suppose $K \in \Complex^{n\times n}$ is positive semidefinite
Hermitian.  Then, we have
\[
\log \det K \le \tr K - n,
\]
with equality if and only if $K = I_n$.
\end{lemma}

\begin{lemma}
\label{lemma:trace}
Suppose $K$ and $\tilde{K}$ are positive semidefinite Hermitian of the
same size.  Then, we have
\[
\tr (K \tilde{K}) \ge 0.
\]
Furthermore, the following statements are equivalent:
\begin{enumerate}[label=(\roman*)]
\item
$\tr (K \tilde{K}) = 0.$
\item
$K \tilde K = 0.$
\item
There exist a unitary matrix $Q$ and diagonal matrices $D$ and
$\tilde{D}$ such that $K = QDQ'$, $\tilde{K} = Q\tilde{D}Q'$,
and $D\tilde{D} = 0$.
\end{enumerate}
\end{lemma}

\vspace*{1em}
\section{Gaussian Nonfeedback Capacity Revisited}
\label{sec:nonfeedback}

Before we set off to a long discussion on the feedback capacity, we
revisit the (nonfeedback) capacity of a stationary Gaussian channel.
In particular, we give a detailed derivation of the water-filling
capacity formula
\begin{align*}
\tag{\ref{eq:c}}
C &= \intt \half \log
\frac{\max\{S_Z(\eit), \lambda\}}{S_Z(\eit)}\dth\\
P &= \intt \max\{0, \lambda -
S_Z(\eit)\}\dth.
\tag{\ref{eq:p}}
\end{align*}
This apparent digression will be rewarded in three ways.  First, we
will present an elementary proof of the capacity theorem that does not
rely on Szeg\H{o}'s theorem on the asymptotics of large Toeplitz
matrices, and hence is interesting on its own.  Secondly, the parallel
development of both feedback and nonfeedback capacities answers
interesting questions such as when feedback increases the capacity.
Thirdly and most importantly, the proof techniques developed for the
nonfeedback problem will be utilized heavily for the case of feedback
in the subsequent sections.

We start with the $n$-block capacity for the Gaussian channel in the
Cover--Pombra sense~\cite{Cover--Pombra1989}.  Define
\begin{equation}
C_n := \max_{K_{X,n}} \half \log \frac{\det(K_{X,n} +
K_{Z,n})^{1/n}}{\det(K_{Z,n})^{1/n}}
\label{eq:n-block-capacity}
\end{equation}
where the maximization is over all $n\times n$ positive semidefinite
symmetric matrices $K_{X,n}$ satisfying the power constraint $\tr
(K_{X,n}) \le nP.$ The coding theorem by Cover and Pombra
\cite[Theorem 1]{Cover--Pombra1989} states that
the rate $C_n$ is achievable, that is, for every $\epsilon > 0$, there
exists a sequence of $(2^{n(C_n-\epsilon)},n)$ codes with $P_e^{(n)}
\to 0$.  Conversely, for $\epsilon > 0$, any sequence of
$(2^{n(C_n+\epsilon)},n)$ codes has $P_e^{(n)}$ bounded away from zero
for all $n$.

The quantity $nC_n$ corresponds to the maximum mutual information
\begin{align*}
I(X^n;Y^n) &= h(Y^n) - h(Y^n|X^n)\\
&= h(Y^n) - h(X^n + Z^n|X^n)\\
&= h(Y^n) - h(Z^n|X^n)\\
&= h(Y^n) - h(Z^n)
\end{align*}
between the channel input $X^n$ and the channel output $Y^n = X^n +
Z^n$, maximized over all Gaussian inputs $X^n \sim N_n(0,K_{X,n})$
with $\tr(K_{X,n})\le nP.$ Since the Gaussian input distribution
maximizes the output entropy $h(Y^n)$ under a given covariance
constraint, $nC_n$ is the mutual information $I(X^n;Y^n)$ maximized
over all input distributions on $X^n$ satisfying the power constraint
$E\sum_{i=1}^nX_i^2 \le nP$.

Now from the stationarity of the noise process $\{Z_i\}_{i=1}^\infty$,
the $n$-block capacity $C_n$ is superadditive in the sense that
\[
mC_m + nC_n \le (m+n)C_{m+n}
\]
for all $m$ and $n$.  Indeed, if ${X}^{m+n} \sim N_{m+n}(0,
K_{X,m}^{\star}\!\oplus{K}_{X,n}^{\star})$ where
$K_{X,m}^{\star}\!\oplus{K}_{X,n}^{\star} = \diag(K_{X,m}^{\star},
{K}_{X,n}^{\star})$ denotes the direct sum of the matrices
$K_{X,m}^{\star}$ and ${K}_{X,n}^{\star}$ that achieve the capacity
for block sizes $m$ and $n$, respectively, under the power constraint
$P$, we have
\begin{align}
\nonumber
mC_m + nC_n
&= I(X_1^m;Y_1^m) + I(X_{m+1}^{m+n};Y_{m+1}^{m+n})\\
\nonumber
&= h(X_1^m) + h(X_{m+1}^{m+n}) 
- h(X_1^m | Y_1^m) - h(X_{m+1}^{m+n}|Y_{m+1}^{m+n})\\
\label{eq:cond-ineq}
&\le h(X_1^m, X_{m+1}^{m+n}) - h(X_1^m, X_{m+1}^{m+n}|Y_1^m, Y_{m+1}^{m+n})\\
\nonumber
&= I({X}^{m+n}; {Y}^{m+n})\\
\label{eq:power-ineq}
&\le (m+n)C_{m+n} 
\end{align}
where \eqref{eq:cond-ineq} follows from nonnegativity of mutual
information and the independence of $X_1^m$ and $X_{m+1}^{m+n}$, and
\eqref{eq:power-ineq} follows since $\tr(K_{X,m}^{\star}\!\oplus
K_{X,n}^{\star}) = \tr(K_{X,m}^\star) + \tr(K_{X,n}^\star) \le (m+n)P$
and thus $K_{X,m}^\star\!\oplus K_{X,n}^\star$ is a feasible solution
to the $(m+n)$-block capacity problem under the power constraint $P$.
Consequently, from a classical result in analysis (see, for example,
Poly\'{a} and Szeg\H{o}~\cite{Polya--Szego1998a}), the superadditivity
of $C_n$ implies that the limit of $C_n$ exists and $\lim_n C_n =
\sup_n C_n$.  Therefore, the capacity $C$ of the Gaussian channel $Y_i
= X_i + Z_i,$ $i=1,2,\ldots,$ is given by
\begin{align}
\nonumber
C &= \lim_{n\to\infty} C_n\\
&= \lim_{n\to\infty} \max_{\tr K_{X,n} \le nP} 
\half \log \frac{\det(K_{X,n} + K_{Z,n})^{1/n}}{\det(K_{Z,n})^{1/n}}.
\nonumber
\end{align}

In order to obtain the parametric characterization of capacity $C$ in
\eqref{eq:c} and \eqref{eq:p}, there is one more step that needs to be
taken.  In the classical approach, the optimization problem for $C_n$
is solved for each $n$ and then the limiting behavior of $C_n$ is
analyzed via Szeg\H{o}'s limit theorem.

For each fixed $n$, the optimization problem for $C_n$ in
\eqref{eq:n-block-capacity} is well-studied; see, for example, Cover
and Thomas~\cite[Section 10.5]{Cover--Thomas1991}.  The optimal
$K_{X,n}^{\star}$ belongs to the same eigenspace as $K_{Z,n}$, that
is, if $K_{Z,n}$ has an eigenvalue decomposition $K_{Z,n} = Q\Lambda
Q'$ with a diagonal matrix $\Lambda =
\diag(\lambda_1,\ldots,\lambda_n)$ and a unitary matrix $Q$, then
$K_{X,n}^\star = QLQ'$ for some diagonal matrix $L =
\diag(l_1,\ldots,l_n)$.  Furthermore, the input eigenvalues
$l_1,\ldots,l_n$ ``water-fill'' the noise eigenvalues
$\lambda_1,\ldots,\lambda_n$ in the sense that
\begin{equation}
\label{eq:waterfilling}
l_i = (\lambda - \lambda_i)^+
= \max\{\lambda - \lambda_i, 0\},\qquad i = 1,\ldots, n
\end{equation}
where $\lambda$ is chosen such that
\[
\tr K_{X,n}^{\star} = \sum_i l_i = \sum_i (\lambda - \lambda_i)^+ = nP.
\]
Plugging $K_{X,n}^{\star} = QLQ'$ into \eqref{eq:n-block-capacity}, we get
\[
C_n = \frac{1}{2n} \sum_{i=1}^n \log \frac{\lambda_i + (\lambda -
\lambda_i)^+}{\lambda_i} = \frac{1}{2n} \sum_{i=1}^n \log
\frac{\max\{\lambda_i,\lambda\}}{\lambda_i}.
\]

In fact, the optimization problem in \eqref{eq:n-block-capacity} is a
simple instance of a \emph{matrix determinant maximization problem}.
(See Vandenberghe, Boyd, and Wu~\cite{Vandenberghe--Boyd--Wu1998} for
an excellent review of the matrix determinant maximization (max-det)
problem with linear matrix inequality constraints.)  Indeed, ignoring
the subscripts, we can reformulate \eqref{eq:n-block-capacity} as
\begin{align}
\label{eq:finite-capacity}
\begin{array}{l@{\quad}l}
\text{maximize\ }&\log\det K_Y\\[.25em]
\text{subject to}&K_Y - K_Z \geql 0\\[.25em]
&\tr (K_Y - K_Z) \le nP.
\end{array}
\end{align}
Now consider any $\nu > 0$ and any positive definite matrix $\Phi$
such that $\Psi := \nu I - \Phi \geql 0$.  From
Lemmas~\ref{lemma:logdet} and \ref{lemma:trace} in the previous
section and constraints on $K_Y$ in \eqref{eq:finite-capacity}, we
have for any feasible $K_Y$ that
\begin{align}
\nonumber
\log\det(K_Y) &\le -\log\det(\Phi) + \tr (K_Y\Phi) - n\\
\nonumber
&= -\log\det(\Phi) + \nu \tr (K_Y) - \tr(K_Y\Psi) - n\\
\nonumber
&\le -\log\det(\Phi) + \nu (\tr (K_Z) + nP) - \tr(K_Z\Psi) - n\\
\label{eq:duality-ineq}
&= -\log\det(\Phi) + \tr(K_Z\Phi) + nP \nu - n.
\end{align}
Thus, we get the following optimization problem as an upper bound on
\eqref{eq:finite-capacity}, which is another max-det problem:
\begin{align}
\label{eq:dual-finite-capacity}
\begin{array}{l@{\quad}l}
\text{minimize\ }&-\log\det \Phi + \tr(\Phi K_Z) + nP \nu - n\\[.2em]
\text{subject to}
& \nu > 0 \\[.2em]
&\Phi \succ 0\\[.2em]
& \nu I - \Phi \geql 0.
\end{array}
\end{align}
Although we have arrived at the problem
\eqref{eq:dual-finite-capacity} from first principles, we can easily
check that this problem is indeed the Lagrange dual to
\eqref{eq:finite-capacity}; see Vandenberghe et al.~\cite[Section
3]{Vandenberghe--Boyd--Wu1998}.  Moreover, both the primal
problem~\eqref{eq:finite-capacity} and the dual
problem~\eqref{eq:dual-finite-capacity} are strictly feasible.  Hence,
from the standard results in convex optimization
(Rockafellar~\cite[Sections 29--30]{Rockafellar1970} and Boyd and
Vandenberghe~\cite[Chapter 5]{Boyd--Vandenberghe2004}), \emph{strong
duality} holds and there exist $K_Y^{\star}, \nu^{\star},
\Phi^{\star}$ satisfying \eqref{eq:duality-ineq} with equality.
Indeed, following the equality conditions for the chain of
inequalities \eqref{eq:duality-ineq}, we find the following
properties of the optimal $K_Y^{\star}$.

\begin{prop}
\label{prop:optimal-cn}
The $n$-block capacity $C_n$ defined in \eqref{eq:n-block-capacity} is
achieved by $K_X^\star$ and the corresponding $K_Y^\star = K_X^\star +
K_Z$ if and only if both of the following conditions are satisfied:
\begin{eqenumerate}
\item \text{Power:}\quad $\tr(K_X^{\star}) = nP.$
\item \text{Water-filling:}\quad $
\tr(K_X^{\star}(K_Y^{\star} - \lambda_{\min}(K_Y^{\star})I)) = 0.$
\label{eq:waterfilling-alt}
\end{eqenumerate}
\end{prop}
Although the water-filling condition \eqref{eq:waterfilling-alt} looks,
at first, quite different from the traditional representation,
Lemma~\ref{lemma:trace} shows that \eqref{eq:waterfilling-alt} is
indeed equivalent to \eqref{eq:waterfilling}.

Once we have the parametric characterization of the capacity as
\begin{align*}
C(\lambda) &= \lim_{n\to\infty} \frac{1}{2n} \sum_{i=1}^n \log
\frac{\max\{\lambda_i(K_{Z,n}),\lambda\}}{\lambda_i(K_{Z,n})}\\
P(\lambda) &= \lim_{n\to\infty} \frac{1}{n} \sum_{i=1}^n (\lambda -
\lambda_i(K_{Z,n}))^+
\end{align*} 
we can apply Szeg\H{o}'s limit theorem and use the continuity of
the capacity $C$ in the power constraint $P$ to obtain the desired
capacity formula:
\begin{align*}
\tag{\ref{eq:c}}
C(\lambda) &= \intt \half \log \frac{\max \{S_Z(\eit),
\lambda\}}{S_Z(\eit)} \dth\\
P(\lambda) &= \intt (\lambda - S_Z(\eit))^+ \dth.
\tag{\ref{eq:p}}
\end{align*}

This standard derivation of the Gaussian channel capacity based on the
first Szeg\H{o} theorem traces back to Tsybakov~\cite{Tsybakov1965,
Tsybakov1970} in the literature.  (See also Gray~\cite[Section
V]{Gray1972} and Blahut~\cite{Blahut1987} for a detailed proof.)  An
alternative proof was given by Hirt and Massey \cite{Hirt--Massey1988}
who approximated a finite impulse response intersymbol interference
channel (or equivalently, a finite-order autoregressive noise channel)
by an intersymbol interference Gaussian channel with circular
convolution, and analyzed the asymptotic eigenvalue distribution of the
resulting circulant matrix.  In the light of the standard technique of
approximating a Toeplitz matrix by circulant matrices (see, for
example, tutorials by Gray~\cite{Gray1972, Gray2002}), the development
by Hirt and Massey is essentially along the line of the traditional
approach based on the asymptotics of large Toeplitz matrices.

Now we give yet another proof of the Gaussian capacity theorem that
does not rely on the asymptotics of large Toeplitz matrices.  (To be
fair, no proof can be totally independent of Szeg\H{o}'s limit
theorem, since the entropy rate of a stationary Gaussian process is
given by Szeg\H{o}--Kolmogorov--Krein formula~\eqref{eq:kolmo}.)
The main idea is very simple.  First we spin off from
\eqref{eq:n-block-capacity} and show that the capacity is achieved by
a stationary Gaussian input process, which gives a variational
formulation of the capacity as
\begin{equation}
C = \sup_{S_X(\eit)} \intt \half\log \frac{S_X(\eit)
+ S_Z(\eit)} {S_Z(\eit)} \dth
\label{eq:variational-capacity}
\end{equation}
where the supremum is taken over all $S_X(\eit) \ge 0$ satisfying
the power constraint
\[
\intt S_X(\eit) \dth \le P.
\]
This characterization states that the capacity of Gaussian channel is
equal to the maximum information rate between a stationary (Gaussian)
input process $\{X_i\}_{i=-\infty}^\infty$ and the corresponding output
process $\{Y_i\}_{i=-\infty}^\infty$, or equivalently, the maximum entropy
rate $h(\mathcal{Y})$ of the output process minus the noise entropy
rate $h(\mathcal{Z})$.  Hence, the variational
characterization~\eqref{eq:variational-capacity} can be viewed as the
justification for the interchange of the order of maximum and limit in
\[
\lim_{n\to\infty} \max_{X^n} \frac{1}{n} I(X^n; Y^n)
= \sup_{\{X_i\}} \lim_{n\to\infty} \frac{1}{n} I(X^n; Y^n)
\]
where the maximum on the left-hand side is over all distributions on
random $n$-vectors 
$X^n$ satisfying $E (\sum_{i=1}^n X_i^2) \le nP$ while
the supremum on the right-hand side is over all stationary processes
$\{X_i\}$ with $EX_i^2 \le P$.

Note that the finite-dimensional water-filling solution
\eqref{eq:waterfilling} does not directly imply the variational
formulation~\eqref{eq:variational-capacity}, for, in general,
the optimal $K_{Y,n}^{\star}$ is not Toeplitz nor is the sequence
$\{K_{Y,n}^{\star}\}_{n=1}^\infty$ consistent.  Once we
establish~\eqref{eq:variational-capacity}, we will show by elementary
arguments that the quantity \eqref{eq:variational-capacity} is indeed
equal to the water-filling capacity formula \eqref{eq:c}.  Details of
the proof follow.

Define
\begin{align}
\label{eq:tilde-c}
\tilde{C} &:= \sup_{S_X(\eit)} \intt \half \log \frac{S_X(\eit)
+ S_Z(\eit)} {S_Z(\eit)} \dth
\end{align}
where the supremum is over all $S_X(\eit) \ge 0$ such that $\intt
S_X(\eit)\dth \le P.$ Since Gaussian processes maximize the entropy
rate under the second moment constraint, we have 
\begin{align}
\nonumber 
\tilde{C} &= \sup_{\{X_i\}} \lim_{n\to\infty}
\frac{1}{n} I(X^n; Y^n) \\
\label{eq:def-tilde-C}
&= \sup_{\{X_i\}} h(\mathcal{Y}) - h(\mathcal{Z}),
\end{align}
where the supremums are over all stationary processes $\{X_i\}$,
independent of $\{Z_i\}$, and satisfying the power constraint $E X_i^2
\le P$.

We first prove
\begin{equation}
\label{eq:tilde-c-ineq}
C_n \le \tilde{C} \le C_n + \frac{h(Z^n)}{n} - h(\mathcal{Z})
\end{equation}
for all $n$, which implies that $\lim_{n\to\infty} C_n = \tilde{C}.$
Fix $n$ and let ${K}_{X,n}^{\star}$ achieve $C_n$.  We consider a
two-sided input process $\{X_i\}_{i=-\infty}^{\infty}$ that is
blockwise stationary (=cyclostationary) with $X_{kn+1}^{(k+1)n},$
$-\infty < k <\infty,$ $\text{i.i.d.} \sim N_n(0,K_{X,n}^{\star})$, and
is independent of the stationary Gaussian noise process
$\{Z_i\}_{i=-\infty}^\infty$.  Let $Y_i = X_i + Z_i,$ $-\infty < i <
\infty,$ be the corresponding output process through the stationary
Gaussian channel.  For each $t = 0,1,\ldots,n-1$, define a
time-shifted process $\{X_{i}{(t)}\}_{i=-\infty}^\infty$ as
$X_{i}{(t)} =X_{i+t}$ for all $i$ and similarly define
$\{Y_{i}{(t)}\}_{i=-\infty}^\infty$ and
$\{Z_{i}{(t)}\}_{i=-\infty}^\infty$. Obviously, $Y_i(t) = X_i(t) +
Z_i(t)$ for all $i$. Using the inequality \eqref{eq:cond-ineq} that
was used to prove the superadditivity of $C_n$, we have
\[
C_n \le \frac{1}{kn} I(X^{kn}; Y^{kn}),\qquad k = 1,2,\ldots
\]
and hence for all $m = 1, 2,\ldots,$ and each $t = 0,\ldots, n-1$, we
have
\begin{align}
\nonumber
C_n &\le \frac{1}{m} I(X_{1}^m{(t)}; Y_{1}^m(t)) + \epsilon_m\\
\nonumber
&= \frac{1}{m} \big( h(Y_{1}^m(t)) - h(Z_{1}^m(t)) \big) + \epsilon_m\\
\label{eq:C_n-ineq}
&= \frac{1}{m} \big( h(Y_{1}^m(t)) - h(Z_1^m) \big) + \epsilon_m
\end{align}
for some $\epsilon_m$ that vanishes uniformly in $t$ as $m\to\infty$.
Here the last equality follows from the stationarity of $Z$.

Now let $T$ be a random variable uniformly distributed on
$\{0,1,\ldots,n-1\}$ and independent of $\{X_i\}_{i=-\infty}^\infty$
and $\{Z_i\}_{i=-\infty}^\infty$.  We make the following observations:
\begin{eqenumerate}
\item $\{(X_{i}(T), Y_{i}(T), Z_{i}(T))\}_{i=-\infty}^\infty$ is
stationary with $Y_{i}(T) = X_{i}(T) + Z_{i}(T)$ for all $i$.
\label{eq:xyz-t}

\item $E [X_{i}^2(T)] = E[E(X_{i}^2(T)|T)] \le P$.

\item The autocorrelation function of $\{X_{i}(T)\}$ is banded, and
hence the power spectral distribution of $\{X_{i}(T)\}$ is absolutely
continuous with respect to the Lebesgue measure.

\item The processes $\{X_{i}(T)\}$ and $\{Z_{i}(T)\}$ are orthogonal
in the sense that, for all $i,j$,
\[
E[X_{i}(T)Z_{j}(T)] = E[E(X_{i}(T)Z_{j}(T)|T)] = 0.
\]

\item $\{Z_{i}(T)\}$ has the same distribution as $\{Z_i\}$.
\label{eq:z-t}
\end{eqenumerate}
Finally let $\{\tilde{X}_i, \tilde{Y}_i,
\tilde{Z}_i\}_{i=-\infty}^\infty$ be a jointly Gaussian process with
the same mean and autocorrelation as the stationary process $\{X_i(T),
Y_i(T), Z_i(T)\}_{i=-\infty}^\infty$.  Note that $\{\tilde{X}_i,
\tilde{Y}_i, \tilde{Z}_i\}$ also satisfies the properties
\eqref{eq:xyz-t}--\eqref{eq:z-t}.  In addition, the input process
$\{\tilde{X}_i\}$ is independent of the noise process
$\{\tilde{Z}_i\}$.  Hence, $\{\tilde{X}_i\}$ is feasible for the
maximization in \eqref{eq:def-tilde-C}.  Now that $\{\tilde{Y}_i\}$
has a larger entropy rate than $\{Y_i(T)\}$ and the inequality
\eqref{eq:C_n-ineq} holds uniformly for any $t$, we continue from the
inequality~\eqref{eq:C_n-ineq} to get
\begin{align*}
C_n &\le \frac{1}{m} 
\big ( h(Y_{1}^m(T)|T) - h(Z_1^m) \big) + \epsilon_m\\
&\le \frac{1}{m} 
\big ( h(Y_{1}^m(T)) - h(Z_1^m) \big) + \epsilon_m\\
&\le \frac{1}{m} \big ( h(\tilde{Y}_{1}^m) - h(Z_1^m) \big) + \epsilon_m.
\end{align*}
By letting $m$ tend to infinity, we have
\begin{align*}
C_n \le h(\mathcal{\tilde{Y}}) - h(\mathcal{Z}) \le \tilde{C}
\end{align*}
where the last inequality follows from the definition of $\tilde{C}$
in \eqref{eq:def-tilde-C}.

For the other direction of inequality, fix $\epsilon > 0$ and let the
stationary Gaussian input process $\{X_i\}_{i=-\infty}^\infty$ achieve
$\tilde{C} - \epsilon$.  Let $\{Y_i\}_{i=-\infty}^\infty$ be the
corresponding output process.  Since $X_1^n$ trivially satisfies the
power constraint $E \sum_{i=1}^n X_i^2 \le nP$, we have
\[
C_n \ge \frac{1}{n} I(X^n; Y^n) 
= \frac{1}{n} \left(h(Y^n) - h(Z^n)\right).
\]
But $n^{-1} h(Y^n)$ is decreasing in $n$, with limit
$h(\mathcal{Y})$. Hence,
\[
C_n \ge h(\mathcal{Y}) - \frac{h(Z^n)}{n} = 
\tilde{C} - \epsilon + h(\mathcal{Z}) - \frac{h(Z^n)}{n}.
\]
The desired inequality follows immediately since $\epsilon > 0$ is
arbitrary.  Thus, we have shown that $C = \tilde{C}.$

Now we show that the supremum of \eqref{eq:tilde-c} is attained by
\[
S_X^{\star}(\eit) = \bigl(\lambda - S_Z(\eit)\bigr)^+
= \max\bigl\{\lambda - S_Z(\eit), 0\bigr\}
\]
where $\lambda$ is chosen to satisfy the power constraint with
equality.  For a parallel development with the feedback case in the
subsequent sections, we change the optimization variable to
$S_Y(\eit)$ and show that the infinite-dimensional optimization
problem
\begin{equation}
\label{eq:max-log-S_Y}
\begin{array}{l@{\quad}l}
\text{maximize\ }&\intt \log S_Y(\eit) \dth\\[.75em]
\text{subject to}&S_Y(\eit) - S_Z(\eit) \ge 0,\qquad \text{for all } 
\theta\\[.75em]
&\intt \bigl(S_Y(\eit) -S_Z(\eit)\bigr) \dth \le P
\end{array}
\end{equation}
has the optimal solution
\begin{equation}
\label{eq:optimal-S_Y}
S_Y^{\star}(\eit) = \left(\lambda - S_Z(\eit)\right)^+ + S_Z(\eit) =
\max \{S_Z(\eit), \lambda\}
\end{equation}
with $\lambda > 0$ chosen to satisfy
\begin{equation}
\label{eq:optimal-lambda}
\intt \bigl(\lambda - S_Z(\eit)\bigr)^+ \dth = P. 
\end{equation}

Note that this optimization problem is the infinite-dimensional
analogue of the matrix determinant maximization problem
\eqref{eq:finite-capacity} for the $n$-block capacity $C_n$.  However,
it is often very difficult to establish the strong duality for the
infinite-dimensional optimization problem, even when the problem is
convex.  (See Ekeland and Temam~\cite{Ekeland--Temam1976}.)  Here we
avoid using the general duality theory on topological vector spaces
and take a rather elementary approach to duality, which turns out
to be powerful enough to establish the optimality of
$S_Y^{\star}(\eit)$.

Take any $\nu > 0$ and $\phi(\eit) > 0$ such that $\psi(\eit) :=
\nu - \phi(\eit) \ge 0$ for all $\theta$.  Consider any feasible
$S_Y(\eit)$ satisfying the constraints for the maximization problem
\eqref{eq:max-log-S_Y}.  Since $\log x \le x - 1$ for all $x > 0$,
we have
\begin{align}
\nonumber
\log S_Y(\eit) &\le - \log \phi(\eit) + \phi(\eit)S_Y(\eit) - 1\\
\nonumber
&= - \log \phi(\eit) + \nu S_Y(\eit) - \psi(\eit)S_Y(\eit) - 1\\
\nonumber
&\le - \log \phi(\eit) + \nu S_Y(\eit) - \psi(\eit) S_Z(\eit) - 1.
\end{align}
By integrating both sides of the above inequality with respect to
$\theta$ and applying the constraints on $S_Y(\eit)$ in
\eqref{eq:max-log-S_Y}, we obtain an upper bound of
\eqref{eq:max-log-S_Y} as
\begin{align}
\intt \log S_Y(\eit) \dth
&\le - \intt \log \phi(\eit)\dth 
+ \intt \phi(\eit) S_Z(\eit) \dth + \nu P - 1.
\label{eq:S_Y-ineq}
\end{align}
This upper bound is universal in the sense that the
inequality~\eqref{eq:S_Y-ineq} holds for any feasible $S_Y(\eit)$ and
any $\nu > 0$ and $0 < \phi(\eit) \le \nu$.

Now consider a particular choice of $\nu = \nu^\star$ and $\phi(\eit)
= \phi^\star(\eit)$ with $\nu^\star = {1}/{\lambda} > 0$ and
$\phi^\star(\eit) = {1}/{S_Y^{\star}(\eit)} > 0,$ where
$S_Y^{\star}(\omega)$ and $\lambda$ are given by
\eqref{eq:optimal-S_Y} and \eqref{eq:optimal-lambda}.  It is easy to
check that
\[
\psi(\eit) = \nu^\star - \phi^\star(\eit) =
\frac{1}{\lambda}\frac{S_Y^{\star}(\eit) -\lambda}{S_Y^{\star}(\eit)}
\ge 0,\qquad\text{for all }\theta.
\]
Plugging $(\nu^\star, \phi^\star(\eit))$ into the right-hand side of
\eqref{eq:S_Y-ineq} yields
\begin{align*}
&\intt \log S_Y^\star(\eit) \dth
+ \intt \frac{S_Z(\eit)}{S_Y^\star(\eit)} \dth + \frac{P}{\lambda} - 1\\
&= \intt \log S_Y^\star(\eit) \dth + 
\intt \frac{S_Z(\eit)-S_Y^\star(\eit)}{S_Y^\star(\eit)} \dth +
\intt \frac{S_Y^\star(\eit)-S_Z(\eit)}{\lambda} \dth\\
&=
\intt \log S_Y^\star(\eit) \dth + \intt
\frac{\big(S_Y^\star(\eit)-S_Z(\eit)\big)
\big(S_Y^\star(\eit) - \lambda\big)}{\lambda S_Y^\star(\eit)} \dth \\
&=
\intt \log S_Y^\star(\eit) \dth + \intt
\frac{(\lambda-S_Z(\eit))^+(S_Z(\eit) - \lambda)^+}
{\lambda S_Y^\star(\eit)} \dth \\
&=
\intt \log S_Y^\star(\eit) \dth.
\end{align*}
Thus, we have shown that
\[
\intt \log S_Y(\eit) \dth \le
\intt \log S_Y^\star(\eit) \dth
\]
for any feasible $S_Y(\eit)$.  This establishes the optimality of
$S_Y^\star(\eit)$, whence 
the parametric expression \eqref{eq:c} for
the Gaussian channel capacity $C$.


\section{Variational Characterization of Gaussian Feedback Capacity}
\label{sec:feedback}

Given a stationary Gaussian channel $Y_i = X_i + Z_i,\enspace i =
1,2,\ldots,$ with the noise spectral distribution $d\mu_Z(\theta) =
S_Z(\eit) d\theta,$ we wish to prove that
\[
C_\textit{FB} = \sup_{S_V(\eit), B(\eit)}
\intt \half
\log
\frac{S_V(\eit) + |1 + B(\eit)|^2 S_Z(\eit)}{S_Z(\eit)}\dth
\tag{\ref{eq:cfb_var}}
\]
with the supremum taken over all $S_V(\eit) \ge 0$ and all strictly
causal polynomials $B(\eit) = \sum_{k=1}^m b_k e^{ik\theta}$
satisfying the power constraint
\[
\intt S_V(\eit) + |B(\eit)|^2 S_Z(\eit) \dth \le P.
\]
We will closely follow the derivation of
\eqref{eq:variational-capacity} and \eqref{eq:tilde-c-ineq} for the
nonfeedback case in the previous section. Again we start from the
Cover--Pombra formulation of the $n$-block feedback capacity given by
\[
\tag{\ref{eq:cnfb}}
C_{\textit{FB},n} = \max_{K_{\!V\!,n}, B_n} 
\half \log
\frac{\det(K_{V,n} +
(I + B_n)K_{Z,n}(I + B_n)')^{1/n}}{\det(K_{Z,n})^{1/n}}
\]
where the maximum is over all positive semidefinite $K_{V,n}$ and
strictly lower triangular $B_n$ such that $\tr(K_{V,n} +
B_nK_{Z,n}B_n') \le nP$.  Again the coding theorem by Cover and Pombra
states that for every $\epsilon > 0$, there exists a sequence of
$(2^{n(C_{\textit{FB},n}-\epsilon)},n)$ feedback codes with $P_e^{(n)}
\to 0$.  Conversely, for $\epsilon > 0$, any sequence of
$(2^{n(C_{\textit{FB},n}+\epsilon)},n)$ codes has $P_e^{(n)}$ bounded
away from zero for all $n$.
Tracing the development of Cover and Pombra
backwards, we express $C_{\textit{FB},n}$ as
\begin{align}
\nonumber
C_{\textit{FB},n} &= \max_{V^n + B_n Z^n} \half \log
\frac{\det(K_{Y,n})^{1/n}}{\det(K_{Z,n})^{1/n}}\\
\nonumber
&= \max_{V^n + B_n Z^n} h(Y^n) - h(Z^n)\\
&= \max_{V^n + B_n Z^n} I(V^n;Y^n)
\nonumber
\end{align}
where the maximization is over all $X^n$ of the form $X^n = V^n + B_n
Z^n,$ resulting in $Y^n = V^n + (I+B_n)Z^n$, with strictly
lower-triangular $B_n$ and multivariate Gaussian $V^n,$ independent of
$Z^n$, satisfying the power constraint $E\sum_{i=1}^n X_i^2 \le nP.$

Before we jump into the proof of \eqref{eq:cfb_var} through a detailed
analysis on the asymptotics of the $n$-block feedback capacity
$C_{\textit{FB},n}$, we first explore a few interesting properties of
$C_{\textit{FB},n}$ itself for a finite $n$, which will be useful when
we discuss properties of the (infinite-dimensional) feedback capacity
$C_\textit{FB}$ in subsequent sections.

For a given $n$, finding $C_{\textit{FB},n}$ is equivalent to solving
the following optimization problem:
\begin{align}
\label{eq:fb-max1}
\begin{array}{l@{\quad}l}
\text{maximize\ }& \log\det (K_V + (I+B)K_Z(I+B)')\\[.25em]
\text{subject to}& K_V \geql 0\\[.25em]
& \tr(K_V + BK_Z B') \le nP\\[.25em]
& B \text{ strictly lower triangular.}
\end{array}
\end{align}

Although this problem is not convex in itself (with optimization
variables $K_V$ and $B$), it can be easily reformulated into a convex
problem.  This relatively unknown result is due to Boyd and Ordentlich
(circa 1994), and appears as an example in Vandenberghe et
al.~\cite[Equation (2.16)]{Vandenberghe--Boyd--Wu1998}.

We observe that, given $B$, $K_Y = K_V + (I+B)K_Z(I+B)'$ is one-to-one
mapped to $K_V$.  So we change the variable to $(K_Y, B)$
and rewrite \eqref{eq:fb-max1} as
\begin{align*}
\begin{array}{l@{\quad}l}
\text{maximize\ }& \log\det (K_Y)\\[.25em]
\text{subject to}& K_Y - (I+B)K_Z(I+B)' \geql 0\\[.25em]
& \tr(K_Y - BK_Z - K_ZB' - K_Z) \le nP\\[.25em]
& B \text{ strictly lower triangular.}
\end{array}
\end{align*}
Now the first constraint
\[
K_Y - (I+B)K_Z(I+B)' \geql 0
\]
can be turned into an equivalent linear matrix inequality
\[
\left[
\begin{matrix}
K_Y &I+B\\
(I+B)' &K_Z^{-1}
\end{matrix}
\right]
\geql 0
\]
from Lemma~\ref{lemma:psd}.  (Recall from Section~\ref{sec:prelim}
that $K_Z$ is nonsingular because $\mu_Z$ is nontrivial.)  Hence, we
obtain the Boyd--Ordentlich formulation of the $n$-block feedback
capacity, which is another instance of the matrix determinant
maximization problem with linear matrix inequality constraints:
\begin{align}
\label{eq:boyd-ord}
\begin{array}{l@{\quad}l}
\text{maximize\ }& \log\det (K_Y)\\[.5em]
\text{subject to}& 
\left[
\begin{matrix}
K_Y &I+B\\
(I+B)' &K_Z^{-1}
\end{matrix}
\right] \geql 0\\[1.2em]
& \tr(K_Y - BK_Z - K_ZB' - K_Z) \le nP\\[.25em]
& B \text{ strictly lower triangular.}
\end{array}
\end{align}

As a simple application of the Boyd--Ordentlich reformulation of the
$n$-block feedback capacity, we can easily recover the following
result due to Yanagi, Chen, and Yu~\cite{Yanagi--Chen--Yu2000}.
\begin{prop}[Yanagi--Chen--Yu]
\label{prop:concavity}
For an arbitrary (not necessarily Toeplitz) noise covariance matrix
$K_Z$, the $n$-block feedback capacity $C_{\textit{FB},n}(P)$
is concave in the power constraint $P$.
\end{prop}
\begin{proof}
In the light of the Boyd--Ordentlich formulation, we write
$C_{\textit{FB},n}(P)$ as
\[
f(P) := C_{\textit{FB},n}(P) = \max_{K_Y, B}
\frac{1}{2n}\log\frac{\det(K_Y)}{\det(K_Z)}
\]
where the maximum is taken over all $K_Y$ and $B$ satisfying the
constraints in \eqref{eq:boyd-ord}.  Suppose $(K_Y^{(1)}, B^{(1)})$
and $(K_Y^{(2)}, B^{(2)})$ achieve the feedback capacity under the
power constraints $P_1$ and $P_2$, respectively.  Consider 
\[
(K_Y, B) =
\lambda (K_Y^{(1)}, B^{(1)}) + (1-\lambda) (K_Y^{(2)}, B^{(2)})
\]
for some $\lambda \in [0,1]$.  It is trivial to check that $(K_Y, B)$
satisfies the constraints in \eqref{eq:boyd-ord} under the power
constraint $P = \lambda P_1 + (1-\lambda) P_2$.  Also from the
concavity of $\log\det(\cdot)$, 
\[
\log\det (K_Y) \ge \lambda \log\det (K_Y^{(1)}) + (1-\lambda) \log\det
(K_Y^{(2)}).
\]
Thus, 
\[
C_{\textit{FB},n}(\lambda P_1 + (1-\lambda) P_2) \ge 
\lambda  C_{\textit{FB},n}(P_1) +
(1-\lambda) C_{\textit{FB},n}(P_2).\qedhere
\]
\end{proof}

The convexity of the problem, however, has more interesting
implications.  As an analogue to Proposition~\ref{prop:optimal-cn}, we
give a characterization of the optimal $(K_V^\star, B^\star)$ in the
following statement.

\begin{prop}
\label{prop:optimal-cnfb}
The $n$-block feedback capacity 
\[
\tag{\ref{eq:cnfb}}
C_{\textit{FB},n} = \max_{K_{V,n}, B_n} 
\half \log
\frac{\det(K_{V,n} +
(I + B_n)K_{Z,n}(I + B_n)')^{1/n}}{\det(K_{Z,n})^{1/n}}
\]
is achieved by $(K_V^{\star}, B^{\star})$ if and only if all of the following
conditions are satisfied:
\begin{eqenumerate}
\item \text{Power:}\quad $\tr(K_V^{\star} + B^{\star}K_Z(B^{\star})') = nP$.
\label{eq:fb-power-cond}

\item \text{Water-filling:}\quad The covariance matrix $K_V^{\star}$ water-fills the
modified noise covariance matrix $(I+B^{\star})K_Z(I+B^{\star})'$.  Equivalently,
\[
\tr (K_V^{\star}(K_Y^{\star} - \lambda_{\min}(K_Y^{\star}) I)) = 0
\]
where $K_Y^{\star} = K_V^{\star} + (I+B^{\star})K_Z(I+B^{\star})'.$

\item \text{Orthogonality:}\quad The current input $X_i$ is
independent of the past output $(Y_1,\ldots, Y^{i-1})$, i.e., $E
X_iY_j = 0$ for all $1 \le j < i \le n$.  Equivalently, $K_V^{\star} +
B^{\star}K_Z(I+B^{\star})'$ is upper triangular.
\label{eq:fb-orth-cond}

\end{eqenumerate}
\end{prop}

The necessity of these conditions is somewhat obvious (see
Ihara~\cite{Ihara1979} and Ordentlich~\cite{Ordentlich1994}).  Indeed,
the first two conditions are needed, since for any $B$, the channel
from $V^n$ to $Y^n$ is a Gaussian nonfeedback channel with the noise
covariance $(I+B)K_Z(I+B)'$ and Proposition~\ref{prop:optimal-cn}
applies.  The orthogonality of the current input $X_i$ and the past
output $(Y_1,\ldots,Y_{i-1})$ is also intuitively clear; otherwise, we
can reduce the input power for the same rate by not sending the
projection of $X_i$ onto the linear span of $(Y_1,\ldots,Y_{i-1})$.
(The receiver has that part of the information, anyway.)  More
precisely, we express the channel input as
\begin{align*}
X^n &= V^n + B Z^n\\
&= \tilde{V}^n + \tilde{B} Y^n
\end{align*}
with $\tilde{V}^n = (I+B)^{-1}V^n$ and $\tilde{B} = (I+B)^{-1}B$, and
denote each row of $\tilde{B}$ as $\tilde{B}_1,\ldots,\tilde{B}_n$.  Then,
\begin{align*}
I(V^n; Y^n) &= I(\tilde{V}^n; Y^n)\\
&= \sum_{i=1}^n I(\tilde{V}^n; Y_i|Y^{i-1})\\
&= \sum_{i=1}^n I(\tilde{V}^n; 
\tilde{V}_i + \tilde{B}_i Y^n + Z_i | Y^{i-1})\\
&= \sum_{i=1}^n I(\tilde{V}^n; \tilde{V}_i + Z_i | Y^{i-1}).
\end{align*}
As a consequence, when the distribution on $\tilde{V}^n$ is held
fixed, $I(V^n;Y^n)$ is independent of $\tilde{B}$.  Under this same
rate, the input power is minimized if we take
\[
X_i = \tilde{V}_i + \tilde{B}_i Y^n 
    = \tilde{V}_i - E(\tilde{V}_i|Y^{i-1})
\]
for all $i$.  Clearly, $X_i$ is independent of $Y^{i-1}$.  This simple
observation has been sometimes emphasized as the optimality of Kalman
filter as the feedback information processor (see, for example, Yang,
Kav\v{c}i\'{c}, and Tatikonda~\cite[Theorem
1]{Yang--Kavcic--Tatikonda2004}).

For the sufficiency (and the necessity as well) of the conditions
\eqref{eq:fb-power-cond}--\eqref{eq:fb-orth-cond} in
Proposition~\ref{prop:optimal-cnfb}, consider any $\nu > 0$ and
$n\times n$ matrices $\Phi, \Psi_1, \Psi_2, \Psi_3$ such that $\Phi
\succ 0$, $\Psi_1 = \nu I - \Phi,$ $\Psi_2 + \nu K_Z$ is upper
triangular, and
\begin{gather*}
\left[
\begin{matrix}
\Psi_1&\Psi_2\\
\Psi_2'&\Psi_3
\end{matrix}
\right] \geql 0.
\end{gather*}
Now for any feasible $B$ and $K_Y$ for \eqref{eq:boyd-ord}, we have
from Lemma~\ref{lemma:psd} that
\[
\tr\left(
\left[
\begin{matrix}
K_Y &I+B\\
(I+B)' &K_Z^{-1}
\end{matrix}
\right]
\left[
\begin{matrix}
\Psi_1&\Psi_2\\
\Psi_2'&\Psi_3
\end{matrix}
\right]\right) 
= \tr(K_Y\Psi_1 + (I+B)'\Psi_2 + (I+B) \Psi_2' + K_Z^{-1}\Psi_3) \ge 0
\]
and hence from Lemma~\ref{lemma:trace} that
\begin{align}
\nonumber
\log\det(K_Y) &\le -\log\det(\Phi) + \tr(K_Y\Phi) - n\\
\nonumber
&= -\log\det(\Phi) + \nu \tr(K_Y) - \tr(K_Y\Psi_1) - n\\
\nonumber
&\le -\log\det(\Phi) + \nu \tr(BK_Z + K_ZB' + K_Z + nP)\\
\nonumber
&\qquad + \tr((I+B)'\Psi_2 + (I+B) \Psi_2' + K_Z^{-1}\Psi_3)  - n\\
\nonumber
&= -\log\det(\Phi) + 2\tr(\Psi_2) + \tr(K_Z^{-1}\Psi_3) + 
\nonumber
2 \tr(B(\Psi_2'\!+\! \nu K_Z)) + \nu(\tr(K_Z) + nP) - n\\
&= -\log\det(\Phi) + 2\tr(\Psi_2) + \tr(K_Z^{-1}\Psi_3) 
+ \nu(\tr(K_Z) + nP) - n
\label{eq:fb-duality}
\end{align}
where the last equality follows from the triangularity conditions on
$B$ and $\Psi_2 + \nu K_Z$.  Thus, we have obtained the dual%
\footnote{The optimization problem~\eqref{eq:dual-boyd-ord} is indeed
the Lagrange dual to \eqref{eq:boyd-ord}, which can be readily
verified; see Vandenberghe et al.~\cite[Section
3]{Vandenberghe--Boyd--Wu1998}.}  problem to \eqref{eq:boyd-ord},
which is, once again, a matrix determinant maximization problem with
linear matrix inequality constraints:
\begin{gather}
\label{eq:dual-boyd-ord}
\begin{array}{l@{\quad}l}
\text{minimize}& -\log\det(\Phi) + 2\tr(\Psi_2) + \tr(K_Z^{-1}\Psi_3)
+ \nu(\tr(K_Z) + nP) - n\\[.25em]
\text{subject to}&\nu > 0\\[.25em] 
&\Phi \succ 0\\[.25em]
&\left[
\begin{matrix}
\nu I - \Phi &\Psi_2\\
\Psi_2'&\Psi_3
\end{matrix}
\right] \geql 0\\[1em]
&\Psi_2 + \nu K_Z \text{ upper triangular.}
\end{array}
\end{gather}
As in the nonfeedback case, the optimality of any $(K_V^{\star},
B^{\star})$ satisfying the conditions
\eqref{eq:fb-power-cond}--\eqref{eq:fb-orth-cond} of
Proposition~\ref{prop:optimal-cnfb} follows from \emph{Slater's
condition} (i.e., both primal and dual problems are strictly feasible)
and strong duality; see Vandenberghe et al.'s
review~\cite{Vandenberghe--Boyd--Wu1998} on the max-det problem.  By
checking the equality conditions for the chain of inequalities
\eqref{eq:fb-duality}, we can easily check that the duality gap is
zero with
\begin{align*}
\nu^{\star} &= 1/\lambda_{\min}(K_Y^{\star})\\
\Phi^{\star} &= (K_Y^{\star})^{-1}\\
\Psi_2^{\star} &= -(\nu^{\star} I - (K_Y^{\star})^{-1})(I+B^{\star})K_Z\\
\Psi_3^{\star} &= K_Z(I+B^{\star})'(\nu^{\star}I - (K_Y^{\star})^{-1})(I+B^{\star})K_Z
\end{align*}
and hence that the conditions
\eqref{eq:fb-power-cond}--\eqref{eq:fb-orth-cond} are sufficient and
necessary.

As a historical note, we remark that Ordentlich~\cite{Ordentlich1994}
obtained the necessary conditions
\eqref{eq:fb-power-cond}--\eqref{eq:fb-orth-cond} from a simple but
elegant fixed point argument.  This development, which predates the
Boyd--Ordentlich formulation~\eqref{eq:boyd-ord}, has certain benefits
over the more refined convex optimization approach explained above.
We will use a variant of Ordentlich's method as well as an
infinite-dimensional version of the above convex optimization method
when we characterize the optimal feedback filter in the next section.

The first nontrivial application of the necessity of
\eqref{eq:fb-power-cond}--\eqref{eq:fb-orth-cond} in
Proposition~\ref{prop:optimal-cnfb} is the following structural
result, again, due to Ordentlich~\cite{Ordentlich1994}.
\begin{coro}[{Ordentlich}]
\label{coro:ordentlich}
Suppose $K_Z$ is a covariance matrix corresponding to a stationary
moving average process of order $k$, or equivalently, $K_Z$ is
Toeplitz and banded with bandwidth $2k+1$ (i.e., $K_Z(i,j) = 0$ if
$|i-j| > k$).  Then the optimal $K_V^\star$ for the optimization
problem~\eqref{eq:fb-max1} has rank at most $k$.
\end{coro}
\begin{proof}
Let $K = K_Y^{\star} - \lambda^{\star} I$ where $\lambda^{\star} =
\lambda_{\min}(K_Y^{\star}).$ Then from the orthogonality condition
(iii), $K - K_Z(I+B^{\star})' + \lambda^{\star} I$ is upper
triangular.  In other words, the strictly lower triangular part of $K$
is equal to that of $K_Z(I+B^{\star})'$.  In particular, if $K_Z$ is
Toeplitz and banded with bandwidth $2k+1$, $K$ is also banded with
bandwidth $2k+1$, $K(i,j) = K_Z(i,j)$ if $|i-j| = k$, and thus $K$ has
rank at least $n-k$.  But from the water-filling condition (ii), we
have $ \rank(K_V^{\star}) + \rank(K) \le n. $ Hence, $K_V^{\star}$ has
rank at most $k$.
\end{proof}
An important observation we can draw from the above proof is that the
optimal output covariance matrix $K_Y^\star$ is also banded with the
bandwidth $2k+1$, regardless of the block size $n$.  Later in
Section~\ref{sec:armak}, we will extend this observation to the ARMA
noise channels and characterize the feedback capacity thereof.

Proposition~\ref{prop:optimal-cnfb} also answers the following
question---when does feedback increase the $n$-block capacity?  This
question was completely answered by Baker~\cite{Baker1988} and Ihara
and Yanagi~\cite{Ihara--Yanagi1989, Yanagi1992}, who characterized the
sufficient and necessary condition for the increment, in the context of
\emph{blockwise whiteness} of the noise covariance matrix.  More
specifically, for an arbitrary (not necessarily Toeplitz) noise
covariance matrix $K_Z$ of size $n\times n$, we define $L_k = \{ l \ne
k: K_Z(k,l) \ne 0 \}$.  We say that $K_Z$ is \emph{white} if $L_k =
\varnothing$ for all $k$, and \emph{blockwise white} if $K_Z$ is
nonwhite and $L_k = \varnothing$ for some $k$.  When $K_Z$ is
blockwise white, we denote by $\hat{K}_Z$ the submatrix of $K_Z$
constructed by $\{k: L_k \ne \varnothing\}$.  Now the result by
Baker--Ihara--Yanagi states that feedback does not increase the
$n$-block capacity for the Gaussian channel with noise covariance
matrix $K_Z$ under the power constraint $P$ if and only if
\begin{enumerate}
\item
$K_Z$ is white, or
\item $K_Z$ is nonwhite and $P \le m \lambda_m - (\lambda_1 +
\cdots + \lambda_m)$ where $0 < \lambda_1 \le \lambda_2 \le \ldots \le
\lambda_n$ are eigenvalues of $K_Z$ and $\lambda_m$ is the smallest
eigenvalue of $\hat{K}_Z$.
\end{enumerate}

Here we give an equivalent statement, accompanied with a simple
proof.
\begin{coro}[Baker--Ihara--Yanagi]
\label{coro:cn-incr}
Suppose $K_X^{\star} = K_X^{\star}(K_Z, P)$ achieves the nonfeedback
capacity $C_n = C_n(K_Z, P)$ for a given (not necessarily Toeplitz)
noise covariance matrix $K_Z$ under the power constraint $P$.  Then,
we have
\[
C_n(K_Z, P) = C_{\textit{FB},n}(K_Z, P)
\]
if and only if $K_X^{\star}(K_Z, P)$ is diagonal.  In particular, if the
noise process is stationary and nonwhite, then feedback increases the
$n$-block capacity for all $n$ and all $P > 0$.
\end{coro}
\begin{proof}
Suppose $C_n = C_{\textit{FB},n}$, that is, $(K_V^{\star}, B^{\star}) = (K_X^{\star}, 0)$
achieves the feedback capacity.  Then, from the orthogonality
condition (iii) in Proposition~\ref{prop:optimal-cnfb}, $K_X^{\star} = K_V^{\star}
+ B^{\star}K_Z(I+B^{\star})'$ is upper triangular.  Since $K_X^{\star}$ is symmetric, it
must be diagonal.  Conversely, we see that $(K_V, B) = (K_X^{\star}, 0)$
satisfies the conditions
\eqref{eq:fb-power-cond}--\eqref{eq:fb-orth-cond} in
Proposition~\ref{prop:optimal-cnfb}, whence $C_n = C_{\textit{FB},n}.$
\end{proof}

From a numerical point of view, the duality result developed above
gives the ``solution'' to the $n$-block feedback capacity problem,
since there is a polynomial-time algorithm for the determinant
maximization problem~\eqref{eq:boyd-ord}, based on the interior-point
method.  (See Nesterov and Nemirovskii
\cite{Nesterov--Nemirovskii1994} and Vandenberghe et
al.~\cite{Vandenberghe--Boyd--Wu1998}.)  In fact, Sina
Zahedi~\cite{Zahedi2003} at Stanford University developed a numerical
solver that can handle arbitrary covariance matrices of size, say,
$n=200$, with moderate computing power.

As for the (infinite-block) feedback capacity, however, there is still
much to be done.  First, the above duality theory is for finite
block-size $n$, however large it may be; it is another story to talk
about the limit.  Furthermore, unlike the nonfeedback case, the
complicated optimality condition in
Proposition~\ref{prop:optimal-cnfb} has both temporal and spectral
components, and consequently, it seems very difficult, if not
impossible, to derive an analytic solution for $(K_{V,n}^{\star},
B_n^{\star})$ even for a small $n$. 

Thus motivated, we move on to the main theme of this section---the
variational characterization of the feedback capacity.

\begin{theorem}
\label{thm:fb-capacity}
Suppose that the stationary Gaussian noise process
$\{Z_i\}_{i=1}^\infty$ has the absolutely continuous power spectral
distribution $d\mu_Z(\theta) = S_Z(\eit)d\theta$.  Then, the feedback
capacity $C_\textit{FB}$ of the Gaussian channel $Y_i = X_i + Z_i,
\enspace i = 1,2,\ldots,$ under the power constraint $P$, is given by
\[
C_\textit{FB} = \sup_{S_V(\eit), B(\eit)} 
\intt \frac{1}{2} \log
\frac{S_V(\eit) + |1+B(\eit)|^2 S_Z(\eit)}{S_Z(\eit)} \dth
\]
where the supremum is taken over all $S_V(\eit) \ge 0$ and all strictly
causal polynomials $B(\eit) = \sum_{k=1}^m b_k e^{ik\theta}$
satisfying the power constraint $\intt (S_V(\eit) + |B(\eit)|^2
S_Z(\eit)) \dth \le P.$
\end{theorem}

\begin{proof}
Define 
\[
\tilde{C}_\textit{FB} = \sup_{S_V(\eit), B(\eit)} 
\intt \frac{1}{2} \log
\frac{S_V(\eit) + |1+B(\eit)|^2 S_Z(\eit)} {S_Z(\eit)} \dth
\]
where the supremum is taken over all $S_V(\eit) \ge 0$ and all
$B(\eit) = \sum_{k=1}^m b_k e^{ik\theta}$ such that the power
constraint $\intt (S_V(\eit) + |B(\eit)|^2 S_Z(\eit)) \dth \le P$ is
satisfied.  In the light of Szeg\H{o}--Kolmogorov--Krein theorem, we
can express $\tilde{C}_\textit{FB}$ also as
\[
\tilde{C}_\textit{FB} = \sup_{\{X_i\}} h(\mathcal{Y}) - h(\mathcal{Z})
\]
where the supremum is taken over all stationary Gaussian processes
$\{X_i\}_{i=-\infty}^\infty$ of the form $X_i = V_i +
\sum_{k} b_k Z_{i-k}$ where $\{V_i\}_{i=-\infty}^\infty$ is
stationary and independent of $\{Z_i\}_{i=-\infty}^\infty$ such that
$E X_i^2 \le P$.

We first show that 
\begin{equation}
\label{eq:c-n-tilde-ineq}
C_{\textit{FB},n} \le \tilde{C}_\textit{FB}
\end{equation}
for all $n$, which is not so difficult thanks to our exercise on the
nonfeedback case in the previous section.  First fix $n$ and let
$(K_{V,n}^{\star}, B_n^{\star})$ achieve $C_{\textit{FB},n}$.  Consider
a process $\{V_i\}_{i=-\infty}^{\infty}$ that is independent of
$\{Z_i\}_{i=-\infty}^\infty$ and blockwise white with
$V_{kn+1}^{(k+1)n},$ $-\infty < k <\infty,$ \iid $\sim
N_n(0,K_{V,n}^\star)$.  Define a process $\{X_i\}_{i=-\infty}^\infty$
as $X_{kn+1}^{(k+1)n} = V_{kn+1}^{(k+1)n} + B_n^\star
Z_{kn+1}^{(k+1)n}$ for all $k$.  And similarly, let $Y_i = X_i + Z_i,$
$-\infty < i < \infty,$ be the corresponding output process through
the stationary Gaussian channel.  Note that $Y_{kn+1}^{(k+1)n} =
V_{kn+1}^{(k+1)n} + \,(I + B_n^\star) Z_{kn+1}^{(k+1)n}$ for all $k$.
For each $t = 0,1,\ldots,n-1$, define the time-shifted process
$\{V_{i}(t)\}_{i=-\infty}^\infty$ as $V_{i}(t) =V_{t+i}$ for all $i$,
and similarly define $\{X_i(t)\}_{i=-\infty}^\infty$,
$\{Y_i(t)\}_{i=-\infty}^\infty$, and $\{Z_i(t)\}_{i=-\infty}^\infty$.
Note that $Y_i(t) = X_i(t) + Z_i(t)$ for all $i$ and all $t =
0,1,\ldots,n-1$, but $X_1^n(t)$ is not equal to $V_1^n(t) +
B_n^{\star} Z_1^n(t)$ in general.

Now we focus on $X_1^{2n} = V_1^{2n} + (B_n^\star \oplus B_n^\star)
Z_1^{2n}$.  Then,
\begin{align*}
2 C_{\textit{FB},n} &= I(V_1^{n}; Y_1^{n}) + I(V_{n+1}^{2n}; Y_{n+1}^{2n})\\
&= h(V_1^n) + h(V_{n+1}^{2n}) - h(V_1^n|Y_1^n) -
h(V_{n+1}^{2n}|Y_{n+1}^{2n})\\ &\le h(V_1^{2n}) - h(V_1^{2n}|Y_1^{2n})\\
&= I(V_1^{2n}; Y_1^{2n})\\ &= h(Y_1^{2n}) - h(Z_1^{2n}).
\end{align*}
By repeating the same argument, we have
\[
C_{\textit{FB},n} \le \frac{1}{kn} I(V_1^{kn}; Y_1^{kn}),
\]
for all $k$.  Hence, for all $m = 1, 2,\ldots,$ and each $t =
0,\ldots, n-1$, we have
\begin{align*}
C_n &\le \frac{1}{m} \big( h(Y_{1}^m(t)) - h(Z_1^m(t)) \big) +
\epsilon_m\\
&= \frac{1}{m} \big( h(Y_{1}^m(t)) - h(Z_1^m) \big) + \epsilon_m
\end{align*}
where $\epsilon_m$ absorbs the edge effect and vanishes uniformly in
$t$ as $m\to\infty$.

As before, we introduce a random variable $T$ uniform on
$\{0,1,\ldots,n-1\}$ and independent of everything else.  It is easy
to check the followings:
\begin{eqenumerate}
\item
\label{eq:fb-joint-stat}
$\{V_i(T), X_i(T), Y_i(T), Z_i(T)\}_{i=-\infty}^\infty$ is stationary with
$Y_i(T) = X_i(T) + Z_i(T)$.
\item
$\{X_i(T)\}_{i=-\infty}^\infty$ satisfies the power constraint
\[E [X_{i}^2(T)] = E[E(X_{i}^2(T)|T)] = \frac{1}{n} \tr (K_{V,n}^\star 
+ B_n^\star K_{Z,n}(B_n^\star)') \le P.\]
\item
$\{V_i(T)\}_{i=-\infty}^\infty$ and $\{Z_i(T)\}_{i=-\infty}^\infty$
are orthogonal; that is, for all $i, j$,
\[
E [V_{i}(T)Z_{j}(T)] = E[E(V_{i}(T)Z_{j}(T)|T)] = 0.
\]
\item
Although there is no linear relationship between $\{X_i(T)\}$ and
$\{Z_i(T)\}$, $\{X_i(T)\}$ still depends on $\{Z_i(T)\}$ in a
strictly causal manner.  More precisely, for all $i, j$,
\begin{align*}
E [X_{i}(T) &Z_{j}(T) | Z_{i-n+1}^{i-1}(T)] \\
&= E \bigl[
E (X_{i}(T) Z_{j}(T) | Z_{i-n+1}^{i-1}(T), T) \,\big|\, Z_{i-n+1}^{i-1}(T) 
\bigr]\\
&= E \bigl[
E (X_{i}(T) | Z_{i-n+1}^{i-1}(T), T) \cdot 
E (Z_{j}(T) | Z_{i-n+1}^{i-1}(T), T) 
\,\big|\, Z_{i-n+1}^{i-1}(T) 
\bigr]\\
&= E \bigl[
E (X_{i}(T) | Z_{i-n+1}^{i-1}(T), T) \cdot
E (Z_{j}(T) | Z_{i-n+1}^{i-1}(T)) 
\,\big|\, Z_{i-n+1}^{i-1}(T) 
\bigr]\\
&= 
E [X_{i}(T) | Z_{i-n+1}^{i-1}(T)] \cdot
E [Z_{j}(T) | Z_{i-n+1}^{i-1}(T)],
\end{align*}
and for all $i$,
\[
\var \bigl(X_{i}(T)-V_{i}(T) | Z_{i-n+1}^{i-1}(T)\bigr)
= E \Bigl[
\var \bigl(X_{i}(T)-V_{i}(T) | Z_{i-n+1}^{i-1}(T), T\bigl) 
\,\Big|\, Z_{i-n+1}^{i-1}(T)
\Bigr]
= 0.
\]
Roughly speaking, $X_i(T) = V_i(T) +
f(Z_{i-n+1}^{i-1}(T))$ almost surely for some $f$.
\item
Since $\{V_i(T)\}$ has an absolutely continuous power spectral
distribution, so does $\{X_i(T)\}$.
\item
\label{eq:fb-noise-same-dist}
$\{Z_i(T)\}$ has the same distribution as $\{Z_i\}$.
\end{eqenumerate}

Finally, define $\{\tilde{V}_i, \tilde{X}_i, \tilde{Y}_i,
\tilde{Z}_i\}_{i=-\infty}^\infty$ to be a jointly Gaussian process
with the same mean and autocorrelation as the stationary process
$\{V_i(T), X_i(T), Y_i(T), Z_i(T)\}_{i=-\infty}^\infty$.  It is easy
to check that $\{\tilde{V}_i, \tilde{X}_i, \tilde{Y}_i, \tilde{Z}_i\}$
also satisfies the properties
\eqref{eq:fb-joint-stat}--\eqref{eq:fb-noise-same-dist} and hence that
$\{\tilde{V}_i\}$ and $\{\tilde{Z}_i\}$ are independent.  It follows
from these properties and the Gaussianity of $\{\tilde{V}_i,
\tilde{X}_i, \tilde{Y}_i, \tilde{Z}_i\}$ that there exists a sequence
$\{b_k\}_{k=1}^{n-1}$ so that $\tilde{X}_i = \tilde{V}_i + \sum_{k}
b_k\tilde{Z}_{i-k}.$ Thus we have
\begin{align*}
C_{\textit{FB},n} 
&\le \frac{1}{m} \bigl( h(Y_{1}^m(T)|T) - h(Z_1^m) \bigr) + \epsilon_m\\
&\le \frac{1}{m} \bigl( h(Y_{1}^m(T)) - h(Z_1^m) \bigr) + \epsilon_m\\
&\le \frac{1}{m} \bigl( h(\tilde{Y}_{1}^m) 
                      - h({Z}_1^m) \bigr) + \epsilon_m.
\end{align*}
By letting $m\to\infty$ and using the definition of
$\tilde{C}_\textit{FB}$, we obtain
\begin{align*}
C_{\textit{FB},n} \le h(\mathcal{\tilde{Y}}) - h(\mathcal{{Z}}) \le
\tilde{C}_\textit{FB}.
\end{align*}

For the other direction of the inequality, we use the notation
$\tilde{C}_\textit{FB}(P)$ and $C_{\textit{FB},n}(P)$ to stress the
dependence of feedback capacity on the power constraint $P$.  Given
$\epsilon > 0$, let $\{\tilde{X}_i = {V}_i + \sum_{k=1}^m b_k
Z_{i-k}\}_{i=-\infty}^\infty$ achieve $\tilde{C}_\textit{FB}(P) -
\epsilon$ under the power constraint $P$. The corresponding channel
output is given as
\[
\tilde{Y}_i = {V}_i + Z_i + \sum_{k=1}^m b_k Z_{i-k}.
\]

Now, we define a single-sided nonstationary
process $\{X_i\}_{i=1}^\infty$ as
\begin{align*}
X_i &= 
\begin{cases}
U_i + V_i + \sum_{k=1}^{i-1} b_k Z_{i-k},& i \le m\\
U_i + V_i + \sum_{k=1}^{m} b_k Z_{i-k},& i > m
\end{cases}
\end{align*}
where $U_1,U_2,\ldots$ are \iid $\sim N(0,\epsilon)$, independent of
$\{Z_i\}$ and $\{V_i\}$.  Thus, $X_i$
depends causally on $Z_1^{i-1}$ for all $i$. Let
$\{Y_i\}_{i=1}^\infty$ be the corresponding channel output $Y_i = X_i
+ Z_i.$ Since $E X_i^2 < \infty$ for $i \le m$
and 
\[
E X_i^2 = E \tilde{X}_i^2 + E U_i^2 = P + \epsilon
\]
for $i > m$, we have
\[
\lim_{n\to\infty} \frac{1}{n} \sum_{i=1}^n E X_i^2 = P + \epsilon.
\]
Also, since $h(Y_1^m|Y_{m+1}^n) \ge h(U_1^m|Y_{m+1}^n) = h(U_1^m) >
-\infty$ and ${Y}_i = \tilde{Y}_i + U_i$ for $i > m$,
\[
\lim_{n\to\infty} \frac{1}{n} h(Y_1^n)
=
\lim_{n\to\infty} \frac{1}{n} h(Y_{m+1}^n)
\ge h(\tilde{\mathcal{Y}}) = h(\mathcal{Z}) 
+ \tilde{C}_\textit{FB}(P) - \epsilon.
\]
Consequently, for $n$ sufficiently
large,
\[
\frac{1}{n} \sum_{i=1}^n E X_i^2 \le  (P + 2\epsilon)
\]
and
\[
\frac{1}{n} h(Y_1^n) - h(Z_1^n) \ge \tilde{C}_\textit{FB}(P) - 2\epsilon.
\]
Therefore, we can conclude that
\[
C_{\textit{FB},n}(P+2\epsilon) \ge \tilde{C}_\textit{FB}(P) - 2\epsilon
\]
for $n$ sufficiently large, whence
\[
\liminf_{n\to\infty} C_{\textit{FB},n}(P+2\epsilon) \ge
\tilde{C}_\textit{FB}(P) - 2\epsilon.
\]
But as a function of $P$, $C_{\textit{FB},n}(P)$ is concave on
$[0,\infty)$ and hence continuous on $(0,\infty)$.  (Recall
Proposition~\ref{prop:concavity}.) In fact, $C_{\textit{FB},n}(P)$
is continuous on $[0,\infty)$ because $C_n(P) \le C_{\textit{FB},n}(P)
\le 2 C_n(P)$ as shown by Cover and Pombra~\cite[Theorem
3]{Cover--Pombra1989} and $C_n(P) \to C_n(0) = 0$ as $P \to 0$.  For
the same reason, $\liminf_{n\to\infty} C_{\textit{FB},n}(P)$ is also
continuous in $P$.  Hence, by taking $\epsilon \to 0$, we can get
\[
\liminf_{n\to\infty} C_{\textit{FB},n}(P) \ge \tilde{C}_\textit{FB},
\]
which, combined with \eqref{eq:c-n-tilde-ineq}, implies that
\[
\lim_{n\to\infty} C_{\textit{FB},n}(P) = \tilde{C}_\textit{FB}(P).
\]
Incidentally, we have proved that the limit of $C_{\textit{FB},n}$
exists, with no resort to the superadditivity of $C_{\textit{FB},n}$;
cf.\@ \eqref{eq:state-y1}.
\end{proof}

\vspace*{1em}
\section{Optimal Feedback Coding Scheme}
\label{sec:optimal}

In this section, we explore many features of the variational
characterization of the Gaussian feedback capacity we established in
Theorem~\ref{thm:fb-capacity}:
\[
C_\textit{FB} = \sup_{S_V(\eit), B(\eit)}
\intt \half
\log
\frac{S_V(\eit) + |1 + B(\eit)|^2 S_Z(\eit)}{S_Z(\eit)}\dth
\tag{\ref{eq:cfb_var}}
\]
with the supremum taken over all $S_V(\eit) \ge 0$ and all strictly
causal polynomials $B(\eit)$ satisfying the power constraint
\[
\intt S_V(\eit) + |B(\eit)|^2 S_Z(\eit) \dth \le P.
\]
The ultimate goal is to obtain an explicit characterization of
$C_\textit{FB}$ as a function of $S_Z$ and $P$, or equivalently,
to solve the optimization problem
\begin{align}
\label{eq:var-opt1}
\begin{array}{l@{\quad}l}
\text{maximize}&\intt \log \bigl(S_V(\eit) 
+ |1+B(\eit)|^2 S_Z(\eit)\bigr) \dth\\[.75em]
\text{subject to}&S_V(\eit) \ge 0\\[.75em]
&B(\eit)=\sum_{k=1}^m b_k e^{ik\theta} \text{ strictly causal}\\[.75em]
&\intt S_V(\eit) + |B(\eit)|^2 S_Z(\eit) \dth \le P.
\end{array}
\end{align}

Recall that the optimization problem~\eqref{eq:var-opt1} is equivalent
to the maximization of the entropy rate of the stationary process
$\{Y_i = X_i + Z_i\}_{i=-\infty}^\infty$ over all stationary processes
$\{X_i\}_{i=-\infty}^\infty$ of the form $X_i = V_i +
\sum_{k=1}^m b_k Z_{i-k}$.  Whenever necessary, our discussion
will resort to the context of the stationary processes and the
corresponding entropy rate.  We start by studying the properties of
an optimal solution $(S_V^\star(\eit), B^\star(\eit))$ to
\eqref{eq:var-opt1}; cf.\@ Proposition~\ref{prop:optimal-cnfb}.

\begin{prop}[Necessary condition for an optimal $(S_V^\star, B^\star)$]
\label{prop:cfb-necessary}

An optimal solution $(S_V^\star(\eit), B^\star(\eit))$ to
\eqref{eq:var-opt1}, if one exists, must satisfy all of the following
conditions:
\begin{eqenumerate}
\item
\label{eq:var-power-cond}
\text{Power:}\quad $\intt S_V^\star(\eit) + |B^\star(\eit)|^2
S_Z(\eit) \dth = P$.
\item 
\label{eq:var-waterfilling-cond}
\text{Water-filling:}\quad $S_V^\star(\eit)$ water-fills the
modified noise spectrum $|1+B^\star(\eit)|^2 S_Z(\eit)$, that is,
\[
S_V^\star(\eit) ( S_Y^\star(\eit) - \lambda^\star) = 0 \quad\text{a.e.}
\]
where $S_Y^\star(\eit) = S_V^\star(\eit) + |1+B^\star(\eit)|^2
S_Z(\eit)$ and ${\lambda^\star = \essinf_{\theta \in [-\pi,\pi)}
S_Y^\star(\eit)}$.

\item
\label{eq:var-orth-cond}
\text{Orthogonality:}\quad The current input $X_n$ is
independent of the past output $\{Y_i\}_{-\infty}^{n-1}$.
Equivalently, 
\[S_V^\star(\eit) + B^\star(\eit)S_Z(\eit)(1 + B^\star(e^{-i\theta}))
\] is anticausal.
\end{eqenumerate}
Furthermore, if $S_Z(\eit)$ is bounded away from zero, i.e.,
$\essinf_\theta S_Z(\eit) > 0$, then there exist $S_V^\star(\eit) \in
L_1$ and $B^\star(\eit) = \sum_{k=1}^\infty b_k e^{ik\theta} \in
H_2$ attaining the maximum of \eqref{eq:var-opt1}.
\end{prop}
\begin{proof}
Necessity of the first two conditions is obvious; since each fixed $B$
gives a nonfeedback channel $|1+B(\eit)|^2S_Z(\eit)$ with the input
spectrum $S_V(\eit)$, the optimality conditions for the nonfeedback
capacity in Section~\ref{sec:nonfeedback} apply.

For the orthogonality condition~\eqref{eq:var-orth-cond}, we modify a
fixed-point method%
\footnote{This material on the fixed-point characterization of the
optimal $(K_{V,n}^\star, B_n^\star)$ was delivered with rigorous
details at ISIT 1994 by Ordentlich, but it never appeared in the
conference proceedings or other places.}
for the finite-dimensional
case by Ordentlich~\cite{Ordentlich1994}.  Suppose $(S_V, B)$ is
optimal and $S_V + B(1+\overline{B})S_Z$ is \emph{not} anticausal.
Then
\[
\intt \left(S_V + B(1+\overline{B})S_Z\right) e^{-in\theta} \dth
= \gamma \ne 0
\]
for some $n \ge 1$.  Let $A(z) = x z^n$ with $|x| < 1$.  Then
$(\tilde{S}_V, \tilde{B}) = (|1+A|^2 S_V, (1+A)(1+B)-1)$ is another
feasible solution to \eqref{eq:var-opt1}.  Since the corresponding
output spectrum 
\[
\tilde{S}_Y = \tilde{S}_V + |1+\tilde{B}|^2S_Z =
|1+A|^2 S_V + |1+A|^2 |1+B|^2 S_Z = |1+A|^2 S_Y,
\]
the entropy rate stays the same for $\tilde{S}_Y$ by Jensen's formula
\eqref{eq:jensen}.  On the other hand, the power usage becomes
\begin{align*}
\tilde{P}(x) &= \intt \tilde{S}_V + |\tilde{B}|^2 S_Z \dth\\
&= \intt |1+A|^2 S_V + |A(1+B)+B|^2 S_Z \dth\\
&= \intt S_V + |B|^2 S_Z \dth + 2 \intt A (S_V + \overline{B}(1+B)S_Z)\dth
+ \intt |A|^2 (S_V + |1+B|^2 S_Z) \dth\\
&= P + 2\gamma x + P_Y x^2
\end{align*} 
where $P_Y = \intt S_Y\dth$ is the original output power.  Since
$\tilde{P}(x)$ is quadratic in $x$ with the leading coefficient $P_Y >
0$, we can choose $x$ small with appropriate sign so that
$\tilde{P}(x) < P$.  But this implies that $(\tilde{S}_V, \tilde{B})$
achieves the same entropy rate as the original $(S_V, B)$ using
strictly less power.  This contradicts the optimality of $(S_V, B)$
and hence we have the anticausality of $S_V + B(1+\overline{B})S_Z$.

The proof of the existence of the optimal $(S_V^\star, B^\star)$ is rather
technical, so it will be given in the Appendix.
\end{proof}

Unlike the finite-dimensional case, the conditions
\eqref{eq:var-power-cond}--\eqref{eq:var-orth-cond} are not
necessarily sufficient; one can easily construct a suboptimal $(S_V,
B)$ satisfying the above conditions.  Nonetheless, we can deduce many
interesting observations from them.
\begin{coro}
\label{coro:fb-increment}
Feedback does not increase the capacity if and only if the noise spectrum is
white, i.e., $S_Z(\eit)$ is constant.
\end{coro}
\begin{proof}
Shannon's 1956 paper~\cite{Shannon1956} shows that feedback does not
increase the capacity for memoryless channels, taking care of the
sufficiency. 
(See also Kadota, Zakai, and
Ziv~\cite{Kadota--Zakai--Ziv1971a, Kadota--Zakai--Ziv1971b}.)

For the necessity, we assume that $S_Z$ is bounded away from zero
without loss of generality.  Indeed, we can use a small amount of
power to water-fill the spectrum first, then use the remaining power
to code with or without feedback.  If the stated claim is true, then
feedback increases the capacity for the modified channel and hence for
the original channel.  (For the nonfeedback coding, there is no loss
of optimality in dividing the power into two parts and water-filling
successively.)

Proceeding on to the proof of the necessity, suppose $S_X^\star(\eit)$
achieves the nonfeedback capacity and hence $(S_V^\star(\eit),
B^\star(\eit)) = (S_X^\star(\eit),0)$ achieves the feedback capacity.
Then, from the condition~\eqref{eq:var-orth-cond},
\[S_V^\star(\eit) + B^\star(\eit)
S_Z(\eit) (1 + B^\star(e^{-i\theta})) = S_X^\star(\eit)
\]
is anticausal and hence is white.  Therefore,
$S_Z(\eit)$ must be also white.
\end{proof}
\begin{coro}
\label{coro:zero-sv}
Suppose $(S_V^\star(\eit), B^\star(\eit))$ attains the maximum of
\eqref{eq:var-opt1}.  Then,
there exists ${B}^{\star\star}(\eit)$ such
that
\[
S_Y^\star(\eit) = S_V^\star(\eit) + |1+B^\star(\eit)|^2 S_Z(\eit) =
|1+{B}^{\star\star}(\eit)|^2 S_Z(\eit)
\]
and
\[
\intt S_V^\star(\eit) + |B^\star(\eit)|^2 S_Z(\eit) \dth = 
\intt |B^{\star\star}(\eit)|^2 S_Z(\eit) \dth.
\]
In particular, $(0, {B}^{\star\star}(\eit))$ attains the maximum of
\eqref{eq:var-opt1}.
\end{coro}
In order to prove Corollary~\ref{coro:zero-sv}, we need the following
simple result, which essentially establishes the optimality of the
original Schalkwijk--Kailath coding scheme for the additive white
Gaussian noise channel.

\begin{lemma}
\label{lemma:white}
Suppose the noise spectrum is white with $S_Z(\eit) \equiv N$.  Then,
the choice of $S_V^\star(\eit) \equiv 0$ and
\begin{align*}
B^\star(\eit) 
&= \frac{1 - a^{-1}\eit}{1 - a \eit} - 1
\end{align*}
with $a = \sqrt{\frac{N}{P+N}}$ achieves the feedback capacity
$C_\textit{FB} = C = \half \log \left( 1 + \frac{P}{N}\right)$ under the
power constraint $P$.  Furthermore, the resulting output spectrum is
given by
\[
S_Y^\star(\eit) \equiv P+N.
\]
\end{lemma}
\begin{proof}
We first check that
\[
S_Y^\star(z) = N(1+B^\star(z))(1+B^\star(z^{-1})) =
N\cdot\frac{1-a^{-1}z}{1-az}\cdot \frac{1-a^{-1}z^{-1}}{1-az^{-1}} = N
a^{-2} = P+N.
\]
On the other hand, since
\[
B^\star(\eit) = (1-a^{-2})\sum_{k=1}^\infty a^k
e^{ik\theta},
\]
we have
\begin{align*}
\intt |B^\star(\eit)|^2 S_Z(\eit) \dth &= N(1-a^{-2})^2 \intt
\Bigl|\sum_{k=1}^\infty a^k e^{ik\theta}\Bigr|^2 \dth\\
&= N(1-a^{-2})^2\cdot\frac{a^2}{1-a^2}\\
&= a^{-2}-1\\
&= P.
\end{align*}
Clearly, we have achieved $C_\textit{FB}(P) = \half \log \left( 1+
\frac{P}{N}\right).$
\end{proof}

The choice of the feedback filter $B^\star(\eit)$ is far from unique;
for example, we can use any causal filter derived from the normalized
Blaschke product as
\begin{equation}
\label{eq:white_b}
B(z) = \prod_{k=1}^\infty \frac{1-a_k^{-1} z^{j_k}}{1 - a_k z^{j_k}} - 1
\end{equation}
where $\{j_k\}_{k=1}^\infty$ is an arbitrary sequence of positive
integers and $\{a_k\}_{k=1}^\infty$ is a sequence of real numbers such
that $|a_k| < 1$ for all $k$ and $\prod_{k=1}^\infty a_k^2 = N/(P+N).$
(We will prove the optimality of these feedback filters later in the
next section.)  Note that there are filters that are not covered by
the form~\eqref{eq:white_b}, but still achieve the capacity for white
spectrum.

Now we move on to the proof of Corollary~\ref{coro:zero-sv}.

\begin{proof}[Proof of Corollary~\ref{coro:zero-sv}]  
Suppose 
\[
\intt S_V^\star(\eit) \dth = P_1
\]
and
\[
\intt |B^\star(\eit)|^2 S_Z(\eit) \dth = P - P_1.
\]
We assume $P_1 > 0$; otherwise, there is nothing to prove.

We argue that $S(\eit) := |1+B^\star(\eit)|^2 S_Z(\eit)$ must be
white.  Assume the contrary and consider the Gaussian feedback channel
with the noise spectrum $S(\eit)$ under the power constraint $P_1$.
But from Corollary~\ref{coro:fb-increment}, $(S_V^\star(\eit), 0)$ is
strictly dominated by some feedback coding scheme $(S_V(\eit),
B(\eit))$ with nonzero $B(\eit)$.  Hence, for the original channel, we
have a two-stage strategy $(S_V(\eit), (1+B^\star(\eit))(1+B(\eit)) -
1)$ with the corresponding output entropy higher than that of the
original $S_Y^\star(\eit)$, which contradicts the optimality of
$(S_V^\star(\eit), B^\star(\eit))$.

Now suppose the white spectrum $S(\eit)$ has the power, say, $N_1$.
From the water-filling condition~\eqref{eq:var-waterfilling-cond},
$S_V^\star \equiv P_1$ and the resulting output spectrum
$S_Y^\star(\eit) \equiv P_1 + N_1$.  On the other hand, from
Lemma~\ref{lemma:white}, we can achieve the feedback capacity $\half
\log(1+\frac{P_1}{N_1})$ for the new channel $S(\eit)$ by using
$B(\eit) = (1 - a^{-1}\eit)/(1 - a\eit)$, $a = \sqrt{N_1}/\sqrt{P_1 +
N_1}$.  Consequently, we can achieve the feedback capacity of the
original channel $S_Z(\eit)$ through a two-stage strategy: first
transform the channel into $S(\eit)$ using $B^\star(\eit)$, and then
use $B(\eit)$ for the channel $S(\eit)$.  The corresponding one-stage
filter is given by
\[
B^{\star\star}(\eit) = (1+B^\star(\eit))(1+B(\eit)) - 1
\]
and $(0, B^{\star\star}(\eit))$ achieves the feedback capacity with
the same output spectrum $S_Y^\star(\eit)$.
\end{proof}

\begin{remark}
\label{remark:sv}
We can make a somewhat stronger statement---if $S_Z$ is nonwhite,
then $S_V^\star$ must be zero.  To see this, first note from the
above proof that, if $S_V^\star$ is nonzero, then $|1+B^\star|^2S_Z$ and
$S_V^\star$, as well as $S_Y^\star$ should be white.  Now from the
orthogonality condition~\eqref{eq:var-orth-cond}, $S_V^\star +
B^\star S_Z (1+\overline{B^\star})$ is anticausal, or
equivalently, $S_Z (1+\overline{B^\star})$ is anticausal,
which is true only if $S_Z$ is white.
\end{remark}

The essential content of Corollary~\ref{coro:zero-sv} is that we can
restrict attention to the solutions of the form $(0, B(\eit))$,
even in the case the supremum in \eqref{eq:var-opt1} is not
attainable.  Indeed, we can easily modify the proof of
Corollary~\ref{coro:zero-sv} to show that for any solution $(S_V, B)$,
there exists another solution $(0, \tilde{B})$ such that the
corresponding output entropy rate is no less than the original under
the same power usage.  This observation yields a simpler
characterization of the feedback capacity.

\begin{theorem}
\label{thm:fb-capacity2}
Suppose that the stationary Gaussian noise process
$\{Z_i\}_{i=1}^\infty$ has the absolutely continuous power spectral
distribution $d\mu_Z(\theta) = S_Z(\eit)d\theta$.  Then, the feedback
capacity $C_\textit{FB}$ of the Gaussian channel $Y_i = X_i + Z_i,
\enspace i = 1,2,\ldots,$ under the power constraint $P$, is given by
\begin{equation}
C_\textit{FB} 
= \sup_{B(\eit)}
\intt \frac{1}{2} \log
\frac{|1+B(\eit)|^2 S_Z(\eit)}{S_Z(\eit)} \dth
= \sup_{B(\eit)}
\intt \frac{1}{2} \log {|1+B(\eit)|^2} \dth
\label{eq:cfb-no-sv}
\end{equation}
where the supremum is taken over all strictly causal polynomials
$B(\eit) = \sum_{k=1}^m b_k e^{ik\theta}$ 
satisfying the power constraint $\frac{1}{2\pi}\intt |B(\eit)|^2
S_Z(\eit) \,d\theta \le P.$
\end{theorem}

Although Proposition~\ref{prop:cfb-necessary} and its corollaries
reveal the structure of the capacity-achieving feedback filter, it is
still short of characterizing the capacity-achieving feedback filter
itself.  For example, we can show that there are more than one
feedback filter $B$ satisfying the orthogonality condition.  We remedy
the situation by deriving a universal upper bound on the feedback
capacity and finding the condition under which this upper bound is
tight.

We begin with a program similar to the one at the end of
Section~\ref{sec:nonfeedback}.  We will assume that $S_Z(\eit)$ is
bounded away from zero, which does not incur much loss of generality,
for we can always perturb the noise spectrum with little power without
changing the output entropy rate by much.  (Also recall that the
condition for existence of an optimal solution in
Proposition~\ref{prop:cfb-necessary}.)  From the canonical spectral
factorization theorem, we write $S_Z(\eit) =
H_Z(\eit)H_Z(e^{-i\theta})$ with $H_Z \in H_2$.  Since $S_Z(\eit)$ is
bounded away from zero, $1/H_Z \in H_\infty$.

Under the change of variable $
S_Y(\eit) = S_V(\eit) +|1+B(\eit)|^2S_Z(\eit),$
we rewrite the optimization problem \eqref{eq:var-opt1} as
\begin{align}
\label{eq:var-opt2}
\begin{array}{l@{\quad}l}
\text{maximize}&{\displaystyle \intt \log S_Y(\eit) \dth}\\[1.25em]
\text{subject to}&{\displaystyle S_Y(\eit) \ge |1+B(\eit)|^2 S_Z(\eit)}\\[.25em]
&{\displaystyle \intt S_Y(\eit) - (B(\eit) + B(e^{-i\theta}) + 1)S_Z(\eit) \dth \le P}\\[.8em]
&B(\eit) \in H_2 \text{ strictly causal.}
\end{array}
\end{align}
Take any $\nu \!>\! 0, \enspace \phi, \psi_1 \in L_\infty,$ and
$\psi_2, \psi_3 \in L_1$ such that $\phi > 0,\enspace\log \phi \in
L_1,\enspace\overline{\psi_2} {H_Z^{-1}} \in L_2, \enspace \psi_1 =
\nu - \phi \ge 0,\enspace A := {\psi_2} + \nu S_Z$ is anticausal, and
\begin{gather*}
\left[\hspace{0.3em}
\begin{matrix}
\psi_1(\eit) &\psi_2(\eit)\\[0.4em]
\overline{\psi_2(\eit)} & \psi_3(\eit)
\end{matrix}
\hspace{0.3em}\right] \geql 0.
\end{gather*}
Now that any feasible $B(\eit)$ and $S_Y(\eit)$ satisfy
\[
\left[\hspace{0.3em}
\begin{matrix}
S_Y(\eit) &1+B(\eit)\\[0.4em]
1+\overline{B(\eit)} & S_Z^{-1}(\eit)
\end{matrix}
\hspace{0.3em}\right] \geql 0,
\]
we have from Lemma~\ref{lemma:trace} that
\[
\tr\left(
\left[\hspace{0.3em}
\begin{matrix}
S_Y&1+B\\[0.4em]
1+\overline{B}&S_Z^{-1}
\end{matrix}
\hspace{0.3em}\right]
\left[\hspace{0.3em}
\begin{matrix}
\psi_1 &\psi_2\\[0.4em]
\overline{\psi_2} & \psi_3
\end{matrix}
\hspace{0.3em}\right] \right) = \psi_1 S_Y + \psi_2 (1+\overline{B}) +
\overline{\psi_2}(1 + B) + \psi_3 S_Z^{-1} \ge 0.
\]
Proceeding as in the nonfeedback case, we invoke the inequality $\log
x \le x - 1$ for all $x > 0$ with $x = \phi S_Y$ to get
\begin{align}
\nonumber
\log S_Y &\le -\log \phi + \phi S_Y - 1\\
\nonumber
&= -\log \phi + \nu S_Y - \psi_1 S_Y - 1\\
\label{eq:fb-var-ineq}
&\le -\log \phi + \nu S_Y + \psi_2(1+\overline{B}) 
+ \overline{\psi_2}(1+B) + \psi_3 S_Z^{-1} -1.
\end{align}
Furthermore, since $A = \psi_2 + \nu S_Z$ is anticausal and $B$ is
strictly causal, $A\overline{B} \in L_1$ is strictly anticausal;
recall Lemma~\ref{lemma:causality}.  (Indeed, $\overline{\psi_2} B =
(\overline{\psi_2} H_Z^{-1})\cdot(H_Z B) \in L_1$ since the first
factor is $L_2$ while the second factor is $H_2$.)  Hence
\begin{equation}
\label{eq:fb-causal-int}
\intt A(\eit)\overline{B(\eit)} \dth = \intt
\overline{A(\eit)}B(\eit)\dth = 0.
\end{equation}
By integrating both sides of \eqref{eq:fb-var-ineq}, we get
\begin{align}
\nonumber
\intt \log S_Y &\le \intt - \log \phi +
\nu S_Y + \psi_2(1+\overline{B}) + \overline{\psi_2}(1+B) 
+ \psi_3 S_Z^{-1} - 1\\[.25em]
\nonumber
&\le \intt - \log \phi +
\nu \bigl((B + \overline{B} + 1) S_Z + P\bigr)
+ \psi_2(1+\overline{B}) + \overline{\psi_2}(1+B) 
+ \psi_3 S_Z^{-1} - 1\\[.25em]
\nonumber
&=  \intt - \log \phi +
\psi_2 + \overline{\psi_2} + \psi_3 S_Z^{-1} + \nu (S_Z + P) -1
+ A\overline{B} + \overline{A}B \\[.25em]
\nonumber
&= \intt -\log \phi + \psi_2 + \overline{\psi_2} 
+ \psi_3 S_Z^{-1} + \nu(S_Z +  P) -1
\end{align}
where the second inequality follows from the power constraint
in~\eqref{eq:var-opt2} and the last equality follows from
\eqref{eq:fb-causal-int}.

In summary, we have derived a general upper bound
on the feedback capacity:
\begin{prop}
\label{prop:upper}
Suppose the noise power spectral density $S_Z(\eit)$ is bounded away
from zero and has the canonical spectral factorization $S_Z(\eit) =
|H_Z(\eit)|^2$.  Then, the feedback capacity $C_\textit{FB}$ under the
power constraint $P$ is upper bounded by
\begin{equation}
\label{eq:inf-dual}
C_\textit{FB} \le  
\half \intt \left[ -\log \!\left(\frac{\phi(\eit)}{S_Z(\eit)}\right)
+ \psi_2(\eit) + \overline{\psi_2(\eit)} 
+ \left(\frac{\psi_3(\eit)}{S_Z(\eit)}\right) 
+ \nu(S_Z(\eit) \!+\!  P) -1\right] \!\dth
\end{equation}
for any $\nu > 0, \enspace \phi, \psi_1 \in L_\infty,$ and $\psi_2,
\psi_3 \in L_1$ such that 
\begin{gather*}
\phi > 0\\
\log \phi \in L_1\\
\overline{\psi_2} H_Z^{-1} \in L_2\\
\psi_1 = \nu - \phi \ge 0\\
\psi_2 + \nu S_Z \text{ is anticausal}
\intertext{and}
\left[\hspace{0.3em}
\begin{matrix}
\psi_1(\eit) &\psi_2(\eit)\\[0.4em]
\overline{\psi_2(\eit)} & \psi_3(\eit)
\end{matrix}
\hspace{0.3em}\right] \geql 0.
\end{gather*}
\end{prop}

As before, for us, the major utility of this upper bound lies in the
characterization of the optimal solution $B^\star(\eit)$.  Tracing the
equality conditions in \eqref{eq:inf-dual}, we can establish the
following sufficient condition for the optimality of a specific
feedback filter $B(\eit)$.

\begin{prop}
\label{prop:fb-var-suff}
Suppose $S_Z(\eit)$ is bounded away from zero.  Suppose $B(z) \in
H_2$ is strictly causal (i.e., B(0) = 0) with
\begin{equation}
\label{eq:fb-var-power-eq}
\intt |B(\eit)|^2 S_Z(\eit) \dth = P.
\end{equation}
If there exists $\lambda > 0$ such that
\begin{eqenumerate}
\item
\label{eq:lambda-cond}
$\displaystyle \lambda \le \essinf_{\theta\in[-\pi,\pi)} {|1+B(\eit)|^2 S_Z(\eit)}$
\end{eqenumerate}
and that
\begin{eqenumerate}
\item
\label{eq:causality-cond}
${\displaystyle \frac{\lambda}{1+B(e^{-i\theta})} - B(\eit)S_Z(\eit)}$
is anticausal,
\end{eqenumerate}
then $B(\eit)$ achieves the feedback capacity; that is, $B(\eit)$
achieves the maximum of
\[
C_\textit{FB} =
\max_{B(\eit)} \intt \frac{1}{2} \log {|1+B(\eit)|^2} \dth
\tag{\ref{eq:cfb-no-sv}}
\]
over all strictly causal $B(\eit)$ satisfying $\intt |B(\eit)|^2
S_Z(\eit) \dth \le P$.
\end{prop}

\begin{proof}
Let $S_Y = |1+B|^2 S_Z \ge \lambda > 0 $. 
Let
\begin{align}
\nonumber
\nu &= 1/\lambda > 0\\
\label{eq:fb-var-f}
\phi &= S_Y^{-1} \in L_\infty\\
\nonumber
\psi_1 &= \nu - \phi = \frac{1}{\lambda} - \frac{1}{S_Y} \in L_\infty\\
\nonumber
\psi_2 &= - \psi_1 (1+B)S_Z  \in L_1 \nonumber
\intertext{and}
\nonumber
\psi_3 &= \psi_1^2 |1+B|^2 S_Z^2 \in L_1.
\end{align}
It is straightforward to verify that $\nu, \phi, \psi_1, \psi_2,$ and
$\psi_3$ defined above satisfy the conditions set forth for the upper
bound \eqref{eq:inf-dual}.  Moreover, from the condition
\eqref{eq:causality-cond},
\begin{align*}
\psi_2(\eit) + \nu S_Z(\eit)
&= -\nu (1+B(\eit))S_Z(\eit) + \frac{1}{1+B(e^{-i\theta})} + \nu S_Z(\eit) \\
&= -\frac{1}{\lambda} 
\left ( \frac{\lambda}{1+B(e^{-i\theta})} - B(\eit)S_Z(\eit)\right)
\end{align*}
is anticausal.

Now, it is easy to check that
\[
\psi_1 S_Y + \psi_2(1+\overline{B}) 
= \overline{\psi_2}(1+B) + \psi_3 S_Z^{-1}
= 0
\]
which makes the second inequality of \eqref{eq:fb-var-ineq} an
equality.  On the other hand, \eqref{eq:fb-var-f} makes the first
inequality of \eqref{eq:fb-var-ineq} an equality while the condition
\eqref{eq:fb-var-power-eq} makes the second inequality in
\eqref{eq:inf-dual} an equality.  Combining these three equality
conditions, we have the equality in \eqref{eq:inf-dual}, 
and hence the optimality of $B(\eit)$.
\end{proof}

Note that the causality condition \eqref{eq:causality-cond} in the
above proposition implies the orthogonality condition
\eqref{eq:var-orth-cond} in Proposition~\ref{prop:cfb-necessary}, for,
if $\lambda/(1+B(e^{-i\theta})) - B(\eit)S_Z(\eit)$ is
anticausal, $B(\eit)S_Z(\eit)(1+B(e^{-i\theta}))$ is
anticausal.  The converse is not necessarily true and hence there is a
nontrivial gap between the necessary conditions in
Proposition~\ref{prop:cfb-necessary} and the sufficient condition in
Proposition~\ref{prop:fb-var-suff}.

Although the conditions
\eqref{eq:fb-var-power-eq}--\eqref{eq:causality-cond} give a
characterization of the optimal feedback filter, this characterization
is rather implicit and still falls short of yielding what can be
called a closed-form solution for the feedback capacity
problem~\eqref{eq:cfb-no-sv}.  In the next two sections, we find more
explicit answers by narrowing attention to special classes of
noise spectra.

\vspace*{2em}
\section{First-order ARMA noise spectrum}
\label{sec:arma1}

As a gentle start, we consider the zeroth-order autoregressive
moving average (=white) noise spectrum first.  Since the spectrum is
bounded away from zero, from Theorem~\ref{thm:fb-capacity2} and
Proposition~\ref{prop:fb-var-suff}, the feedback capacity is
characterized in the following variational formula:
\[
C_\textit{FB} 
= \max_{B(\eit)}
\intt \frac{1}{2} \log
{|1+B(\eit)|^2} \dth
\]
where the maximum is taken over all strictly causal filters $B(\eit)
= \sum_{k=1}^\infty b_k e^{ik\theta}$ satisfying the
power constraint $\frac{1}{2\pi}\intt |B(\eit)|^2 S_Z(\eit) \,d\theta
\le P.$ In Lemma~\ref{lemma:white}, we already proved that
\[
B(z) = \frac{1- a^{-1} z} {1 - a z} - 1,\qquad -1 < a < 1
\]
achieves the feedback capacity for the noise spectrum $S_Z \equiv N$
under the power constraint $P = N(1/a^2 - 1)$.  Here we establish the
optimality of filters of the form
\begin{equation}
\label{eq:prod_bz}
B(z) = \prod_{k=1}^\infty \frac{1-a_k^{-1} z^{j_k}}{1 - a_k z^{j_k}} - 1
\end{equation}
where $\{j_k\}_{k=1}^\infty$ is an arbitrary sequence of positive
integers and $\{a_k\}_{k=1}^\infty$ is a sequence of real numbers such
that $|a_k| < 1$ for all $k$ and $\prod_{k=1}^\infty a_k^2 = N/(P+N).$

Although we can employ direct brute-force calculation similar to the
proof of Lemma~\ref{lemma:white}, we apply
Proposition~\ref{prop:fb-var-suff} as an elegant alternative.  Observe
that the resulting output spectrum is
\[
S_Y(z) = N \cdot \prod_{k=1}^\infty \left (\frac{1 - a_k^{-1} z^{j_k}}{1 - a_k z^{j_k}} \cdot
 \frac{1 - a_k^{-1} z^{-j_k}}{1 - a_k z^{-j_k}}\right) 
= N \cdot \prod_{k=1}^\infty \frac{1}{a_k^2} = P+N.
\]
Take $\lambda = \min_{|z| = 1} S_Y(z) = 1/(P+N) = N 
\prod_{k=1}^\infty(1/a_k^2).$  Then,
\begin{align*}
\frac{\lambda}{1 + B(z^{-1})} - B(z)S_Z(z) 
&= \left(N \prod_{k=1}^\infty \frac{1}{a_k^2} \right)\cdot
\left(\,\prod_{k=1}^\infty \frac{1 
- a_k z^{-j_k}}{1-a_k^{-1}z^{-j_k}} \,\right)
- N \left(\,\prod_{k=1}^\infty 
\frac{1 - a_k^{-1}z^{j_k}}{1-a_kz^{j_k}} - 1\,\right)\\[.5em]
&= N \left(\, \prod_{k=1}^\infty \frac{1 - a_k^{-1}z^{j_k}}{1-a_kz^{j_k}} 
- \prod_{k=1}^\infty \frac{1 - a_k^{-1}z^{j_k}}{1-a_kz^{j_k}} 
+ 1\,\right)\\[.5em]
&= N.
\end{align*}
Hence, the feedback filter $B(z)$ given in \eqref{eq:prod_bz}
satisfies the sufficient condition in
Proposition~\ref{prop:fb-var-suff}, which confirms its optimality.

Now we turn our attention to the first-order autoregressive
moving average noise spectrum $S_Z(\eit)$, defined by
\begin{equation}
\label{eq:arma1-spectrum}
S_Z(\eit) = \left|\frac{1 + \alpha \eit}{1 + \beta \eit}\right|^2
\end{equation}
for $\alpha \in [-1, 1]$ and $\beta \in (-1, 1)$.  (The case $|\alpha|
> 1$ can be taken care of by the canonical spectral factorization and
proper scaling.)  This spectral density corresponds to the stationary
noise process given by
\[
Z_i + \beta Z_{i-1}  =  U_i + \alpha U_{i-1}, \qquad i \in \mathbb{Z}
\]
where $\{U_i\}_{i=-\infty}^\infty$ is a white Gaussian process with
zero mean and unit variance.
\begin{theorem}
\label{thm:arma1}
Suppose the noise process $\{Z_i\}_{i=1}^\infty$ has the power
spectral density $S_Z(\eit)$ defined in \eqref{eq:arma1-spectrum}.  Then,
the feedback capacity $C_\textit{FB}$ of the Gaussian channel $Y_i = X_i
+ Z_i, \enspace i=1,2,\ldots,$ under the power constraint $P$, is given
by
\[
C_\textit{FB} = - \log x_0
\]
where $x_0$ is the unique positive root of the fourth-order polynomial
\begin{equation}
\label{eq:arma1-poly}
P\,x^2 = 
\frac{(1 - x^2)( 1 + \sigma \alpha x)^2}{(1 + \sigma \beta x)^2}
\end{equation}
and
\[
\sigma = \sgn(\beta-\alpha) = 
\begin{cases}
1, & \beta > \alpha\\
0, & \beta = \alpha\\
-1, & \beta < \alpha.
\end{cases}
\]
\end{theorem}

\begin{proof}
Without loss of generality, we assume $|\alpha| < 1$; for the
case $|\alpha| = 1$, we can perturb the noise spectrum with small
power to transform it into another ARMA(1) spectrum with $|\alpha| < 1$.
Under the assumption $|\alpha| < 1$, $S_Z(\eit)$ is bounded away from
zero, so we can apply Proposition~\ref{prop:fb-var-suff}.

Here is the bare-bone summary of the proof: We will take the feedback filter of
the form
\begin{equation}
\label{eq:b_z}
B(z) = \frac{1+\beta z}{1+\alpha z} \cdot \frac{yz}{1-\sigma xz}
\end{equation}
where $x \in (0,1)$ is an arbitrary parameter corresponding to each
power constraint $P \in (0,\infty)$ under the choice of 
\begin{equation}
\label{eq:y1}
y = \frac{x^2-1}{\sigma x}\cdot\frac{1+\alpha \sigma x}{1+\beta \sigma x}
= - P \sigma x \left(\frac{1+\beta\sigma x}{1+\alpha\sigma x}\right) .
\end{equation}
Then, we can show that $B(z)$ satisfies the sufficient
condition in Proposition~\ref{prop:fb-var-suff} under the power
constraint
\[
P = \intt |B(\eit)|^2 S_Z(\eit) \dth
= \intt \frac{y^2}{|1-x\eit|^2} \dth
= \frac{y^2}{1-x^2} 
= \frac{1-x^2}{x^2}\left(\frac{1+\alpha\sigma x}{1+\beta\sigma x}\right)^2
\]
with the corresponding information rate given by
\[
h(\mathcal{Y}) - h(\mathcal{Z}) = \intt \half \log |1+B(\eit)|^2 \dth
= -\half \log x^2.
\]
The rest of the proof is the actual implementation of this idea.

Assume $-1 < \alpha < \beta < 1$.  Given $x \in (0,1),$ take 
$y$ as in \eqref{eq:y1}.
Then, we can factor $1+B(z)$ as
\begin{align*}
1 + B(z) &= 1 + \frac{1+\beta z}{1+\alpha z} \cdot \frac{yz}{1-xz}\\
&= \frac{1 - (\alpha - x + y) z + (\beta y - \alpha x) z^2}
{( 1+\alpha z)(1 - x z)}\\
&= \frac{(1 + (\alpha x - \beta y)x z)(1-x^{-1}z)}
{(1+\alpha z)(1 - x z)}\\
&= \frac{(1 - r z)(1-x^{-1}z)}
{(1+\alpha z)(1 - x z)}
\end{align*}
where $r = -(\alpha x - \beta y) x$.
The corresponding output spectrum is given by
\begin{align}
\nonumber
S_Y(z) &= 
(1+B(z))S_Z(z)(1+B(z^{-1}))\\
\nonumber
&= 
\frac{(1 - r z)(1-x^{-1}z)}
{(1+\alpha z)(1 - x z)}
\frac{(1+\alpha z)}{(1+\beta z)}
\frac{(1+\alpha z^{-1})}{(1+\beta z^{-1})}
\frac{(1 - r z^{-1})(1-x^{-1}z^{-1})}
{(1+\alpha z^{-1})(1 - x z^{-1})}\\
\label{eq:s_y}
&=
\frac{1}{x^2}\frac{1-rz}{1+\beta z}\frac{1-rz^{-1}}{1+\beta z^{-1}}.
\end{align}

We first check that $|r| < 1$.  Indeed, from \eqref{eq:y1}, we can
express $r = r(x)$ as
\begin{align}
\label{eq:rx_1}
r(x) &= \frac{(\beta-\alpha) x^2 - \alpha\beta x - \beta}{1 + \beta x}\\
\label{eq:rx_2}
&= -\beta + (\beta-\alpha)\cdot\frac{\beta+x}{\beta + \frac{1}{x}}\\
\nonumber
&= -\beta \left(1 - \frac{\beta +x}{\beta + \frac{1}{x}} \right) 
-\alpha \left(\frac{\beta+x}{\beta + \frac{1}{x}}\right).
\end{align}
When $\beta \ge 0$, $0 < (\beta+x)/(\beta+x^{-1}) < 1$ (recall $0 < x
< 1)$ so that $-r$ is a convex combination of $\alpha$ and $\beta$;
hence $-1 < r < 1$.  When $\beta < 0$, we differentiate
\eqref{eq:rx_1} to find that $r'(0) < 0,$ $r'(1) > 0,$ $\max_{x \in
[0,1]} r(x) = \max\{r(0), r(1)\},$ and that there exists a unique
$x^\star \in (0,1)$ attaining the minimum of $r(x)$ on $[0,1]$.  Since
$r(0) = -\beta$ and $r(1) = -\alpha$, it sufficies to check that
$r(x^\star) > -1$.  We have
\begin{align*}
0 &= \left(\frac{\partial}{\partial x}
\bigl((\beta-\alpha) x^2 - \alpha\beta x -
\beta\bigr)\right)(1+\beta x) - \bigl((\beta-\alpha) x^2 - \alpha\beta x -
\beta\bigr)\frac{\partial}{\partial x}(1+\beta x)\\
&= \bigl(2(\beta-\alpha)x - \alpha
\beta)(1+\beta x) - \beta \bigl((\beta-\alpha) x^2 - \alpha\beta x -
\beta\bigr)\\
&= (\beta-\alpha)(\beta x^2 + 2x + \beta)
\end{align*}
at $x = x^\star$, whence
\[
r(x^\star) = \frac{2(\beta-\alpha) x^\star - \alpha \beta}{\beta}
= 2 x^\star + \alpha (x^\star)^2 \ge 0.
\]
Therefore, $|r| < 1$.

Now let
\[
\lambda = S_Y(x) = 
\frac{1}{x^2}\left(\frac{(1-rx)(1-rx^{-1})}{(1+\beta x)(1+\beta x^{-1})}\right)
= \frac{1}{x^2}\left(\frac{(1-rx)(x-r)}{(1+\beta x)(x+\beta)}\right). 
\]
We will show that 
\[
0 < \lambda \le \min_{\theta\in[-\pi,\pi)} S_Y(\eit).
\]
For the positivity of $\lambda$, it suffices to show that
$(x-r)/(x+\beta)$ is positive.  From \eqref{eq:rx_2}, we have
\begin{equation}
\label{eq:x+r}
x - r = (x + \beta) \left( 1 - \frac {\beta - \alpha}{\beta + x^{-1}}\right)
= (x + \beta) \left(\frac{\alpha + x^{-1}}{\beta + x^{-1}}\right)
\end{equation}
so that $(x - r)/(x+\beta)$ is positive.  (The case $x + \beta = 0$ is
trivial since $r(-\beta) = -\beta$.)

The upper bound requires a little more work.  Let
\[
f(u) = \frac{(1 +r^2) - 2 r u}{(1 + \beta^2) + 2\beta u}
\]
for $-\infty < u < \infty$.  Then, we can express
\[
S_Y(\eit) = \frac{(1 - r \eit)(1-r e^{-i\theta})}
{(1 + \beta \eit)(1 + \beta e^{-i\theta})}
= f(\cos\theta)
\]
for $\theta \in [-\pi,\pi)$ and similarly express
\[
S_Y(x) = f\left(\frac{x+x^{-1}}{2}\right).
\]
Since the linear fractional function $f(u)$ does not have a
singularity in $[-1,1]$, the minimum occurs at one of the end points
and
\[
\min_\theta S_Y(\eit) = \min_\theta f(\cos\theta) =
\min\{f(1),f(-1)\}.
\]
We consider different cases.
\setenumerate{topsep=.5em, partopsep=0em, leftmargin=4\parindent,
rightmargin=0\parindent, labelsep=1em, itemsep=3pt, itemindent=1.06\parindent}
\begin{enumerate}
\item[Case 1:]
$\beta \ge 0$.  Then, $f(u)$ is decreasing on $(-\half(\beta + \beta^{-1}), \infty)$
since
\[
f'(u) = -\frac{(\beta + r)(1+\beta r)}{(1+\beta^2) + 2\beta u}
\]
and 
\[
\beta + r = (\beta - \alpha)\left(\frac{\beta + x}{\beta + x^{-1}}\right)
\]
is positive.  (Recall the standing assumption $\beta - \alpha > 0$.)
By Jensen's inequality, 
\[
\frac{x + x^{-1}}{2} > 1
\]
so that
\[
f(-1) \ge f(1) \ge f\left(\frac{x+x^{-1}}{2}\right) = \lambda.
\]

\item[Case 2:] $0 < -\beta < x$.  Same as the previous
case since $\beta + r$ is positive.

\item[Case 3:] $0 < -\beta  = x$.  As we saw before, $r =
  -\beta$ so that $f(u)$ is constant for all $u$.  

\item[Case 4:] $0 < x < -\beta$.  Since $f'(u) > 0$ with a singularity
at $-(\beta+ \beta^{-1})/2 > 1$ and 
\[
\frac{x + x^{-1}}{2} > -\frac{{\beta} + \beta^{-1}}{2},
\]
we have
\[
f\left(\frac{x+x^{-1}}{2}\right) \le \inf \{f(u): u < -(\beta+ \beta^{-1})/{2}\}
\le \min\{f(1), f(-1)\}.
\]
\end{enumerate}
Therefore, $B(\eit)$ and $\lambda$ satisfy the condition
\eqref{eq:lambda-cond} in Proposition~\ref{prop:fb-var-suff}.

Finally we check the condition~\eqref{eq:causality-cond}, namely,
anticausality of 
\begin{align}
\nonumber
\frac{\lambda}{1 + B(z^{-1})} - B(z) S_Z(z)
&= \lambda
\left(\frac{(1+\alpha z^{-1})(1-x z^{-1})}
{(1- r z^{-1})(1-x^{-1}z^{-1}}\right)
- \frac{(1+\alpha z^{-1})(yz)}{(1+\beta z^{-1})(1 - xz)}\\
\label{eq:anticausal-eq}
&= (1 + \alpha z^{-1}) \left[
\frac{\lambda(1+\beta z^{-1})(x^2 - x z) - (yz)(1-rz^{-1})}
{(1-rz^{-1})(1+\beta z^{-1})(1-xz)}\right].
\end{align}  
From \eqref{eq:y1} and \eqref{eq:x+r}, we have
\begin{align*}
\lambda (1+ \beta x) (x^2 - 1) - (y x^{-1}) (1 - rx)
&= \frac{1-rx}{x^2} \left( \frac{(x-r)(x^2-1)}{x+\beta} 
- \frac{(x^2-1)(1+\alpha x)}{1+\beta x}\right)\\
&= \frac{(1-rx)(x^2-1)}{x^2}\left( \frac{x-r}{x+\beta} - \frac{1+\alpha x}{1+\beta x}\right)\\
&= 0.
\end{align*}
Hence, the numerator of \eqref{eq:anticausal-eq} has a factor
$(1-xz)$, so that \eqref{eq:anticausal-eq} is anticausal and the
condition \eqref{eq:causality-cond} is satisfied.  This establishes
the optimality of $B(z)$ defined in \eqref{eq:b_z} with $0< x <1$ and
$y$ satisfying \eqref{eq:y1}.  From Jensen's formula~\eqref{eq:jensen},
we see that the corresponding feedback capacity is given by
\[
C_\textit{FB}(x) = \intt \half \log S_Y(\eit) \dth = - \half \log x^2
\]
under the power constraint
\[
P(x) = \frac{y^2}{1-x^2} = \frac{(1-x^2)(1+\alpha x)^2}{x^2(1+ \beta x)^2}.
\]

The case $\beta < \alpha$ can be treated similarly with $x < 0$, while
the case $\beta = \alpha$ (i.e., $S_Z \equiv 1$) is trivial.  This completes
the proof of Theorem~\ref{thm:arma1}.
\end{proof}

Now we interpret in several ways the optimal feedback filter
$B^\star(z)$ we found in \eqref{eq:b_z}.  First, we show that the
celebrated Schalkwijk--Kailath signaling scheme is asymptotically
equivalent to our feedback filter $B^\star$, establishing the
optimality of the Schalkwijk--Kailath coding scheme for the
ARMA(1) noise spectrum.

Consider the following coding scheme.  Let $V \sim N(0,1)$.  Over the
Gaussian channel $Y_i = X_i + Z_i$ with the noise spectral density
\[
S_Z(\eit) = \left|\frac{1 + \alpha \eit}{1 + \beta \eit}\right|^2,
\]
the transmitter initially sends
\begin{equation}
\label{eq:x1}
X_1 = V
\end{equation}
and subsequently sends
\begin{equation}
\label{eq:xn}
X_n = (\sigma x)^{-(n-1)} (V - \hat{V}_{n-1}),\qquad n = 2,3,\ldots
\end{equation}
where $\sigma = \sgn(\beta - \alpha),$ $x$ is the unique positive root
of the fourth-order polynomial \eqref{eq:arma1-poly}, and
\[
\hat{V}_{n} = \hat{V}_n(Y^n) = E(V|Y_1,\ldots,Y_{n})
\]
is the minimum
mean-squared error estimate of $V$ given the channel output signals
$Y^n = (Y_1,\ldots,Y_{n})$.

For all $m < n$, we have
\begin{align}
\nonumber
X_n &= (\sigma x)^{-(n-1)}\bigl(
V - E(V|Y^{m-1}) + E(V|Y^{m-1}) - E(V|Y^{n-1})\bigr)\\
\label{eq:x_m}
&= (\sigma x)^{m-n} (X_m - E(X_m|Y^{n-1})) \\
\nonumber
&= (\sigma x)^{m-n} (Y_m - Z_m - E(Y_m - Z_m | Y^{n-1})) \\
\label{eq:z_m}
&= -(\sigma x)^{m-n} (Z_m - E(Z_m | Y^{n-1})).
\end{align}
Furthermore, since $Z_n = -\beta Z_{n-1} + U_n + \alpha U_{n-1}$ with
white $\{U_i\}$, we can show that
\[
Z_n \approx -(\beta - \alpha) \sum_{k=1}^{n-1} (-\alpha)^{k-1} Z_{n-k} + U_n
\]
for large $n$, which, combined with \eqref{eq:z_m}, implies that
\begin{equation}
\label{eq:z_n-approx}
Z_n - E(Z_n|Y^{n-1}) \approx 
\left(\frac{\beta-\alpha}{\alpha + (\sigma x)^{-1}}
\right)X_n + U_n
\end{equation}
for large $n$.  When $\alpha \ne \beta$, that is, when the noise
spectrum is nonwhite, \eqref{eq:z_n-approx} is equivalent to
\begin{equation}
\label{eq:z_n}
X_n \approx \frac{\alpha + (\sigma x)^{-1}}{\beta - \alpha}
(E(Z_n|Z^{n-1}) - E(Z_n|Y^{n-1})).
\end{equation}
Now from \eqref{eq:x_m} with $m = n-1$ and the orthogonality of
$X_{n-1}$ and $Y^{n-2}$, 
\begin{align}
\nonumber
X_n &= (\sigma x)^{-1} \bigl(X_{n-1} - E(X_{n-1}|Y^{n-1})\bigr)\\
\nonumber
&= (\sigma x)^{-1} \bigl(X_{n-1} - E(X_{n-1}|Y^{n-2}) 
   - E(X_{n-1}|\tilde{Y}_{n-1})\bigr)\\
\label{eq:x_n-approx}
&= (\sigma x)^{-1} \bigl(X_{n-1} - E(X_{n-1}|\tilde{Y}_{n-1})\bigr)
\end{align}
where $\tilde{Y}_{n-1} := Y_{n-1} - E(Y_{n-1}|Y^{n-2})$ is the
innovation of the output process at time $n-1$.  Also from
\eqref{eq:z_n-approx} and the orthogonality of $X_{n-1}$ and
$Y^{n-2}$, we have
\begin{align*}
\tilde{Y}_{n-1} &= X_{n-1} + Z_{n-1} - E(X_{n-1} + Z_{n-1} | Y^{n-2})\\
&= X_{n-1} + Z_{n-1} - E(Z_{n-1}| Y^{n-2})\\
&\approx
c X_{n-1} + U_{n-1}
\end{align*}
where
\[
c = 1 + \frac{\beta-\alpha}{\alpha + (\sigma x)^{-1}}
= \frac{1+\beta\sigma x}{1+\alpha \sigma x}.
\]
Finally, returning to \eqref{eq:x_n-approx}, we can easily see that
\begin{align*}
X_n &\approx \frac{(\sigma x)^{-1}}{c^2 P + 1} (X_{n-1} - c P U_{n-1})\\
&= \sigma x X_{n-1} - y U_{n-1},
\end{align*}
where $x$ and $y$ are the constants given by \eqref{eq:arma1-poly} and
\eqref{eq:y1}.  Therefore, the feedback coding scheme given by
\eqref{eq:x1} and \eqref{eq:xn} is asymptotically equivalent to
filtering the noise through the feedback filter
\[
B(z) = \frac{1+\beta z}{1+\alpha z} \cdot \frac{yz}{1-\sigma xz},
\]
which is exactly equal to the optimal feedback filter \eqref{eq:b_z}
we found in the proof of Theorem~\ref{thm:arma1}.

For a more rigorous analysis, we can also show that 
\[
\liminf_{n\to\infty} \frac{1}{2n} I(V; \hat{V}_{n}) \ge \half \log
\left(\frac{1}{x^2}\right)
\]
while 
\[
\limsup_{n\to\infty} \frac{1}{n} \sum_{i=1}^n X_i^2 \le P
\]
under the coding scheme \eqref{eq:xn}.  
Recall that
\begin{align*}
Y_n &= (\sigma x)^{-(n-1)}(V - \hat{V}_{n-1}(Y^{n-1})) + Z_n 
\end{align*}
and define
\begin{align*}
Y'_1 &= Y_1,\\
Y'_n &= (Y_n + (\sigma x)^{-(n-1)} \hat{V}_{n-1}) + \beta (Y_{n-1}
+ (\sigma x)^{-(n-2)} \hat{V}_{n-2}), \qquad n \ge 2,
\intertext{and}
Y''_n &= \sum_{k=1}^n (-\alpha)^{n-k} Y'_k, \qquad n \ge 1.
\end{align*}
Clearly, $Y''_n$ can be represented as a linear combination of
$Y_1,\ldots, Y_n$ and therefore, for any $c_2, c_3, \ldots$,
\begin{equation}
\label{eq:ck}
E (V - \hat{V}_n)^2 \le
E \left(V - \left(\sum_{k=2}^n c_k Y''_k\right)\right)^2.
\end{equation}
Now we express
\begin{align*}
Y'_n 
&= (\sigma x)^{-(n-1)}(1+\sigma \beta x) V + Z_n + \beta Z_{n-1}\\
&= (\sigma x)^{-(n-1)}(1+\sigma \beta x) V + U_n + \alpha U_{n-1}
\end{align*}
and
\begin{align*}
Y''_n = d_n V + U_n + (-\alpha)^{n-1} U'
\end{align*}
where $U' = \alpha U_0 - \beta Z_0$ and
\begin{align*}
d_n &= (1+\sigma \beta x) \left(\sum_{k=1}^n (-\alpha)^{n-k} 
(\sigma x)^{-(k-1)}\right)\\
&= \left(\frac{1+\sigma \beta x}{1+\sigma \alpha x}\right)
 ( 1 - (-\sigma \alpha x)^n)
(\sigma x)^{-(n-1)}.
\end{align*} 
By taking $c_k = d_k$ in \eqref{eq:ck}, we can easily verify that
\[
\frac{E (V - (\sum_{k=2}^{n} d_k Y''_k))^2}
{E (V - (\sum_{k=2}^{n-1} d_k Y''_k))^2} \longrightarrow \frac{1}{x^2},
\]
whence
\[
\limsup_{n\to\infty} \frac{1}{n} \log E (V - \hat{V}_{n})^2 \le \log
\left(\frac{1}{x^2}\right)
\]
and
\[
\liminf_{n\to\infty} \frac{1}{2n} I(V; \hat{V}_{n}) \ge \half \log
\left(\frac{1}{x^2}\right).
\]
On the other hand, 
\begin{align*}
E X_n^2 &= x^{-2(n-1)} E (V - \hat{V}_{n-1})^2\\
&\le x^{-2(n-1)}{E \left(V - \left( \sum_{k=2}^{n-1} 
d_k Y''_k\right)\right)^2},
\end{align*}
which converges to
\begin{align*}
\lim_{n\to\infty} \frac{x^{-2(n-1)}}{\sum_{k=2}^{n-1} d_k^2}
= \frac{(1+\sigma \alpha x)^2}{(1 + \sigma \beta x)^2} \cdot (x^{-2} - 1)
= P.
\end{align*}

The coding scheme described above uses the minimum mean-square error
decoding of the message $V$, or equivalently, the joint typicality
decoding of the Gaussian random codeword $V$, based on the general
asymptotic equipartition property of Gaussian processes shown by Cover
and Pombra~\cite[Theorem 2]{Cover--Pombra1989}.  It is fairly
straightforward to transform the Gaussian coding scheme to the
original Schalkwijk--Kailath coding scheme.  Here we sketch the
standard procedure.  A detailed analysis is given in
Butman~\cite{Butman1969, Butman1976}.

Instead of the Gaussian codebook $V$, the transmitter initially sends
a real number $\theta$ that is chosen from some equally spaced signal
constellation $\Theta$, say,
\[
\Theta = \{-1, -1+\delta, -1+2\delta, \ldots, 1-2\delta, 1-\delta, 1\},\qquad
\delta = 2/(2^{nR} - 1)
\]
and subsequently sends $\theta - \hat{\theta}_n$ (up to the same
scaling as before) at time $n$, where $\hat{\theta}_n$ is the
\emph{minimum variance unbiased linear} estimate of $\theta$ given
$Y^{n-1}$.  Now we can verify that the optimal
\emph{maximum-likelihood} decoding is equivalent to find $\theta^\star
\in\Theta$ that is closest to $\hat{\theta}_n$, which results in the
error probability
\[
P_e^{(n)} \le \erfc\sqrt{c_0 x_0^{-2n} / 2^{2nR}}
\]
where $x_0$ is the unique positive root of \eqref{eq:arma1-poly},
$c_0$ is a constant independent of $n$, and
\[
\erfc(x) = \frac{2}{\sqrt{\pi}} \int_{x}^\infty \exp(-t^2) dt
\]
is the complementary error function. 
Now we can easily see that $P_e$ decays doubly
exponentially fast as far as $R < - \log x_0 = C_\textit{FB}.$ Finally
note that the doubly exponential decay of error probability can be
raised to an arbitrary higher order by modifying the adaptive power
allocation scheme by Pinsker~\cite{Pinsker1968},
Kramer~\cite{Kramer1969}, and Zigangirov~\cite{Zigangirov1970}.  Also
note that \eqref{eq:x_m}, \eqref{eq:z_m}, and \eqref{eq:z_n} give
interesting alternative interpretations of the Schalkwijk--Kailath
coding scheme; the optimal transmitter refines the receiver's
knowledge of any past input \eqref{eq:x_m}, or equivalently, any past
noise \eqref{eq:z_m}.  Also asymptotically, the optimal transmitter
sends the difference between what he knows about the upcoming noise
and what the receiver knows about it \eqref{eq:z_n}.

Before we move on to a more general class of noise spectra, we provide
another angle on the optimal coding scheme by considering the
following state-space model of the ARMA(1) noise process:
\begin{align*}
\label{eq:state-z}
\begin{array}{r@{\;}l}
S_{n+1} &= -\beta S_n + U_n\\[.25em]
Z_n &= (\alpha-\beta) S_n + U_n
\end{array}
\end{align*}
where $\{U_i\}_{i=0}^\infty$ are independent and identically
distributed zero-mean unit-variance Gaussian random variables, and the
\emph{state} $S_n$ is independent of $U_n$ for each $n$.  It is easy
to check that this state-space model represents the noise
spectrum
\[
S_Z(\eit) = |H_Z(\eit)|^2 = 
\left|\frac{1 + \alpha \eit}{1 + \beta \eit}\right|^2.
\tag{\ref{eq:arma1-spectrum}}
\]
For simplicity, we consider a
slightly nonstationary noise model by assuming that $S_0 = U_0 = 0.$
One can prove that this does not change the feedback
capacity~\cite[Appendix]{Kim2004}, which implies that the Gaussian
feedback channel $Y_i = X_i + Z_i$ with the noise
spectrum~\eqref{eq:arma1-spectrum} is asymptotically equivalent to the
intersymbol interference channel
\[
Y'_k = \sum_{j=1}^k g_{k-j} X_j + U_k
\]
where $\{g_k\}_{k=0}^\infty$ is the Fourier coefficient of the
whitening filter 
\[
G(\eit) = \frac{1}{H_Z(\eit)} = \frac{1+\beta \eit}{1+\alpha \eit}
\]
and $\{U_k\}_{k=1}^\infty$ is the white innovations process.

Consider the following coding scheme, which is ``stationary'' from
time 2.  At time $1$, the transmitter sends $X_1 = V \sim N(0,
\sigma_V^2)$ to learn $U_1 = Y_1 - X_1$ and subsequently sends
\begin{equation}
\label{eq:coding-s}
X_n = \chi (S_n - E (S_n | Y_1^{n-1}) ),\qquad n = 2,3,\ldots
\end{equation}
where 
\[
\chi = \frac{1-\sigma \alpha x}{\sigma x}
\] 
and $x$ is the unique positive parameter satisfying the capacity
polynomial
\[
P\,x^2 = 
\frac{(1 - x^2)( 1 + \sigma \alpha x)^2}{(1 + \sigma \beta x)^2}.
\tag{\ref{eq:arma1-poly}}
\]

We can easily prove the optimality of this coding scheme from our
previous analysis of the coding scheme \eqref{eq:xn}. Indeed, it is
straightforward to transform the refinement of the message $V$ in
\eqref{eq:xn} to the refinement of the noise state $S_n$ in
\eqref{eq:coding-s} and vice versa.  However, the direct analysis has
two important benefits.  First, as we will see in the next section,
the optimal feedback coding scheme for a general finite-order ARMA
channel can be represented most naturally as the refinement of current
noise state.  Second, we can interpret the role of the message bearing
signal $V$ as a perturbation to boost the output entropy rate; see
Subsection~\ref{subsec:dare}.

For the analysis of the coding scheme \eqref{eq:coding-s}, we
introduce the notation
\begin{align*}
\hat{S}_n &= E(S_n|Y_1^{n-1})\\
\tilde{S}_n &= S_n - E(S_n|Y_1^{n-1}) = S_n - \hat{S}_n
\end{align*}
and similarly define $\hat{Y}_n = E(Y_n|Y_1^{n-1})$ and $\tilde{Y}_n =
Y_n - E(Y_n|Y_1^{n-1})$.
Under this notation, we can express the channel output as
\begin{align}
\nonumber
Y_n &= X_n + Z_n\\
\nonumber
&= \chi \tilde{S}_n + (\alpha-\beta) S_n + U_n\\
&= (\alpha-\beta+\chi) \tilde{S}_n + (\alpha-\beta) \hat{S}_n + U_n.
\nonumber
\end{align}
Let $\sigma_n^2 = E\tilde{Y}_n^2$ and $s_n^2 = E\tilde{S}_n^2.$
Then, we have
\begin{align*}
\hat{S}_{n+1} &= E(S_{n+1}|Y_1^n)\\
&= E(S_{n+1}|Y_1^{n-1}, \tilde{Y}_n)\\
&= E(S_{n+1}|Y_1^{n-1}) + E(S_{n+1}|\tilde{Y}_n)\\
&= E(-\beta S_n + U_n | Y_1^{n-1})
  + E(-\beta S_n + U_n |\tilde{Y}_n)\\
&= -\beta \hat{S}_n + \gamma_n \tilde{Y}_n,
\end{align*}
where
\[
\gamma_n = \frac{1}{\sigma_n^{2}}
(-\beta (\alpha-\beta+\chi)s_n^2 + 1).
\]
From this we get the state-space model for $\tilde{Y}_n$ as
\begin{align*}
\tilde{S}_{n+1} &= (-\beta - \gamma_n(\alpha-\beta+\chi))
\tilde{S}_n + (1-\gamma_n) U_n\\
\tilde{Y}_n &= (\alpha-\beta+\chi) \tilde{S}_n + U_n,
\end{align*}
which implies the following recursive relationship 
for $\sigma_n^2$ and $s_n^2$ for $n \ge 2$:
\begin{align}
\nonumber
\sigma_n^2 &= 1+(\alpha-\beta+\chi)^2 s_n^2
\intertext{and}
\nonumber
s_{n+1}^2 &= 
(\beta + \gamma_n(\alpha-\beta+\chi))^2 s_n^2 + (1-\gamma_n)^2\\
\nonumber
&= \beta^2 s_n^2 + 1 - \frac{(-\beta (\alpha - \beta + \chi) s_n^2 + 1)^2}
{1 + (\alpha-\beta+\chi)^2 s_n^2}. 
\end{align}
It is easy to recall from Subsection~\ref{subsec:dare} that the above
recursion for $s_n^2$ is nothing but the one-dimensional discrete
Riccati recursion.

Suppose we have $V = 0$.  Then $s_n^2 \equiv 0$ for all $n$ and
$\sigma_n^2 =1$ for all $n$.  In other words, the information rate
$h(\mathcal{Y}) - h(\mathcal{Z}) = 0$; obviously, if we send nothing,
the information rate should be zero.

Now take any $\epsilon > 0$.  If $V \sim N(0,\epsilon),$
Lemma~\ref{lemma:riccati-recursion}\ref{property:convergence} shows
that $s_n^2 \to s^2$ where $s^2$ is the positive solution to the
one-dimensional Riccati equation
\[
s^2 = \beta^2 s^2 + 1 - \frac{(-\beta (\alpha - \beta + \chi) s^2 + 1)^2}
{1 + (\alpha-\beta+\chi)^2 s^2},
\]
so that $\sigma_n^2 \to 1 + (\alpha - \beta + \chi)^2 s^2$.  
With a little algebra, we can solve the Riccati equation to get
\[
s^2 = \frac{(\chi + \alpha)^2 - 1}{(\chi+\alpha-\beta)^2},
\]
which, combined with our choice of $\chi = (1+\sigma\beta x)/(\sigma
x)$, implies that $1 + (\alpha - \beta + \chi)^2 s^2 = 1/x^2$.  On the
other hand,
\[
E X_n^2 = \chi^2 s_n^2 \to \chi^2 s^2 = P.
\]
Hence, the coding scheme given by \eqref{eq:coding-s} achieves the
information rate $-\log x$ under the power constraint $P$, and hence
is optimal.

The above analysis gives two complementary interpretations for the
role of the signal $V$.  Most naturally, we view the feedback capacity
problem as that of maximizing the information rate and $V$ obviously
has the role of carrying the information we wish to transmit.  On the
other hand, if we view the feedback capacity problem as that of
maximizing the output entropy rate, then $V$ has the role of
perturbing the (nonstationary) output process so that the resulting
perturbed output process has the same entropy rate as its stationary
version.  This second interpretation leads to the following
observation in the spectral domain.

In the notation of Cover--Pombra's $n$-block capacity, let $B_n^\star$
denote the ``almost Toeplitz'' feedback matrix corresponding to the
optimal coding scheme and $K_{V,n}^\star$ denote the message
covariance matrix of rank one.  If $\{\lambda_{1}, \ldots,
\lambda_{n}\}$ denote the eigenvalues of $(I+B_n^\star)
K_{Z,n}(I+B_n^\star)'$, then the asymptotic distribution of
$\{{\lambda}_i\}_{i=1}^n$ follows the optimal output spectrum
$S_Y^\star$ in \eqref{eq:s_y}.

Now we argue that there must be one eigenvalue, say ${\lambda}_1$,
that goes down to zero exponentially fast (as $n\to\infty$) and the
rate of decay is in fact the feedback capacity.  Why?  The rank of
$K_{V,n}^\star$ is 1.  Hence, roughly speaking, $K_{V,n}^\star$ is
water-filling the eigenmode corresponding to $\lambda_1$ with small
power $\epsilon$.  This results in
\[
\det(K_{Y,n}^\star) \doteq ({\lambda}_1 + \epsilon)
\prod_{i=2}^n {\lambda}_i \doteq x^{-2n} \det(K_{Z,n}).
\]
But we have
$1 \doteq \det(K_{Z,n}) =
\det((I+B_n^\star)K_{Z,n}(I+B_n^\star)') = \prod_{i=1}^n {\lambda}_i;
$
thus $\lambda_1 \doteq x^{2n}$.  Therefore, we can view the role of
the rank-one $K_{V,n}^\star$ as the tiny drop of water that fills the
modified terrain $(I+B_n^\star) K_{Z,n}(I+B_n^\star)'$ shaped by the
optimal feedback filter $B_n^\star$.

\vspace*{1em}
\section{General Finite-Order ARMA Noise Spectrum}
\label{sec:armak}

We turn our focus to the general autoregressive moving average noise
spectrum with finite order, say, $k$.  We assume that the noise power
spectral density $S_Z(\eit)$ has the canonical spectral factorization
$S_Z(\eit) = H_Z(\eit)H_Z(e^{-i\theta})$ where
\begin{align}
\label{eq:armak-canon}
H_Z(z) &= \frac{P(z)}{Q(z)} = 
\frac{1+\sum_{n=1}^k p_n z^n}{1+\sum_{n=1}^k q_n z^n}
\end{align}
such that at least one of the monic co-prime polynomials $P(z)$ and
$Q(z)$ has degree $k$ and all zeros of $P(z)$ and $Q(z)$ lie strictly
outside the unit circle (i.e, both $P(z)$ and $Q(z)$ are
\emph{stable}).  In particular, $S_Z(\eit)$ is bounded away from zero.

We first prove a proposition on the structure of the optimal output
spectrum, which is reminiscent of Corollary~\ref{coro:ordentlich}.
\begin{prop}
\label{prop:armak-output-structure}
Suppose that the ARMA($k$) noise process has the rational power spectral
density $S_Z(\eit)) = H_Z(\eit)H_Z(e^{-i\theta}),$ where $H_Z(\eit)$ is
given in \eqref{eq:armak-canon}.  Then, the supremum in the
variational characterization of the feedback capacity problem
\begin{equation}
C_\textit{FB} 
= \sup_{B(\eit)}
\intt \frac{1}{2} \log
{|1+B(\eit)|^2} \dth
\tag{\ref{eq:cfb-no-sv}}
\end{equation}
is attained by a strictly causal $B^\star(\eit) \in H_\infty$ and
the corresponding output spectrum $S_Y^\star(\eit) =
|1+B^\star(\eit)|^2 S_Z(\eit)$ is of the form $S_Y^\star(\eit) =
\sigma^2 H_Y(\eit)H_Y(e^{-i\theta})$ with
\[
H_Y(z) = \frac{R(z)}{Q(z)} = 
\frac{1+\sum_{n=1}^k r_n z^n}{1+\sum_{n=1}^k q_n z^n}.
\]
\end{prop}
\begin{proof}
This is a simple exercise from Proposition~\ref{prop:cfb-necessary}.
Since $S_Z$ is bounded away from zero, the supremum is attainable by
a strictly causal $B^\star \in H_2$.  From the orthogonality
condition~\eqref{eq:var-orth-cond},
\[
B^\star(1+\overline{B^\star}) S_Z = 
S_Y^\star - \frac{|P|^2}{|Q|^2}(1+\overline{B^\star}) =: A
\]  
is anticausal.  Now consider
\[
S = |Q|^2 S_Y^\star = |P|^2 (1+\overline{B^\star}) + |Q|^2 A.
\]
Since $P$ and $Q$ are polynomials of degree at most $k$, and
$(1+\overline{B^\star})$ and $A$ are anticausal, it is easy to see
that $S(z)$ is of the form
\[
S(z) = s_k z^k + s_{k-1} z^{k-1} + \cdots,
\]
that is,
\[
\intt S(\eit)e^{-ij\theta} \dth = 0
\]
for $j \ge k+1$.  But from the symmetry $S(\eit) = S(e^{-i\theta})$,
this implies that $S(z)$ is of the form
\[
S(z) = s_k z^k + \cdots + s_k z^{-k},
\]
or equivalently, $S(z)$ has the canonical factorization $S(z) =
\sigma^2 R(z) R(z^{-1})$ for some polynomial $R$ of degree at most $k$.
Since
\[
S_Y^\star = \sigma^2 \frac{|R|^2}{|Q|^2} = {|1+B|^2}\frac{|P|^2}{|Q|^2},
\]
$1+B^\star$ must be of the form
\[
1+B^\star(z) = b(z) \frac{R(z)}{P(z)} \in H_\infty
\]
for some normalized Blaschke product $b(z)$ with $|b(\eit)|^2 = \sigma^2$.
\end{proof}

As was hinted at the end of the previous section, the state-space
representation leads to a much richer development.  Our result is, in
some sense, expected from a motivating result by Yang, Kav\v{c}i\'{c},
and Tatikonda~\cite{Yang--Kavcic--Tatikonda2005}, which shows that the
feedback-dependent Markov source distribution achieves the maximum of
finite-dimensional Marko--Massey directed mutual
information~\cite{Massey1990} of a finite-state machine channel.
However, the proof technique in~\cite{Yang--Kavcic--Tatikonda2005}
does not seem to be applicable to our situation, so we have to take a
different approach.

We start by introducing the state-space model for the ARMA($k$) noise
spectrum \eqref{eq:armak-canon}.  Given stable monic polynomials
$P(z)$ and $Q(z)$ with coefficients $\{p_n\}_{n=1}^k$ and
$\{q_n\}_{n=1}^k$, respectively, as in \eqref{eq:armak-canon}, we
construct real matrices $F, G,$ and $H$ of sizes $k \times k$, $k
\times 1$, and $1 \times k$ as
\begin{gather*}
F =
\left[
\begin{matrix}
-q_1 &-q_2 &  \ldots &&-q_k\\
1    &0    &   \ldots && 0 \\
0   & 1    &    \ldots && 0 \\
\vdots & \vdots & \ddots && \vdots\\
0  & 0 &  \cdots &1 & 0
\end{matrix}
\right]\\
G = \left[
\begin{matrix}
1 &0 & \cdots & 0
\end{matrix}
\right]'\\
H = \left[
\begin{matrix}
(p_1-q_1) & \cdots & (p_k-q_k)
\end{matrix}
\right].
\end{gather*}
Let $\{U_n\}_{n=-\infty}^\infty$ be independent and identically
distributed normal random variables with zero mean and unit variance.
We introduce a state-space model of a linear system driven by
$\{U_n\}$ as the input:
\begin{align}
\label{eq:state-z}
\begin{array}{r@{\;}l}
S_{n+1} &= F S_n + G U_n\\[.25em]
Z_n &= H S_n + U_n
\end{array}
\end{align}
where the state $S_n$ and the input $U_n$ are independent of each
other.  We can easily check that the output $\{Z_n\}_{-\infty}^\infty$
is a stationary Gaussian process with power spectral density
$S_Z(\eit) = |H_Z(\eit)|^2$, where
\begin{align*}
H_Z(z) &= \frac{P(z)}{Q(z)}\\
&= \frac{\det(I - z(F-GH))}{\det(I - zF)}\\
&= z H (I - zF)^{-1} G + 1.
\end{align*}
Under the above state-space representation, the channel output can be
expressed as
\begin{equation}
\label{eq:y_n}
Y_n = X_n + Z_n = X_n + H S_n + U_n.
\end{equation}

We state our main result in this section.\pagebreak
\begin{theorem}
\label{thm:armak-capacity}
Suppose the stationary Gaussian noise process $\{Z_i\}_{i=1}^\infty$
has the state-space representation~\eqref{eq:state-z}.  Then, the
feedback capacity $C_\textit{FB}$ of the Gaussian channel $Y_i =
X_i+Z_i,\enspace i = 1,2,\ldots,$ under the power constraint $P$, is
given by
\begin{equation}
\label{eq:armak-capacity}
C_\textit{FB} = \max_{X}
\half \log (1+(X+H)\Sigma_+(X)(X+H)')
\end{equation}
where the maximum is taken over all $X \in \mathbb{R}^{1\times k}$
such that $F - G(X+H)$ has no unit-circle zero and $X\Sigma_+(X)X' \le
P$, with $\Sigma_+(X)$ being the maximal solution to the discrete
algebraic Riccati equation
\begin{align}
\label{eq:armak-dare}
\Sigma 
= F \Sigma F' + G G' 
- \frac{(F\Sigma(X+H)' + G)(F\Sigma(X+H)'+G)'}
{1+(X+H)\Sigma(X+H)'}.
\end{align}
\end{theorem}

We prove Theorem~\ref{thm:armak-capacity} in two steps.  The first
step is the following structural result.
\begin{lemma}
\label{lemma:armak-markov}
Suppose the ARM\hspace{-.5pt}A($k$) noise process has the state-space
representation \eqref{eq:state-z}.  Then the feedback capacity is
achieved by the input process $\{X_n\}_{n=-\infty}^\infty$ of the form
\[
X_n = X \bigl(S_n - E(S_n|Y_{-\infty}^{n-1})\bigr)
\]
for some $X \in \mathbb{R}^{1\times k}$ such that $F-G(X+H)$ has no
unit-circle eigenvalue.
\end{lemma}

\begin{proof}
Suppose that $B(z) = \sum_{j=1}^\infty b_j e^{ij\theta}$ achieves the
maximum of the variational problem in \eqref{eq:cfb-no-sv}, or
equivalently, the stationary process $\{X_i\}_{i=-\infty}^\infty$
defined by $X_n = \sum_{j=1}^\infty b_j Z_{n-j}$ achieves the feedback
capacity.  If we regard $X_n$ as a vector in the Hilbert space
generated by linear spans of $\{Z_i\}_{i=\infty}^\infty$, $X_n$ lies
in the closed linear span of all past $Z_i$'s, that is, $X_n \in
\clin\{Z_{-\infty}^{n-1}\}$.  Equivalently,
\[
X_n \in \mathcal{H}_n := \clin\{S^n, Y_{-\infty}^{n-1}\}.
\]
We decompose $X_n$ into two orthogonal parts as
\[
X_n = \xi_n + \zeta_n,
\]
where $\xi_n$ lies in the closed linear span $\mathcal{G}_n$ of $S_n$
and $Y_{-\infty}^{n-1}$, and $\zeta_n$ lies in the orthogonal
complement of $\mathcal{G}_n$ in $\mathcal{H}_n$, namely,
\begin{align*}
\xi_n &\in \mathcal{G}_n := \clin\{S_n, Y_{-\infty}^{n-1}\}\\
\zeta_n &\in (\mathcal{H}_n \ominus\: \mathcal{G}_n).
\end{align*}
Since $\{X_n\}$ achieves the feedback capacity, from the
orthogonality condition \eqref{eq:var-orth-cond} in
Proposition~\ref{prop:cfb-necessary},
\[
\xi_n = X (S_n - E(S_n|Y_{-\infty}^{n-1}))
\]
for some $X \in \mathbb{R}^{1\times k}$.
In other words, for each orthogonal feedback filter $B(z)$, we have
a representation 
\begin{equation}
\label{eq:x_n}
X_n = X(S_n - E(S_n|Y_{-\infty}^{n-1})) + \xi_n
\end{equation}
for some $X \in \mathbb{R}^{1\times k}$.

To ease the notation a little, we shall subsequently write
\begin{align*}
\hat{A}_n &:= E(A_n|Y_{-\infty}^{n-1})\\
\tilde{A}_n &:= A_n - \hat{A}_n
\end{align*}
for a generic random variable (or a random vector) $A_n$.  Under this
notation, we have 
\begin{align}
\label{eq:y_n2}
Y_n &= X \tilde{S}_n + H S_n + \zeta_n + U_n\\
\intertext{so that}
\tilde{Y}_n &= (X+H) \tilde{S}_n + \zeta_n + U_n.
\nonumber
\end{align}

Let $P_\zeta = E\zeta_n^2,$ $\sigma^2 = E \tilde{Y}_n^2,$ and $\Sigma :=
\cov(\tilde{S}_n)$.  Then, from the mutual orthogonality of $\zeta_n,$
$U_n,$ and $\tilde{S}_n$,
\[
\sigma^2 = (X+H)\Sigma(X+H)' + P_\zeta + 1.
\]
On the other hand, it is easy to check that
\begin{align}
\nonumber
\hat{S}_{n+1} &= E(F S_n + G U_n|Y_{-\infty}^n)\\
\nonumber
&= E(F S_n + G U_n|Y_{-\infty}^{n-1}) + 
E(F S_n + G U_n|\tilde{Y})\\ 
&= F \hat{S}_n + \Gamma \tilde{Y}_n,
\label{eq:state-s-hat}
\end{align}
where 
\begin{align*}
\Gamma &:= \frac{1}{\sigma^2} \bigl(
F \Sigma(X+H)' + G \bigr).
\end{align*}
Thus, we have the state-space representation of $\tilde{Y}_n$ as
\begin{align}
\label{eq:state-y-tilde}
\begin{array}{r@{\;}l}
\tilde{S}_{n+1} &= \bigl(F-\Gamma(X+H)\bigr) \tilde{S}_n
- \Gamma \zeta_n + (G - \Gamma) U_n \\[.25em]
\tilde{Y}_n &= (X+H) \tilde{S}_n + \zeta_n + U_n,
\end{array}
\end{align}
which implies that $\Sigma$ satisfies the following discrete algebraic
Riccati equation (DARE):
\begin{align}
\nonumber
\Sigma &= (F-\Gamma(X\!+\!H))\Sigma(F-\Gamma(X\!+\!H))' 
  + (G-\Gamma)(G-\Gamma)' + P_\zeta\Gamma\Gamma'\\
&= F \Sigma F' + G G' 
- (F\Sigma(X\!+\!H)' + G)(1+P_\zeta+(X\!+\!H)\Sigma(X\!+\!H)')^{-1}
(F\Sigma(X\!+\!H)'+G)'.
\label{eq:dare1}
\end{align}

We now ask the question whether there exists a positive semidefinite
solution $\Sigma_+$ to the DARE \eqref{eq:dare1} that stabilizes the
matrix
\begin{align*}
F - \Gamma(X+H) = F -
(F\Sigma_+(X+H)'+G)(1+P_\zeta+(X+H)\Sigma_+(X+H)')^{-1}(X+H),
\end{align*}
that is, all eigenvalues of $F-\Gamma(X+H)$ lie in the unit circle.
Obviously this condition is necessary to make the state-space
equations \eqref{eq:y_n} and \eqref{eq:state-y-tilde} have any meaning
for the stationary output process and its innovations.

We note that the stability of $F$ clearly implies the detectability of
$\{F, X+H\}$ (i.e., there exits a matrix $K$ such that $F-K(X+H)$ is
stable).  In turn, for $P_\zeta > 0$, the stability of $F-GH$ implies
the unit-circle controllability of $\{F_\zeta, G_\zeta\}$, where
\begin{align*}
F_\zeta &= F-\frac{G(X+H)}{1+P_\zeta}\\
G_\zeta &= \sqrt{\frac{P_\zeta}{1+P_\zeta}}G.
\end{align*}
This condition of the unit-circle controllability (or controllability
on the unit circle) means that there exists a matrix $K$ such that
$F_\zeta - G_\zeta K$ has no eigenvalues on the unit circle.  When
$P_\zeta = 0$, the unit-circle controllability of $\{F_\zeta,
G_\zeta\}$ is equivalent to the condition that $F - G(X+H)$ has no
unit-circle controllability.

Turning back to the above question of the existence of the stabilizing
solution $\Sigma_+$ to \eqref{eq:dare1}, we see from a standard result
on DARE \cite[Theorem E.5.1]{Kailath--Sayed--Hassibi2000} that the
detectability of $\{F, X+H\}$ and the unit-circle controllability of
$\{F_\zeta, G_\zeta\}$ is \emph{equivalent} to the existence of a
stabilizing solution $\Sigma_+$.  Moreover, this stabilizing solution
is unique and positive semidefinite.  Therefore, the input process
described by \eqref{eq:x_n} and the corresponding output process
\eqref{eq:y_n2} are well-defined and uniquely determined by $(X,
P_\zeta)$.

Now we prove that $P_\zeta$ is necessarily zero.  We first observe
that the derivation of the state-space equation
\eqref{eq:state-y-tilde} depends on the fact that $\zeta_n \in
\mathcal{H}_n = \clin\{S^n, Y_{-\infty}^{n-1}\}$ only via the
orthogonality of $\zeta_n$ and $(S_n, Y_{-\infty}^{n-1})$.  Therefore,
if the input process
\[
X_n = X \bigl(S_n  - E(S_n|Y_{-\infty}^{n-1})\bigr) + \zeta_n
\]
achieves the feedback capacity, inducing the output distribution
uniquely defined by \eqref{eq:y_n}--\eqref{eq:dare1}, any other input
process of the form
\[
X_n = X \bigl(S_n  - E(S_n|Y_{-\infty}^{n-1})\bigr) + W_n
\]
achieves the feedback capacity with the same output distribution,
provided that $E W_n^2 = P_\zeta$ and $W_n$ is orthogonal to $(S_n,
Y_{-\infty}^{n-1}, U_n)$.%
\footnote{Although $E({S}_n|Y_{-\infty}^{n-1})$ is symbolically the
same for any choice of $W_n$, each could result in different output
processes defined recursively by $ Y_n = X(S_n -
E(S_n|Y_{-\infty}^{n-1})) + W_n + U_n.  $ However, our analysis of the
Riccati equation shows that the output process is uniquely defined for
any choice of $W_n$.}  In particular, we can take $W_n = V_n,$ where
$\{V_n\}_{n=-\infty}^\infty$ is a white Gaussian process with power
spectral density $S_V(\eit) \equiv P_\zeta$, independent of
$\{Z_n\}_{n=-\infty}^\infty$.

But as Remark~\ref{remark:sv} shows, a nonzero white $S_V^\star$
achieves the feedback capacity only if the noise spectrum itself is
white.  Since $S_Z$ is nonwhite, $P_\zeta$ must be zero.  Therefore,
the optimal input process must be of the form
\[
X_n = X \bigl(S_n  - E(S_n|Y_{-\infty}^{n-1})\bigr)
\]
for some $X$ such that $F - G(X+H)$ has no unit-circle eigenvalue.
\end{proof}

Equipped with Lemma~\ref{lemma:armak-markov}, the proof of
Theorem~\ref{thm:armak-capacity} is straightforward.

\begin{proof}[Proof of Theorem~\ref{thm:armak-capacity}]
We know that the capacity achieving input process is of the form
\[
X_n = X \bigl(S_n  - E(S_n|Y_{-\infty}^{n-1})\bigr).
\]
From \eqref{eq:state-y-tilde}, the state-space equation for
$\tilde{Y}_n$ becomes
\begin{align}
\label{eq:state-y-tilde2}
\begin{array}{r@{\;}l}
\tilde{S}_{n+1} &= \bigl(F-\Gamma(X+H)\bigr) \tilde{S}_n
(G - \Gamma) U_n \\[.25em]
\tilde{Y}_n &= (X+H) \tilde{S}_n + U_n,
\end{array}
\end{align}
where
\begin{align*}
\Gamma &= \Gamma(X) = (F\Sigma_+(X+H)'+G)(1+(X+H)\Sigma_+(X+H)')^{-1}
\end{align*}
and $\Sigma_+ = \Sigma_+(X)$ is the unique positive semidefinite stabilizing
solution to the DARE
\begin{align*}
\Sigma &= F \Sigma F' + G G' -
(F\Sigma(X+H)' +
G)(1+(X+H)\Sigma(X+H)')^{-1}(F\Sigma(X+H)'+G)'.
\end{align*}
Since $\tilde{Y}_n$ is a white process with variance 
\begin{align*}
\sigma^2 &= 1 + (X+H)\Sigma_+(X)(X+H)'\\
&= \frac{\det(F-G(X+H))}{\det(F-\Gamma(X+H))}
\end{align*}
and $h(\tilde{\mathcal{Y}}) = h(\mathcal{Y})$, the corresponding
information rate is
\[
\half\log\bigl(1+(X+H)\Sigma_+(X)(X+H)'\bigr)
\]
under the power consumption $X\Sigma_+(X)X'$.  Clearly, the feedback
capacity $C_\textit{FB}(P)$ is the maximal information rate over all
$X$'s satisfying the power constraint $X\Sigma_+(X)X' \le P$.
\end{proof}

The proofs of Lemma~\ref{lemma:armak-markov} and
Theorem~\ref{thm:armak-capacity} reveal the structure of the optimal
output spectrum once again (cf.\@
Proposition~\ref{prop:armak-output-structure}).  Indeed, we have
\begin{align*}
Y_n &= \hat{Y}_n + \tilde{Y}_n\\
&= H \hat{S}_n + \tilde{Y}_n,
\end{align*}
which, combined with \eqref{eq:state-s-hat}, implies
\begin{equation}
\label{eq:armak-opt-sy}
S_Y(z) = \sigma^2 \frac{\det(I - z(F-\Gamma H))}{\det(I-zF)}
\frac{\det(I - z^{-1}(F-\Gamma H))}{\det(I-z^{-1}F)},
\end{equation}
which is bounded away from zero~\cite[Lemma
  8.3.1]{Kailath--Sayed--Hassibi2000}.  Furthermore, since the optimal
input can be expressed as $X_n = X \tilde{S}_n$, we can easily check
from \eqref{eq:state-y-tilde2} that the corresponding feedback filter
is given as
\begin{align}
\nonumber
B(z) &= zX (I-z(F-\Gamma(X+H)))^{-1}(G-\Gamma)
\frac{\det(I-zF)}{\det(I-z(F-GH))}\\
\label{eq:optimal-bz}
&= \frac{\det(I-z(F-G(X+H)))}{\det(I-z(F-\Gamma(X+H)))}
\frac{\det(I-z(F-\Gamma H))}{\det(I-z(F-GH))} - 1.
\end{align}
From Lemma~\ref{lemma:dare}\ref{property:zero-canc}, it is easy to see
that
\[ 
\frac{\det(I-z(F-G(X+H)))}{\det(I-z(F-\Gamma(X+H)))}
\]
is a normalized Blaschke product whose zeros determine the entropy
rate of the output process.

Now we can easily relate Theorem~\ref{thm:armak-capacity} to the
Schalkwijk--Kailath coding scheme.  Since we already went through
detailed discussions of the Schalkwijk--Kailath coding for the
first-order ARMA spectrum in the previous section, we give here a
rather sketchy argument.  For simplicity, assume the state-space
representation \eqref{eq:state-z} of the noise process
$\{Z_i\}_{i=1}^\infty$ with $S_0 = 0$ and $U_0 = 0$.  For the initial
$k$ transmissions, the transmitter sends $X_n = V_n,\enspace
n=1,\ldots, k,$ with $V^k \sim N_k(0, K_{V})$ and subsequently,
\begin{equation}
\label{eq:coding-armak}
X_n = X (S_n - E (S_n | Y^{n-1}) ),\qquad n = k+1,k+2,\ldots
\end{equation}
where $X \in \mathbb{R}^{1\times k}$ achieves the maximum in
\eqref{eq:armak-capacity}.  In other words, after the initial $k$
transmissions, the transmitter refines the receiver's error of the
current noise state.  Since the error is $k$-dimensional, one must
project it down in the direction $X$.

Lemma~\ref{lemma:riccati-recursion} shows that, as far as $K_{V}$ is
positive definite, or equivalently, as far as $\cov(S_{k+1}|Y_1^k)$ is
positive definite, $\cov(S_n|Y_1^{n-1})$ converges to the unique
stabilizing solution $\Sigma_+(X)$ of the DARE \eqref{eq:armak-dare}
and thus $h(Y_n|Y_1^{n-1})$ converges to $C_\textit{FB}$.  It is also
straightforward to rewrite the coding scheme \eqref{eq:coding-armak}
as the successive refinement of the message-bearing signal $(V_1,
\ldots, V_k)$, from which we can generalize the original
one-dimensional Schalkwijk--Kailath coding scheme into the
$k$-dimensional one with $(\theta_1,\ldots, \theta_k)$ in some equally
spaced constellation.  (Instead of using the minimum mean square error
estimate of $V^k$, we use the minimum variance unbiased estimate of
$\theta^k$; both estimates are linearly related~\cite[Section
3.4]{Kailath--Sayed--Hassibi2000}~\cite[Section 4.5]{Luenberger1969}.)
As before, we can also interpret the role of $V^k$ as tiny drops of
water that fill the noise terrain modified by the optimal feedback
filter $B^\star(z)$.

Finally, we give a more explicit characterization of the optimal
direction $X$.
\begin{prop}
\label{prop:armak-suff}
Suppose $X \in \mathbb{R}^{1\times k}$ satisfies the following
conditions.
\begin{eqenumerate}
\item
\label{cond3:power}
\text{Power:}\quad $X \Sigma_+(X) X' = P,$ where $\Sigma_+(X)$
is the unique stabilizing solution to the DARE \eqref{eq:armak-dare}.

\item
\label{cond3:ev}
\text{Eigenvalues:}\quad $F - G(X+H)$ has distinct eigenvalues
$\alpha_1,\ldots, \alpha_k$ outside the unit circle.  In particular,
$\Sigma_+ \succ 0$.

\item
\label{cond3:spectrum}
\text{Spectrum:}\quad The corresponding output spectrum $S_Y(z)$ in
\eqref{eq:armak-opt-sy} is such that 
\[
0 < S_Y(\alpha_1) = S_Y(\alpha_2) = \cdots = S_Y(\alpha_k) \le 
\min_{|z|=1} S_Y(z).
\]
\end{eqenumerate}
Then, $X$ achieves the maximum in \eqref{eq:armak-capacity}.
\end{prop}
\begin{proof}
We show that the conditions
\eqref{cond3:power}--\eqref{cond3:spectrum} implies the
conditions~\eqref{eq:fb-var-power-eq}--\eqref{eq:causality-cond} in
Proposition~\ref{prop:fb-var-suff}, which in turn implies that the
corresponding feedback filter achieves the feedback capacity.  The
power condition~\eqref{eq:fb-var-power-eq} is satisfied by
\eqref{cond3:power}.  For the other two conditions, take $\lambda =
S_Y(\alpha_1) \le \min_{|z|=1} S_Y(z),$ which satisfies
\eqref{eq:lambda-cond} automatically.  We use the notation (see
\eqref{eq:optimal-bz}):
\begin{align*}
A(z) &= \det(I-z(F-G(X+H))) = (1-\alpha_1^{-1}z)\cdots(1-\alpha_k^{-1}z)\\
A^\#\!(z) &= \det(I - z(F-\Gamma(X+H))) = 
(1- \alpha_1 z)\cdots(1-\alpha_k z)\\
P(z) &= \det(I - z(F-GH))\\
Q(z) &= \det(I - zF)
\intertext{and}
R(z) &= \det(I- z(F -\Gamma H)).
\end{align*}
Note that
\[
\frac{A(z)}{A^\#\!(z)}\frac{A(z^{-1})}{A^\#\!(z^{-1})} = \sigma^2.
\]
Under this notation, the noise spectrum $S_Z(z)$,
the feedback filter $B(z)$,
and the corresponding output spectrum $S_Y(z)$ can be written as
\begin{align}
\nonumber
S_Z(z) &= \frac{P(z)}{Q(z)}\frac{P(z^{-1})}{Q(z^{-1})},\\
\label{eq:bz2}
B(z) &= \frac{A(z)}{A^\#\!(z)}
\frac{R(z)}{P(z)} - 1,
\intertext{and}
\nonumber
S_Y(z) &= \sigma^2 \frac{R(z)}{Q(z)}\frac{R(z^{-1})}{Q(z^{-1})}.
\end{align}

We now consider
\[
f(z) = A(z)(\lambda Q(z)Q(z^{-1}) - \sigma^2 R(z)R(z^{-1})) + \sigma^2
A^\#\!(z)P(z)R(z^{-1}).
\]
We will show that $f(z)/(A^\#\!(z)Q(z))$ is anticausal by showing that
$f(z)$ has factors $Q(z)$ and $A^\#\!(z)$. Indeed, since $X_n \perp
Y^{n-1}$, we have the anticausality of $B(z) S_Z(z) (1+{B(z^{-1})})$,
or equivalently,
\[ 
\left(\frac{A(z)}{A^\#\!(z)} \frac{R(z)}{P(z)} - 1 \right)
\frac{P(z)}{Q(z)}\frac{A(z^{-1})}{A^\#\!(z^{-1})}
= 
\sigma^2 \left(\frac{R(z)}{Q(z)} - 
\frac{A^\#\!(z)}{A(z)}\frac{P(z)}{Q(z)}\right)
\]
is anticausal, which implies that $A(z)R(z) - A^\#(z)P(z)$ and thus
$f(z)$ have a factor $Q(z)$.  On the other hand, $\lambda
Q(z)Q(z^{-1}) - \sigma^2 R(z)R(z^{-1}) = 0$ for each
$\alpha_i,\enspace i=1,\ldots, k,$ which implies that $f(z)$ has a
factor $A^\#\!(z)$.

Finally, the anticausality of $f(z)/(A^\#\!(z)Q(z))$ implies the
anticausality of
\begin{align*}
\lambda \frac{A(z)}{A^\#\!(z)} &Q(z^{-1})
- \sigma^2 \frac{A(z)}{A^\#\!(z)}\frac{R(z)}{Q(z)}R(z^{-1})
+ \sigma^2 \frac{P(z)}{Q(z)} R(z^{-1})\\
&=
\sigma^2
\left(
\lambda \frac{A^\#\!(z^{-1})}{A(z^{-1})}
\frac{P(z^{-1})}{R(z^{-1})}
-
\left(\frac{A(z)}{A^\#\!(z)} \frac{R(z)}{P(z)} - 1\right)
\frac{P(z)}{Q(z)}\frac{P(z^{-1})}{Q(z^{-1})}\right)\\
&= \sigma^2 \left(
\frac{\lambda}{1+B(z^{-1})} - B(z) S_z(z) \right).
\end{align*}
Thus, the feedback filter $B(z)$ in \eqref{eq:bz2} satisfies the
causality condition \eqref{eq:causality-cond} and achieves the
feedback capacity.  In particular, $X$ that satisfies the conditions
set forth in \eqref{cond3:power}--\eqref{cond3:spectrum} maximizes
\eqref{eq:armak-capacity}.
\end{proof}

\vspace*{1em}
\section{Concluding Remarks}
\label{sec:conc}

We have given an attempt to solve the Gaussian feedback capacity
problem in a closed form.  A variational characterization of the
feedback capacity was found (Theorems~\ref{thm:fb-capacity} and
\ref{thm:fb-capacity2}):
\begin{align*}
C_\textit{FB} &= \sup_{S_V(\eit), B(\eit)} 
\intt \frac{1}{2} \log
\frac{S_V(\eit) + |1+B(\eit)|^2 S_Z(\eit)}{S_Z(\eit)} \dth\\
&= \sup_{B(\eit)}
\intt \frac{1}{2} \log
{|1+B(\eit)|^2} \dth,
\end{align*}
which was subsequently simplified into a more explicit form when the
Gaussian noise process has a finite-order autoregressive
moving average noise spectrum (Theorem~\ref{thm:armak-capacity}):
\begin{gather*}
C_\textit{FB} = \max_{X}
\half \log (1+(X+H)\Sigma(X+H)')\\
\Sigma = F \Sigma F' + G G' 
- \frac{(F\Sigma(X+H)' + G)(F\Sigma(X+H)'+G)'}
{1+(X+H)\Sigma(X+H)'}
\end{gather*}
and was solved completely in a closed form when the noise spectrum is the
first-order autoregressive moving average (Theorem~\ref{thm:arma1}):
\begin{gather*}
C_\textit{FB} = - \log x\\
P\,x^2 = 
\frac{(1 - x^2)( 1 + \sigma \alpha x)^2}{(1 + \sigma \beta x)^2}.
\end{gather*}
The optimal coding scheme was interpreted as a natural extension of
the Schalkwijk--Kailath linear signaling scheme:
\begin{align*}
X_n 
&= X (S_n - E(S_n|Y^{n-1}))\\
&\propto \,\theta - \hat{\theta}(Y^{n-1})
\end{align*}
which strongly confirms the common belief that the stationary
Wiener/Kalman filter is the optimal feedback processor.

In some sense, our development can be viewed as an asymptotic analysis
of the sequence of convex optimization problems
\[
C_\textit{FB} = \lim_{n\to\infty} \max_{K_{\!V\!,n}, B_n} \half \log
\frac{\det(K_{V,n} +
(B_n+I_n)K_{Z,n}(B_n+I_n)')^{1/n}}{\det(K_{Z,n})^{1/n}}.
\tag{\ref{eq:cfb}}
\]
Thus, it is refreshing to note that the pivotal proof ingredients, not
to mention the origination of the problem and the interpretations of
the solution, have information theoretic flavors.  Indeed, the proof
of Theorem~\ref{thm:fb-capacity} relies heavily on the maximum entropy
argument, while the proof of Theorem~\ref{thm:armak-capacity} uses
Shannon's water-filling solution for the Gaussian nonfeedback capacity
problem to reach a certain contradiction.

Even in its current intermediate form, the solution to the Gaussian
feedback capacity problem reveals a rich connection between control,
estimation, and communication; roughly speaking, the communication
problem over the Gaussian feedback channel is equivalent to a
stochastic control problem of the receiver's estimation error, which
is, in turn, equivalent to the maximum entropy problem of the output
spectrum.  We conclude by posing a few remaining questions that will
invite further investigations to illuminate a complete picture of this
fascinating interplay between control, estimation, and communication.

\setenumerate{parsep=6pt}
\begin{enumerate}
\item 
From Theorem \ref{thm:fb-capacity2} and Szeg\H{o}--Kolmogorov--Krein
theorem, we get the following max--min characterization of the feedback
capacity:
\begin{equation}
\label{eq:maximin}
C_{\textrm{FB}} = \sup_{\{b_k\}} \inff_{\{a_k\}}
\half \log \left(
\intt \Big|1-\sum_{k} a_k e^{ik\theta}\Big|^2
\Big|1-\sum_{k} b_k e^{ik\theta}\Big|^2
S_Z(\eit) \dth\right)
\end{equation}
where the infimum is taken over all finite sequences $\{a_k\}$ and the
supremum is over all finite sequences $\{b_k\}$ satisfying
\[
\intt \Big|\sum_{k} b_k e^{ik\theta}\Big|^2 S_Z(\eit) \dth
\le P.
\]
Thus, the feedback capacity problem can be viewed as a game
between the controller (feedback filter) $B(z) = \sum_k b_k z^k$
and the estimator $A(z) = \sum_k a_k z^k$.  Does this game has a
saddle point?  If so, can we get an explicit characterization of the
saddle point and the associated value of the game?  The objective of
the optimization problem \eqref{eq:maximin} is not
quasi-convex-concave in $(A, B)$ and the usual Fan--Sion minimax
theorems~\cite{Fan1952}~\cite{Sion1958} do not apply.  Nonetheless,
the problem is quadratic, so a careful application of the
S-procedure (see Yakubovich~\cite{Yakubovich1992}) might lead to an
interesting answer.

\item
We wish to claim that Proposition~\ref{prop:fb-var-suff} gives a
``characterization'' of the optimal feedback filter $B(z)$.
Unfortunately, there are two important links missing to fully justify
this claim.  First, we should prove the existence%
\footnote{We do know that, if the noise spectrum is bounded away from
zero, there exists an optimal filter $B$ that achieves the feedback
capacity; see Proposition~\ref{prop:cfb-necessary}.  The question here
is, roughly speaking, whether the sufficient condition in
Proposition~\ref{prop:fb-var-suff} is necessary as well.}  of a filter
$B$ that satisfies the
conditions~\eqref{eq:fb-var-power-eq}--\eqref{eq:causality-cond}.  The
similarity of the finite-dimensional dual optimization
problem~\eqref{eq:dual-boyd-ord} and its infinite-dimensional
version~\eqref{eq:inf-dual} suggests that strong duality may continue
to hold for the infinite-dimensional problem \eqref{eq:var-opt1}.  It
should be noted, however, that because the sequence
$\{C_{\textit{FB},n}\}$ of the finite-dimensional feedback capacities
is superadditive, the proof technique for the convergence of the
finite-dimensional primal optimization problem~\eqref{eq:boyd-ord} to
its infinite-dimensional version~\eqref{eq:var-opt1} is not directly
applicable in the dual case.  More refined tools from convex analysis
in topological vector spaces (see Ekeland and
Temam~\cite{Ekeland--Temam1976} and Young~\cite{Young1977}) might be
useful in proving the strong duality directly, but little progress has
been made in this direction.

Even with the existence proof, however, the
conditions~\eqref{eq:fb-var-power-eq}--\eqref{eq:causality-cond} are
still very complicated, so their utility looks somewhat limited.  So
the natural question is---can we characterize the optimal feedback
filter $B(z)$ in a more explicit manner, at least at the conceptual
level as in the Wiener--Hopf factorization?

\item
One way of remedying the problem mentioned above is restricting
attention to a limited class of noise spectra.  This is what we did in
Sections~\ref{sec:arma1} and \ref{sec:armak} with moderate success.
However, our characterization of the ARMA($k$) feedback capacity in
Theorem~\eqref{thm:armak-capacity}, while conceptually appealing in
the context of the Schalkwijk--Kailath coding scheme, falls short of a
\emph{numerically tractable} solution.  Although the algebraic Riccati
equation for a fixed projection direction $X$ can be solved
efficiently, for example, by the invariant subspace method, it seems
that finding the optimal direction $X^\star$ under the given power
constraint $P$ is a difficult nonconvex optimization problem.

In this regard, the sufficient condition for the optimal direction
$X^\star$ in Proposition~\ref{prop:armak-suff} has a rather
interesting implication---it indirectly characterizes the solution of
the nonconvex optimization problem~\eqref{eq:armak-capacity}, which is
difficult to solve even numerically.  Can we give a more explicit
characterization of the optimal direction $X^\star$ that satisfies the
conditions \eqref{cond3:power}--\eqref{cond3:spectrum}?

\item
There are two more possible connections to optimal control theory.  First,
the last condition~\eqref{cond3:spectrum} is reminiscent of the
classical interpolation problem studied by Pick and Nevanlinna (see,
for example, Ball, Gohberg, and
Rodman~\cite{Ball--Gohberg--Rodman1990}).  Second, our variational
characterization of the feedback capacity problem may have some
relevance to the risk-sensitive or minimum-entropy control/estimation
problem (see Whittle~\cite{Whittle1990} and Mustafa and
Glover~\cite{Mustafa--Glover1990}) as was pointed out by Babak
Hassibi, Stephen Boyd, and Sanjoy Mitter in private communications.
Indeed, the dual \eqref{eq:inf-dual} to the feedback capacity problem
has the leading entropy term $\intt \log (\nu -\psi_1(\eit))\dth$ that
looks similar to the one in the minimum-entropy control problem.  Can
these connections be made more clear and precise?

\item
Finally, Sergio Verd\'u posed the following question in the context of
(the growth of) spectral efficiency in the wideband
regime~\cite{Verdu2002}: As a function of the power constraint, is
$C''_\textit{FB}(0)$ strictly larger than $C''(0)$?  In contrast to
Dembo's result~\cite{Dembo1989} on the first derivative
$C'_\textit{FB}(0) = C'(0)$, we can show that $C''_\textit{FB}(0) >
C''(0)$ and even surprisingly that $C''_\textit{FB}(0) - C''(0) =
\infty$ for rational noise spectra.  However, in order to answer the
real question whether feedback increases the spectral efficiency in
the wideband regime, we need to better understand the physics of the
Gaussian feedback channel.  Maybe it is too late to ask this question,
but where does the discrete-time Gaussian feedback channel come from?

Usually the physical model for the discrete-time Gaussian nonfeedback
channel comes from the corresponding continuous-time Gaussian
nonfeedback channel; see Gallager~\cite[Chapter 8]{Gallager1968} and
Wyner~\cite{Wyner1966} for details on slightly different alternatives.
For two reasons, unfortunately, the usual method fails to yield our
feedback channel model.  First, the usual approach is based on the
Karhunen--Lo\`eve expansion of the continuous-time waveform filtered
noise process $Z(t)$, which gives a parallel Gaussian channel in which
the noise of the $n$th orthogonal channel has variance that
corresponds to the eigenvalue of the Karhunen--Lo\`eve expansion.  On
the other hand, our discrete-time channel model is defined through a
temporally correlated stationary Gaussian noise process.  One may
argue that this discrepancy is of minor importance in the nonfeedback
case, since the capacity is determined solely from the eigenvalues of
the noise spectrum, not the eigenvectors.

The second reason, however, is much more fundamental to the nature of
causal feedback.  Indeed, the ``time'' indices for the components of
the parallel channel coming from the Karhunen--Lo\`eve expansion do
not correspond to the physical time, and hence they have no causal
relationship among them.  How can we \emph{causally} code over
components of the Karhunen--Lo\`eve expansion?

Schalkwijk's original paper~\cite{Schalkwijk1966} considers the
following ``physical'' model for the discrete-time white Gaussian
noise channel: The transmission take at integer time values with the
unit of time being $1/2W$.  Numbers are sent by amplitude modulation
of ``some basic waveform'' of bandwidth $W$.  The disturbance is white
Gaussian noise and the received output comes from a matched filter.

But it is the very paradox of time-limited \emph{and} band-limited
signals that leads to the rigorous treatments by Wyner and Gallager,
based on the Karhunen--Lo\`eve expansion!  For the particular case of
the strictly band-limited channel, prolate spheroidal functions (see
Slepian, Landau, and Pollak~\cite{Slepian--Pollak1961,
Landau--Pollak1961, Landau--Pollak1962}) form a basis for the
Karhunen--Lo\`eve expansion, which destroys the time causality of
feedback.

If we allow the amplitude-modulating waveform $s(t)$ to span arbitrary
bandwidth, in particular, if $s(t)$ is a rectangular pulse, then the
resulting discrete-time channel is the usual additive white Gaussia
noise channel, where the noise process $\{Z_n\}$ at the matched filter
output is given by
\[
Z_n \propto \int_{n-1}^n d W(t)
\]
and $W(t)$ is the standard Brownian motion.  Of course, we have lost
the tight connection to the continuous-time \emph{band-limited}
Gaussian channel (and thus we can no longer talk about the feedback
capacity of the continuous-time {band-limited} channel), yet we have a
physically plausible model for the discrete-time Gaussian feedback
channel.

In the same vein, we can model the discrete-time first-order
autoregressive Gaussian noise channel from an appropriate
continuous-time channel, via a slightly different path.  Consider the
stationary Ornstein--Uhlenbeck process $Z(t)$ in It\^o representation
(see, for example, Karatzas and Shreve~\cite{Karatzas--Shreve1991}):
\[
d Z(t) = - a Z(t)dt + d W(t)
\]
for some $a > 0$.  Equivalently, 
\[
Z(t) = \int_{-\infty}^t e^{-a(t-s)} d W(s).
\]
If we sample $Z(t)$ to obtain the discrete-time noise process $Z_n =
Z(nT)$, the covariance sequence $R(k) = EZ_nZ_{n+k}$ is given as
\[
R(k) = \frac{1}{2a} e^{-aT|k|}
\] 
which implies that $\{Z_n\}$ is the first-order autoregressive
Gaussian process with parameter $\alpha = \exp(-aT)$.  Thus, the
discrete-time first-order autoregressive discrete-time channel arises
naturally from sampling of the continuous-time first-order
autoregressive waveform channel.  Is our channel model the right one
to consider?  If so, how can we extend it to the general noise
spectrum?

Back to our original question of the spectral efficiency, first note
that the parameter $\alpha$ of the above channel model changes as the
sampling period $T$ changes. Moreover, we use the waveform that is
almost band-limited, but still with infinite bandwidth.  Hence, it
seems quite challenging to give a proper definition of the spectral
efficiency, let alone a rigorous analysis.
\end{enumerate}

\vspace*{1em}
\appendix[Existence of an optimal $(S_V^\star, B^\star)$]

Suppose that the noise spectrum $S_Z(\eit)$ is lower bounded by some
$\delta > 0$.  We write
\[
f(S_V, B) = 
\intt \half
\log
\frac{S_V(\eit) + |1 + B(\eit)|^2 S_Z(\eit)}{S_Z(\eit)}\dth.
\]
Then $C_\textit{FB} = \sup_{S_V, B} f(S_V,B)$ where the supremum is
taken over all $S_V \ge 0$ and strictly causal polynomials $B(\eit)$
with
\[
\intt S_V(\eit) + |B(\eit)|^2 S_Z(\eit) \dth \le P.
\]

By change of variable $S_Y = S_V + |1+B|^2 S_Z$, we write
\[
g(S_Y, B) = f(S_V, B) = \intt \half \log(S_Y) \dth.
\]
Let $H_2(\mu_Z)$ denote the space of analytic functions
square-integrable with respect to the noise spectral distribution
$d\mu_Z = S_Z(\eit)d\theta$.  Since polynomials are dense in
$H_2(\mu_Z)$, it is natural to ask whether the maximum of $g(S_Y, B)$
is achieved by an $(S_Y^\star, B^\star)$ in
\[
K = \left\{(S_Y, B) \in L_1\times H_2(\mu_Z):
S_Y - |1+B|^2 S_Z \ge 0,
B(0) = 0, {\textstyle \intt S_Y - (2 B +1)S_Z \dth \le P}\right\}.
\]
Here the last constraint comes from the facts that
\[
S_V + |B|^2 S_Z = S_Y - B S_Z - \overline{B}S_Z - S_Z
\]
and that $B$ and $S_Z$ have real Fourier coefficients.  Note that we
have $B \in H_2$ whenever $B \in H_2(\mu_Z)$, since
\[
\intt \delta |B|^2 d\theta \le \intt |B|^2 d\mu_Z.
\]

The rest of the proof relies on functional analysis on topological
vector spaces.  See, for example, Megginson~\cite{Megginson1998} and
Dunford and Schwartz~\cite{Dunford--Schwartz1958} for terminology and
proofs of classical theorems we refer to in the following discussion.

First, we relax the constraint set $K$ by embedding the space of $S_Y$
in $L_1$ into the space $M_+$ of positive measures $\mu_Y$ on $[-\pi,
\pi)$.  Noting from Lemma~\ref{lemma:kolmo} that
\[
g(S_Y,B) = \inf_{\{a_k\}} \half \log \left(
\intt |1-\sum_k a_k e^{ik\theta}|^2 S_Y(\eit) \dth \right)
\]
with the infimum over all polynomials with coefficients $\{a_k\}$,
we define 
\[
\tilde{g}(\mu_Y,B) = \inf_{\{a_k\}} \frac{1}{4\pi} \log \left(
\intt |1-\sum_k a_k e^{ik\theta}|^2 d\mu_Y \right).
\]
If $\mu_Y$ is decomposed into absolutely continuous and singular parts
as $d\mu_Y = S_Y(\eit) d\theta + d\mu_{Y,s}$, Lemma~\ref{lemma:kolmo}
shows that
\[
\tilde{g}(\mu_Y,B) = g(S_Y,B),
\]
independent of the singular part $\mu_{Y,s}$.

Now we prove that the maximum of $\tilde{g}(\mu_Y,B)$ is attained in
\begin{align*}
\tilde{K} &= \left\{(\mu_Y,B) \in M_+\times H_2(\mu_Z) : 
(d\mu_Y - |1+B|^2 d\mu_Z) \in M_+,\right.\\
&\qquad\qquad\qquad\qquad
\left.B(0) = 0, {\textstyle \frac{1}{2\pi}\big(\intt d\mu_Y 
- \intt (2 B +1)\, d\mu_Z \big)\le P}\right\}.
\end{align*}
Recall that $M_+$ is a subset of the space $M$ of signed measures and
$M$ is isomorphic to the space of linear functionals on continuous
functions on $[-\pi, \pi)$, that is, $M \simeq C[\pi,\pi)^*$.  Also
$H_2(\mu_Z)$ is a Hilbert space and the dual of itself.  We will show
that the constraint set $\tilde{K}$ is compact in the product topology
of weak$^*$ topology on $M_+$ and weak (=weak$^*$ because $H_2(\mu_Z)$
is a Hilbert space) topology on $H_2(\mu_Z)$.  And then we show that
$\tilde{g}$ is upper semicontinuous under the same topology.  This
clearly implies that the maximum of $\tilde{g}$ is attained in
$\tilde{K}$.  (That the maximum of an upper semicontinuous function is
attained on a compact domain is well-known.  For the proof, see, for
example, Luenberger~\cite[Sections 2.13, 5.10]{Luenberger1969}.)
Finally, because $\tilde{g}(\mu_Y)$ depends only on the absolutely
continuous part of $\mu_Y$, if the maximum of $\tilde{g}$ is attained
by $(\mu_Y^\star,B^\star) \in \tilde{K}$, there exists
$(S_Y^\star,B^\star) \in K$ that attains the same maximum of $g$;
clearly, any singular part of the spectral distribution wastes the
power.  The details of the proof follow.

All topological properties such as compactness, closedness, and
continuity will be used with respect to the product topology of
weak$^\star$ topologies on $M_+$ and $H_2(\mu_Z)$, unless noted otherwise.

For compactness, we observe that $K_1 = \{\mu_Y \in M_+:
\frac{1}{2\pi} \intt d\mu_Y \le P \}$ and $K_2 = \{B \in H_2(\mu_Z):
\frac{1}{2\pi}\intt |B|^2 \,d\mu_Z \le P\}$ are norm balls in respective norm
topologies; thus both are weak$^*$ compact by Alaoglu--Banach theorem,
and so is $K_1 \times K_2$.  Since $\tilde{K} \subset K_1 \times K_2$,
closedness of $\tilde{K}$ will guarantee its compactness.  Since $B(0)
= 0$ if and only if $\intt B(\eit) d\theta = 0$.  Now that $T_1(B) :=
\intt B(\eit) \dth = \intt (1/S_Z(\eit)) \cdot B(\eit)
\,d\mu_Z(\theta)$ is bounded and linear, and thus weakly$^*$
continuous, $\{B(0) = 0\} = T_1^{-1}(\{0\})$ is closed.  Similarly,
\[
T_2(\mu_Y,B) := \intt d\mu_Y - \intt (2B+1)d\mu_Z
\]
is continuous, so the set
\[
\left\{ \frac{1}{2\pi}\intt d\mu_Y -
\frac{1}{2\pi}\intt (2 B +1) d\mu_Z \le P\right\}
\]
is closed.  Finally, $d\mu_Y -
|1+B|^2 d\mu_Z$ is a positive measure if and only if 
\[
T_\phi(\mu_Y, B) := \intt \phi\,d\mu_Y - \intt \phi\,|1+B|^2 d\mu_Z \ge 0
\]
for all $0\le \phi \in C[-\pi,\pi)$.  But for each $\phi \ge 0$, $\intt
  \phi |1+B|^2 d\mu_Z$ is (strongly) continuous and convex.
  Therefore, it is also weakly (=weakly$^*$) lower semicontinuous;
  see Ekeland and Temam~\cite[Section 2.2]{Ekeland--Temam1976}.  This
  implies that $T_\phi$ is upper semicontinuous and
  $T_\phi^{-1}([0,\infty))$ is closed.  Since the intersection of an
  arbitrary collection of closed sets is closed, $\{(\mu_Y,B): d\mu_Y -
  {|1+B|^2 d\mu_Z} \in M_+ \} = \cap_\phi
  T_\phi^{-1}([0,\infty))$ is closed.  For the same reason,
  $\tilde{K}$ is closed, and as a closed subset of a compact set, it
  is compact as well.

For weak$^*$ upper semicontinuity of $\tilde{g}(\mu_Y)$, we first
fix $\mu_Y \in M_+$ and 
note from the definition of weak$^*$ convergence that
\[
\alpha_n(p) := \intt |1-p(\eit)|^2 d\mu_{Y,n} \to \intt |1-p(\eit)|^2 d\mu_Y
=: \alpha(p)
\]
for any fixed strictly causal polynomial $p$ and any sequence
$\mu_{Y,n}$ weakly$^*$ convergent to $\mu_Y$.  Hence
\[
\inf_p \lim_n \alpha_n(p) = \inf_p \alpha(p) = \tilde{g}(\mu_Y).
\]
Now for each $n$, we can find a strictly causal polynomial $p_n$ such that
\[
\alpha_n(p_n) \le \inf_p\alpha_n(p) + \frac{1}{n}.
\]
By taking limits on both sides, we get
\[
\varlimsup_n \inf_p \alpha_n(p) \le \varlimsup_n \alpha_n(p_n)
\le \inf_p\alpha_n(p).
\]
In other words, 
\[
\varlimsup_n \inf_p \intt |1-p|^2d\mu_{Y,n} 
\le \inf_p \intt |1-p|^2 d\mu_{Y} = \tilde{g}(\mu_Y),
\]
for any $\mu_{Y,n}$ weakly$^\star$ convergent to $\mu_Y$.  Thus,
$\tilde{g}(\mu_Y)$ is weakly$^*$ upper semicontinuous.  This completes
the proof that the maximum of the variational characterization of the
feedback capacity is achievable.

Finally we remark that the condition that $S_Z$ is bounded away from
zero is necessary.  As a simple example, if $S_Z(\eit) = |1+\eit|^2$,
it is shown in Section~\ref{sec:arma1} that the feedback capacity of
this noise spectrum corresponds to the output spectrum $S_Y$ of
the form
\[
S_Y(z) = \frac{1}{x^2}(1+x^2z)(1+x^2z^{-1}).
\] 
But we can easily check that there is no $(S_V, B)$ resulting in this
output spectrum.

\vspace*{1em}
\section*{Acknowledgment}
The author wishes to express his deepest gratitude towards Tom Cover
for his continual inspiration and encouragement.  He thanks Erik
Ordentlich, Stephen Boyd, and Persi Diaconis for enlightening
discussions and Sina Zahedi for his numerical optimization program
which was instrumental in the initial phase of this work.  He
also acknowledges interesting conversations with Sergio Verd\'u,
Sanjoy Mitter, Babak Hassibi, and Seung Jean Kim.

\def\cprime{$'$} \def\cprime{$'$} \def\cprime{$'$} \def\cprime{$'$}

\end{document}